\documentclass[12pt,draftclsnofoot,journal,onecolumn]{IEEEtran}
\pdfoutput = 1
\usepackage[T1]{fontenc}
\usepackage[latin9]{inputenc}
\usepackage{textcomp}
\usepackage{amsmath}
\usepackage{amsthm}
\usepackage{amssymb}
\usepackage{graphicx}
\usepackage{bm}
\usepackage{booktabs}
\usepackage{multirow}
\usepackage{rotating}
\usepackage{subfigure}
\usepackage{stfloats}
\usepackage[vlined,boxed,ruled,linesnumbered,resetcount]{algorithm2e}
\usepackage{xcolor}
\usepackage[noadjust,sort,compress]{cite}
\usepackage{setspace}
\usepackage{caption}
\begin{document}
\title{\huge Quasi-Synchronous Random Access for Massive MIMO-Based LEO Satellite Constellations}	
\author{Keke Ying, Zhen Gao, Sheng Chen,~\IEEEmembership{Fellow,~IEEE},\\
Mingyu Zhou, Dezhi Zheng,
Symeon Chatzinotas,~\IEEEmembership{Fellow,~IEEE},\\ Björn Ottersten,~\IEEEmembership{Fellow,~IEEE}, 
and H. Vincent Poor,~\IEEEmembership{Life Fellow,~IEEE}	
\thanks{Keke Ying, Zhen Gao, and Dezhi Zheng are with the School of Information and Electronics, Beijing Institute of Technology, Beijing 100081, China, and also with the Advanced Research Institute of Multidisciplinary Science (ARIMS), Beijing Institute of Technology, Beijing 100081, China (E-mail:ykk@bit.edu.cn; gaozhen16@bit.edu.cn; zhengdezhi@bit.edu.cn).} %
\thanks{Sheng Chen is with the School of Electronics and Computer Science, University of Southampton, Southampton SO17 1BJ, U.K. (E-mail: sqc@ecs.soton.ac.uk).} %
\thanks{Mingyu Zhou is with Baicells Technologies Co. Ltd., Beijing 100089, China (E-mail: zhoumingyu@baicells.com).}
\thanks{Symeon Chatzinotas and Björn Ottersten are with the Interdisciplinary Centre for Security, Reliability and Trust (SnT), University of Luxembourg, 1855 Luxembourg City, Luxembourg (E-mail: symeon.chatzinotas@uni.lu; bjorn.ottersten@uni.lu).}
\thanks{H. Vincent Poor is with the Department of Electrical and Computer Engineering, Princeton University, Princeton, NJ 08544 USA (E-mail: poor@princeton.edu).}
}
\maketitle

\begin{abstract}
Low earth orbit (LEO) satellite constellation-enabled communication networks are expected to be an important part of many Internet of Things (IoT) deployments due to their unique advantage of providing seamless global coverage. In this paper, we investigate the random access problem in massive multiple-input multiple-output-based LEO satellite systems, where the multi-satellite cooperative processing mechanism is considered. Specifically, at edge satellite nodes, we conceive a training sequence padded multi-carrier system to overcome the issue of imperfect synchronization, where the training sequence is utilized to detect the devices' activity and estimate their channels. Considering the inherent sparsity of terrestrial-satellite links and the sporadic traffic feature of IoT terminals, we utilize the orthogonal approximate message passing-multiple measurement vector algorithm to estimate the delay coefficients and user terminal activity. To further utilize the structure of the receive array, a two-dimensional estimation of signal parameters via rotational invariance technique is performed for enhancing channel estimation. Finally, at the central server node, we propose a majority voting scheme to enhance activity detection by aggregating backhaul information from multiple satellites. Moreover, multi-satellite cooperative linear data detection and multi-satellite cooperative Bayesian dequantization data detection are proposed to cope with perfect and quantized backhaul, respectively. Simulation results verify the effectiveness of our proposed schemes in terms of channel estimation, activity detection, and data detection for quasi-synchronous random access in satellite systems.
\end{abstract}
	
	
\section{Introduction}\label{S1}
As an indispensable component of the space-air-ground integrated network, low earth orbit (LEO) satellites have received extensive attention in the research of beyond fifth generation (B5G) and sixth generation (6G) mobile communication systems \cite{SAGIN_survey,ShanzhiChen, NTN, 3gpp.36.300}. Extensive efforts have been devoted to the construction of satellite constellations over the past few decades, for example, the Iridium system in the 1990s and Starlink LEO constellation projects more recently \cite{LSC_Mag}. With the evolution of space and communication technologies, satellite communication (SatCom) has extended from its original narrowband voice service to broadband multimedia service, which also brings more opportunities to the ubiquitous space-air-ground integrated connectivity.

As an early prototype system of global connectivity, narrowband IoT \cite{nb-iot} has enjoyed significant success in the long-term evolution of terrestrial networks (TNs), which has motivated considerable interest in moving from human-type communication to machine-type communication (MTC). Given the explosive growth of data traffic, advanced MTC applications will be more data-intensive, and the demands of advanced IoT-enabled applications will shift from low-rate short packet transmission to more rigorous low-latency, broadband, and reliable information interaction such as industrial Internet, smart cities, intelligent transportation, industrial metaverse, holographic communications, and so on \cite{IoT_app1}. On the other hand, remote and disaster areas also still face challenges in accessing the network due to the high cost of terrestrial infrastructure. Therefore, it is also of interest to promote broadband LEO satellite-enabled non-terrestrial networks (NTNs) \cite{LEO_IoT_Access} as a part of the communication infrastructure.

Due to the massive number of potential MTC user terminals (UTs) and the long propagation delay between ground and LEO satellites, it is inefficient to coordinate the channel resources for uplink access through traditional handshaking protocols. Grant free-random access (GF-RA) is a compelling paradigm for massive access in MTC since it allows UTs to directly transmit their respective preamble and payload data to base stations (BSs) without the aforehand handshaking process. Such a mechanism leads to the signal superimposition of a large number of potential UTs. In this process, identifying the active devices and estimating corresponding channels are important preconditions to assure successful data detection at the receiver. Therefore, performing the joint activity detection and channel estimation (JADCE) in the preamble stage and data detection (DD) in the data stage are two important problems.
	
	
\subsection{Related Works}\label{S1.1}
In the framework of GF-RA, the JADCE problem has been widely discussed for TNs. The non-orthogonal preamble transmission is an attractive solution in the design of GF-RA, as it reduces the long preamble overhead caused by a large number of potential UTs. By leveraging the sporadic transmission feature of UTs, \cite{LL_TSP} proposed to use the analytic framework of approximate message passing (AMP) in the JADCE problem, where the length of preamble can be reduced through compressed sensing. In \cite{Ke_JSAC}, the authors expanded the original JADCE problem from the single-cell narrowband case to the cell-free wideband case, where a structured-sparsity-based generalized AMP (SS-GAMP) algorithm was utilized to reduce the pilot overhead. An orthogonal AMP (OAMP) with multiple measurement vector (OAMP-MMV) algorithm was further proposed in \cite{OAMP_MMV}. By relaxing the assumption of an independently and identically distributed sensing matrix to a general unitary invariant one, the OAMP-MMV algorithm can be applied to a wider range of compressed sensing problems. Although a correlation-aware group Gram-Schmidt orthogonalization procedure was utilized to capture the spatial and temporal correlations in \cite{OAMP_MMV}, the matrix inversion in the corresponding nonlinear detection step is more computationally demanding. An alternative form of the OAMP-MMV algorithm was proposed in \cite{YK_Mei}, where a structured learning method was utilized to capture the activity state shared by different data symbols. Without the inversion operation, the OAMP-MMV algorithm of \cite{YK_Mei} is computationally simpler to implement. 

On the other hand, some works attempted to migrate existing 5G technologies from TNs to NTNs. In \cite{Zhang_IOTJ}, the JADCE was addressed for LEO satellite-based scenarios. By leveraging the sparsity of UTs' activities, the algorithm based on Bernoulli-Rician message passing with expectation-maximization was proposed to reduce the overhead. Different from \cite{Zhang_IOTJ} which assumed the single receive antenna at the satellite, the authors of \cite{YouLi_JSAC, KXLi_Tcomm} proposed to utilize massive multiple-input multiple-output (mMIMO) for enhancing the transmission performance in LEO satellite communications. More specifically, in \cite{YouLi_JSAC}, a low-complexity statistical channel state information (CSI)-based downlink precoder and uplink receiver were designed to improve the data rate of LEO satellite communication systems. Instead of employing a single antenna at the UT, the authors of \cite{KXLi_Tcomm} further employed a uniform planar array (UPA) at the UT, and they demonstrated that the single-stream precoding for each UT can maximize the ergodic sum rate for the linear transmitters. Furthermore, the densely deployed LEO satellite constellations can provide additional benefits, such as cooperative processing \cite{XJ_Ding}, data offloading \cite{JH_Zhao,offload}, and increased degrees of freedom of observations. In \cite{MSC_RA}, a multi-satellite cooperative random access scheme was proposed for the asynchronous system, where a wide range of ALOHA-based protocols can be applied. Nevertheless, the number of collided packets in one slot was restricted by the number of cooperative satellite nodes, and multiple repetitive packets were transmitted in the time domain, which reduced the transmission efficiency. \cite{DM_MIMO} proposed a distributed mMIMO scheme where multiple satellite access points cooperate to serve the ground UTs, and one central super satellite with advanced computing capabilities coordinated resources for throughput and handover rate optimization. In general, \cite{Zhang_IOTJ, MSC_RA,DM_MIMO} assumed a single antenna for the satellites during multi-user access, which can not fully leverage the spatial domain advantages of LEO constellations. Although sophisticated algorithms have been designed in \cite{LL_TSP,Ke_JSAC, OAMP_MMV, YK_Mei, Zhang_IOTJ} for random access, the common assumption of synchronous arrival is no longer valid in LEO-based communication systems due to the vast distributed areas of UTs. The authors of \cite{XY_Zhou} considered an orthogonal time frequency space paradigm in a single LEO satellite system, where the UTs can access the satellite with imperfect frame synchronization, and the problems of activity detection (AD), channel estimation (CE), and data detection (DD) were comprehensively investigated. However, the angular domain feature of TSL was not exploited and an additional user grouping mechanism was required to combat the high correlation among UTs' channels when performing DD in a single satellite system.  
	
To avoid troublesome scheduling mechanisms among groups \cite{UserGrouping} and to exploit the advantages of mMIMO in LEO satellite constellations, we proposed to coordinate multiple mMIMO-based LEO satellites for improved system performance. In our paper, we investigate the JADCE problem in a quasi-synchronous random access mechanism. Furthermore, considering the limited computation capability of satellites and the need to feed back information to the core network, we will also discuss the multi-satellite cooperative DD mechanism.

\subsection{Our Contributions}\label{S1.2}
	
Encouraged by the aforementioned developments in TNs and NTNs, we investigate the JADCE and DD problems in LEO satellite constellation networks. Motivated by \cite{Zhen_TDS, Zhen_BC, MSC_RA} We propose a training sequence padded (TSP) multi-carrier system (TSP-MCS), where the training sequence (TS) is utilized to perform multipath interference cancellation (MIC) and JADCE. Specifically, the sporadic transmission feature of uplink packets and sparsity of TSLs in the delay domain are exploited by an OAMP-MMV algorithm to obtain an initial estimate of the channel impulse response (CIR) and active user set (AUS). Then, a two-dimensional estimation of signal parameters via rotational invariance technique (2D-ESPRIT) algorithm \cite{2DESPRIT} is further applied to obtain a high-resolution estimate of the angle of arrivals (AoAs), which can be used for subsequent parametrized CE refinement. Finally, by aggregating the processing results from edge satellite nodes, the central server node performs cooperative AD and DD for enhanced performance. 

The main contributions of this paper can be summarized as follows.
	
$\bullet$~\textbf{TSP-MCS for LEO satellite-based IoT:} We use the TSP frame structure to perform MIC and JADCE. Specifically, a TS is designed to be longer than the maximum relative channel delay (RCD) among different UTs. The front part of the TS is used to combat multipath interference (caused by imperfect synchronization) among different UTs, while the remaining non-inter-symbol-interference (non-ISI) region of the TS is utilized to perform JADCE.

$\bullet$~\textbf{OAMP-MMV algorithm for JADCE and parameterized CE enhancement:} We use the OAMP-MMV algorithm for initial JADCE, where the AUS and CIR are acquired by exploiting the delay domain sparsity and sporadic traffic feature. Since the TSLs also exhibit high directivity in the angular domain, a 2D-ESPRIT algorithm is further applied to capture the spatial structure of the receiver antenna array. Benefiting from the accurate estimation of AoAs, a parametrized estimation and the reconstruction of the channel are realized with enhanced performance.

$\bullet$~\textbf{Cooperative multi-satellite AD and DD:} We adopt a diversity transmission scheme to address the adverse TSL circumstances, where multi-stream transmission is adopted at the UT side and an augmented AD is performed at the central server node \footnote{To distinguish different functionalities, we refer to satellites as edge nodes and the server as the central node. To emphasize the uplink transmission, we use the `backhaul link' to equivalently represent the term `feeder link' in traditional SatCom.} by aggregating the estimated AUSs from edge LEO satellites. We propose to use observations from multiple LEO satellites and adopt a cooperative DD at the central node to alleviate the high channel correlation problem in a single satellite system. Furthermore, considering the signaling cost of backhaul links, we propose a multi-satellite cooperative linear DD algorithm and a multi-satellite cooperative Bayes dequantization DD algorithm to cope with perfect and quantized backhaul, respectively.
	
The rest of this paper is organized as follows. Section~\ref{S2} describes the channel model and transmission process. Section~\ref{S3} delves into the JADCE and DD in a single satellite receiver, which includes the OAMP-MMV algorithm-based JADCE, 2D-ESPRIT-based CE refinement, and single satellite DD algorithm. The cooperative AD and DD for multi-satellite systems are covered in Section~\ref{S4}, where both perfect feedback and quantized feedback are discussed. The complexities of proposed schemes are analyzed in Section~\ref{S_add} and simulation results are presented in Section~\ref{S5}. Finally, our conclusions are summarized in Section~\ref{S6}.
	
\textit{Notation}: Matrices and column vectors are denoted by uppercase and lowercase boldface letters, respectively. $\bm{I}_{N}$ is the $N\times N$ identity matrix, and $\bm{0}_{N\times M}$ is the $N\times M$ all-zero matrix. $(\cdot)^{\rm T}$, $(\cdot)^{*}$, $(\cdot)^{\rm H}$, $(\cdot)^{-1}$, $\mathrm{E}\{\cdot\}$, and $\mathrm{Var}\{\cdot\}$ denote the transpose, conjugate, Hermitian transpose, inversion, expectation, and variance operations, respectively. The $\left(i,j\right)$-th entry of matrix $\bm{X}$ is denoted as $[\bm{X}]_{i,j}$. The $i$-th row and $j$-th column of $\bm{X}$ are denoted by $[\bm{X}]_{[i,:]}$ and $[\bm{X}]_{[:,j]}$, respectively. $[\bm{X}]_{[\mathcal{I},:]}$ denotes the sub-matrix consisting of the rows of $\bm{X}$ whose indexes are specified by the index set $\mathcal{I}$. The $i$-th element of $\bm{a}$ is denoted by $[\bm{a}]_i$. $\left\Vert \bm{X}\right\Vert _{\rm F}$ denotes the Frobenius norm of $\bm{X}$. Vector convolution operation is denoted by $\star$, $\otimes$ denotes the Kronecker product operator and $\circ$ represents the outer product operator. $\mathrm{supp}\{\cdot\}$ is the support set of a vector. $\mathrm{diag}(\bm{a})$ transforms vector $\bm{a}$ into the corresponding diagonal matrix, and $\langle \bm{A}\rangle$ computes the mean of the diagonal entries of matrix $\bm{A}$. $|\mathcal{A}|$ is the cardinality of the set $\mathcal{A}$. $\mathcal{A}\backslash\mathcal{B} = \left\{x|x\in\mathcal{A}, x \notin \mathcal{B}\right\}$ is the difference set operation. $\mathcal{N}_{c}\left(\bm{x};\bm{a},\bm{A}\right)$ denotes the probability density function (PDF) of a Gaussian random vector $\bm{x}$ with mean $\bm{a}$ and covariance matrix $\bm{A}$. 
	
\section{Preliminaries}\label{S2}

In this section, we first illustrate the processing framework in LEO constellation-based random access. Then the system model and sparsity properties of TSLs are illustrated. Finally, the detailed transmission procedure in the LEO constellation network is explained. 
	
\subsection{Processing Framework}\label{sec.framework} 
	
Depending on the signal processing ability, the satellites in an LEO constellation can be classified as bent-pipe satellites or regenerative satellites \cite{ShanzhiChen}. A bent-pipe satellite serves as a relay to transparently forward signals from UTs to the central server node, while a regenerative satellite acts as a BS with signal modulation and demodulation ability. With the development of satellite technology and space communication networks, regenerative satellites with signal processing ability are preferred in LEO constellation, so that the computation task can be offloaded to satellites \cite{offload} and the communication delay can be decreased correspondingly. According to different levels of cooperation, the processing framework in a regenerative satellites-enabled communication network can be further divided into three categories. 
	
$\bullet$~\textit{Single satellite non-cooperative processing (SSNCP):} With adequate computing resources, each satellite performs JADCE and DD independently.
	
$\bullet$~\textit{Multi-satellite cooperative terrestrial processing (MSCTP):} After performing CE and AD at each satellite, necessary information (e.g., the pre-processed received signal, estimated CIR, and AUS) is fed back to the terrestrial server for further DD processing.
	
$\bullet$~\textit{Multi-satellite cooperative onboard processing (MSCBP):} The satellite constellation is heterogeneous, and one specific satellite with high computation power can act as an onboard server. After performing CE and AD at each edge satellite, necessary information (e.g., the pre-processed received signal, estimated CIR, and AUS) is fed back to this onboard server for further DD processing.
	
As mentioned later in Section~\ref{S4}, SSNCP suffers from performance deterioration when the channel correlation (caused by similar AoAs) between two geographically adjacent active UTs is sufficiently high. By contrast, the cooperative processing schemes, including MSCTP and MSCBP, can alleviate this problem through diversity gain. In the sequel, we focus on the cooperative processing mechanism, where the JADCE with channel estimation refinement is performed at edge nodes to provide a reliable CE and an initial AUS estimation. Then enhanced AD and DD are performed at the central server by fusing the information from multiple edge nodes. 

\begin{figure*}[!tp]
\centering
\captionsetup{font={footnotesize}, singlelinecheck = off,name={Fig.},justification=centering, labelsep=period}
\includegraphics[width=0.6\textwidth]{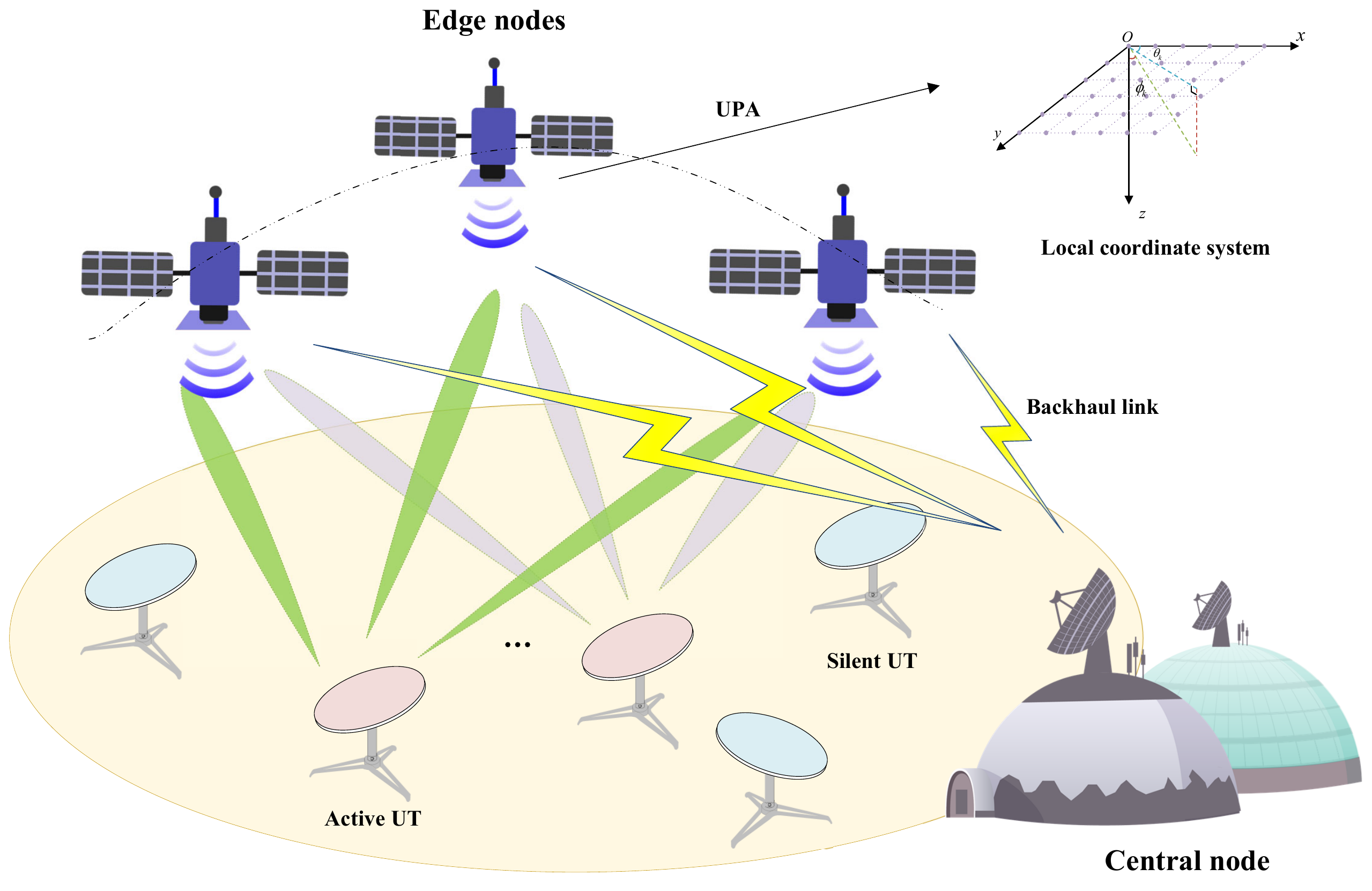}
\caption{Schematic diagram of random access in mMIMO-based LEO constellation for MTC.}
\label{fig1_scene} 
\vspace{-6mm}
\end{figure*}
	
\subsection{System Model}\label{S2.2}

Consider the uplink random access scenario in an LEO constellation network, where $Q$ LEO satellites simultaneously provide access service to $K$ UTs in the same coverage area. As shown in Fig.~\ref{fig1_scene}, each LEO satellite is equipped with an $N_r = N_x \times N_y$ UPA with $N_x$ and $N_y$ antenna elements on the x-axis and y-axis, respectively, and the target UTs are distributed at the intersection area of the LEO satellites' footprint. Target devices can access the satellite network via very small aperture terminals (VSATs). Advanced VSAT may adopt a phased array antenna with multi-stream beamforming capability, e.g., the Starlink terminal \cite{Starlink}. Hence, we assume that each UT is equipped with $Q$ radio frequency (RF) chains and $N_t$ antennas to provide multi-stream beamforming capability, and the same data packets are forwarded through different RF chains to provide diversity gain. For a specific coherence time interval, only $K_a$ UTs are active owing to the sporadic traffic feature in MTC.
	
We focus on the channel model between a specific UT and an LEO satellite. The target UT on the ground is usually surrounded by limited scatterers within a radius of $r_d$. Let the distance between the UT and the LEO satellite be $R_d$. The angle spread $\Delta = \arctan \left(r_{d}/R_{d}\right)$ observed from the LEO satellite is very small since $R_d \gg r_d$. Consequently, multipath components in TSL share the same AoA but different delays. The delay-domain mMIMO channel matrix $\bm{G}_{k,q}[l] \in \mathbb{C}^{N_r\times N_t}$ between the $k$-th UT and the $q$-th satellite at the $l$-th tap can be expressed by \cite{3gpp.38.811} 
\begin{equation}\label{Channel0} 
 \bm{G}_{k,q}[l] = \alpha_k \sum_{p=0}^{P-1} \beta_{k,p,q}[l] \bm{a}_{r}\left(\theta_{k,q},\phi_{k,q}\right) \bm{a}_{t}^{\rm H}\left(\vartheta_{k,p,q},\varphi_{k,p,q}\right) ,
\end{equation}
where $P$ is the number of paths, and the binary indicator $\alpha_k \in \{0,1\}$ defines the active status of the UT. By defining $\mathcal{A} = \{k\, \vert \, \alpha_k = 1, 1\le k \le K \}$, the number of active UTs is given by $K_a = |\mathcal{A}|$. Typically, $K_a \ll K$, since MTC exhibits a sporadic traffic feature. In (\ref{Channel0}),	$\beta_{k,p,q}[l] = \varrho_{k,p,q} r\big(l T_s-\tau_{k,p,q}\big)$ is the delay-domain channel coefficient between the $k$-th UT and the $q$-th satellite in the $p$-th path, in which $\varrho_{k,p,q}$, $\tau_{k,p,q}$, and $r(\tau)$ are the complex path gain, RCD\footnote{Timing advance mechanisms \cite{3gpp.38.811} are often used to ensure the transmit packets from all active UTs in the same cell are approximately synchronized when received by the BS. Generally, the timing advance cannot be perfectly performed for SatCom, and hence a residual delay exists.}, and pulse shaping filter with $T_s$-spaced signaling, respectively. Furthermore, $\bm{a}_{r}\left(\theta_{k,q},\phi_{k,q}\right)\in \mathbb{C}^{N_r}$ and $\bm{a}_{t}\left(\vartheta_{k,p,q},\varphi_{k,p,q}\right)\in \mathbb{C}^{N_t}$ are the steering vectors at the $q$-th satellite and the $k$-th UT, respectively, which are detailed as follows based on some reasonable assumptions.
	
$\bullet$~\textit{Array manifold at the satellite side:} For the $k$-th UT and $q$-th satellite, we denote the steering vectors with respect to $x$-axis and $y$-axis by $\bm{v}_k^{s}\left(\theta_{k,q},\phi_{k,q}\right) =\left[1 ~ e^{-\textsf{j}\mu_{k,q}^{s}} \cdots e^{-\textsf{j}\left(N_s-1\right)\mu_{k,q}^{s}}\right]^{\rm T}$, $s\in \{x,y\}$. For the coordinate system in Fig.~\ref{fig1_scene}, $\mu_{k,q}^{x} = \frac{2\pi d}{\lambda}\cos\theta_{k,q} \sin\phi_{k,q}$ and $\mu_{k,q}^{y} = \frac{2\pi d}{\lambda}\sin\theta_{k,q} \sin\phi_{k,q}$, where $\lambda$ is the wavelength of the carrier frequency, $d = {\lambda}/{2}$ is the adjacent antenna spacing, and $\theta_{k,q}$ and $\phi_{k,q}$ are the azimuth AoA and elevation AoA from the $k$-th UT to $q$-th satellite, respectively. Therefore, the steering vector at the satellite side can be expressed by
\begin{equation}\label{SteerVector}	
 \bm{a}_{r}\left(\theta_{k,q},\phi_{k,q}\right) = \bm{v}_{k}^{y}\left(\theta_{k,q},\phi_{k,q}\right) \otimes \bm{v}_{k}^{x}\left(\theta_{k,q},\phi_{k,q}\right) .
\end{equation}
	
$\bullet$~\textit{Array manifold at the UT side:} Let $\vartheta_{k,p,q}$ and $\varphi_{k,p,q}$ be the azimuth angle of departure (AoD) and elevation AoD from the $k$-th UT to the $q$-th satellite in the $p$-th path, respectively. Similar to (\ref{SteerVector}), the transmit steering vector is defined as $\bm{a}_{t}\left(\vartheta_{k,p,q},\varphi_{k,p,q}\right) = \bm{v}_{k}^{y}\left(\vartheta_{k,p,q},\varphi_{k,p,q}\right) \otimes \bm{v}_{k}^{x}\left(\vartheta_{k,p,q},\varphi_{k,p,q}\right)$. Since the number of RF chains for each UT is equal to the number of satellites,  we assume that the signal of the $q$-th stream is transmitted to the $q$-th satellite. The beamforming vector for the $k$-th UT at the $q$-th RF chain is denoted by $\bm{w}_{k,q}\in \mathbb{C}^{N_t}$. Considering the known UT's position and deterministic LEO's trajectory, we further assume that $\vartheta_{k,0,q}$ and $\varphi_{k,0,q}$ for the line-of-sight (LOS) path are available at the transmitter. By constructing a match filtering beamformer $\bm{w}_{k,q}\left(\vartheta_{k,0,q},\varphi_{k,0,q}\right) = \bm{a}_{t}\left(\vartheta_{k,0,q},\varphi_{k,0,q}\right)$, the equivalent channel vector $\bm{g}_{k,q}[l] = \bm{G}_{k,q}[l] \bm{w}_{k,q}\left(\vartheta_{k,0,q},\varphi_{k,0,q}\right)\in \mathbb{C}^{N_r}$ from the $k$-th UT's $q$-th RF chain to the $q$-th satellite can be obtained as
\begin{equation}\label{Channel1} 
 \bm{g}_{k,q}[l] = \alpha_k \sum_{p=0}^{P-1} \tilde{\beta}_{k,p,q}[l] \bm{a}_{r}\left(\theta_{k,q},\phi_{k,q}\right),
\end{equation}
where $\tilde{\beta}_{k,p,q}[l]  = \tilde{\varrho}_{k,p,q} r\big(l T_s-\tau_{k,p,q}\big)$, and $\tilde{\varrho}_{k,p,q} = \varrho_{k,p,q} \bm{a}_{t}^{\rm H}\left(\vartheta_{k,p,q},\varphi_{k,p,q}\right) \bm{w}_{k,q}\left(\vartheta_{k,0,q},\varphi_{k,0,q}\right)$. Generally, the value of $\tilde{\varrho}_{k,p,q}$ is decided by the array manifold at the UT side and the difference between the angle pair $\left(\vartheta_{k,p,q},\varphi_{k,p,q}\right)$ and $\left(\vartheta_{k,0,q},\varphi_{k,0,q}\right)$. The phased array antenna adopted by UT usually has a large number of antennas, e.g., the VSAT for Starlink has nearly one thousand antenna elements. Therefore, the transmit beam for each data stream can exhibit high spatial directionality. Without loss of generality, we simplify the modeling of the equivalent channel coefficients as $\tilde{\varrho}_{k,0,q} = \sqrt{K_{f}/\left(K_{f}+1\right)}$ for the LOS path, and $\tilde{\varrho}_{k,p,q} \sim \sqrt{1/\left[\left(K_{f}+1\right)\left(P-1\right)\right]} \mathcal{N}_{c}(\tilde{\varrho};0,1)$, $\forall p\ne 0$, for the non-line-of-sight (NLOS) paths, where $K_{f}$ is the power distribution factor and $P>1$. When $P = 1$, it can be equivalently seen as only the LOS path exists and the $K_f$ goes to infinity. An increase in the number of transmit antennas will lead to a higher $K_{f}$. Furthermore, the interference to undesired satellites also decreases to zero due to the asymptotic orthogonality of mMIMO channels, i.e., $\lim\limits_{N_t\rightarrow \infty}\bm{a}_{t}^{\rm H}\left(\vartheta_{k,p,q^{\prime}},\varphi_{k,p,q^{\prime}}\right)\bm{w}_{k,q}\left(\vartheta_{k,0,q},\varphi_{k,0,q}\right) = 0$, $\forall q^{\prime} \neq q$.
	
Let the maximum RCD be $L$. The overall CIR matrix (CIRM) between the $k$-th UT and the $q$-th satellite can be constructed as $\bm{H}_{k,q} = \left[\bm{g}_{k,q}[0] \cdots \bm{g}_{k,q}[L-1]\right]^{\rm T}\in \mathbb{C}^{L\times N_r}$ or equivalently as
\begin{equation}\label{Channel2} 
 \bm{H}_{k,q} = \alpha_k \bm{h}_{k,q} \circ \bm{a}_{r}\left(\theta_{k,q},\phi_{k,q}\right) , 
\end{equation}
where $\bm{h}_{k,q}\in \mathbb{C}^{L}$ is the CIR from the $k$-th UT to the $q$-th satellite's first receive antenna (reference antenna), whose $l$-th tap is given by ${h}_{k,q}\left[l\right] = \sum_{p=0}^{P-1}\tilde{\beta}_{k,p,q}[l]$. Note that (\ref{Channel2}) is an accurate approximation of the TSL channel due to the facts that: 1)~All the paths from the same UT have almost the same AoAs at the satellite; 2)~All the antennas nearly receive the signal simultaneously since the maximum arrival time difference between different receive antennas is far less than one symbol duration; 3)~Other impairments such as Doppler and coarse delay difference are compensated.
	
\subsection{Transmission Procedure}\label{S2.3}
\begin{figure*}[!htp]
	\centering
	\captionsetup{font={footnotesize}, singlelinecheck = off,name={Fig.},justification=centering, labelsep=period}
	\includegraphics[width=0.9\textwidth]{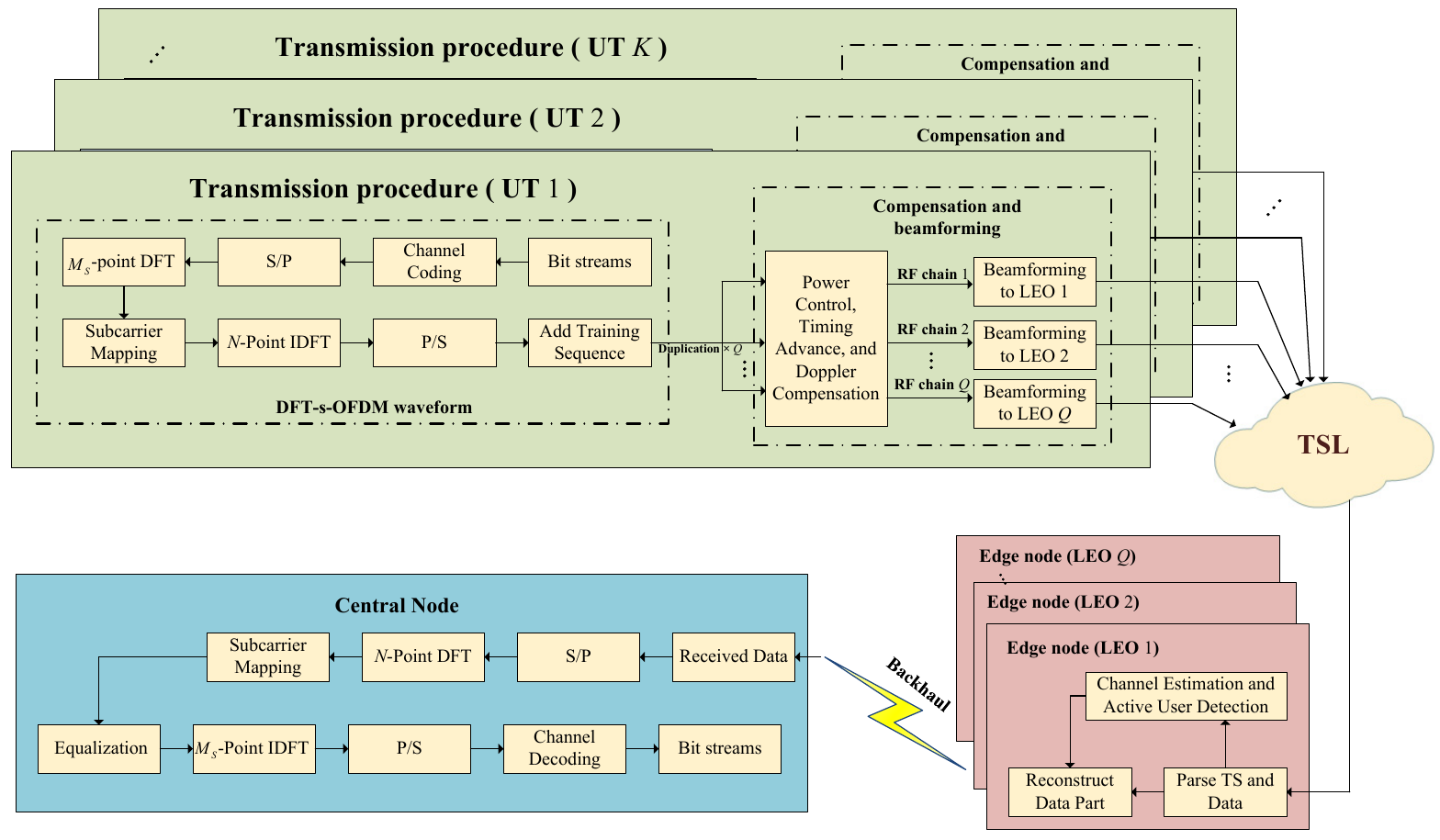}
	\caption{Block diagram of the proposed diversity transmission and receiving scheme.}
	\label{fig2_diagram} 
\end{figure*}

In the proposed TSP-MCS, a typical time-domain TSP frame consists of a known pseudo-noise TS $\bm{c}=\left[c_0 ~ c_1 \cdots c_{M-1} \right]^{\rm T}$ followed by a data block $\bm{x} =\left[x_0 ~ x_1 \cdots x_{N-1}\right]^{\rm T}$. To achieve a low peak-to-average-power-ratio, DFT-s-OFDM waveform is utilized to assemble the data block $\bm{x}$, as shown in Fig.~\ref{fig2_diagram}. During the transmission, the TS and data block constitute a TSP frame $\tilde{\bm{s}} = \left[\bm{c}^{\rm T} ~ \bm{x}^{\rm T} \right]^{\rm T}$ in the time domain. Before transmitting the source packets to different target LEO satellites, the TSP frame is duplicated to multiple copies, with each stream performing individual timing advance and power control so that balanced received energy can be achieved at the receiver. Besides, the Doppler is assumed to be compensated at the transmitter \footnote{The Doppler shifts can be compensated at the UT side due to the known UT's position and deterministic LEO's trajectory\cite{Doppler_Comp1,Doppler_Comp2}. If the residual Doppler shift can be controlled within a certain range, the equivalent channel coherence time can be very long, which means the effects on JADCE and DD are small.} and its residual effect can be small. Then, analog beamforming is performed at each RF chain independently to produce a directional beam for each LEO satellite. Specifically, for the $q$-th satellite, the received signal at the $j$-th antenna from the $k$-th user is the convolution between the $q$-th transmit stream $\bm{s}_{k,q}=\left[\tilde{\bm{s}}_{k,q,t}^{\rm T} ~ \tilde{\bm{s}}_{k,q,t+1}^{\rm T} \cdots \right]^{\rm T}$ and the corresponding CIR vector $\left[\bm{H}_{k,q}\right]_{[:,j]} \in \mathbb{C}^{L}$, where $\tilde{\bm{s}}_{k,q,t} = \left[\bm{c}_{k}^{\rm T} ~ \bm{x}_{k,q,t}^{\rm T} \right]^{\rm T}\in\mathbb{C}^{M+N}$ is the $t$-th TSP frame at the $q$-th RF chain of the $k$-th UT. By summing up the signals from all the UTs, the received signal stream $\bm{y}_{q}^{j}$ is given by
\begin{equation}\label{RxS} 
 \bm{y}_q^j = \sum_{k=1}^{K} \bm{s}_{k,q} \star \left[{\bm{H}}_{k,q}\right]_{[:,j]} + \bm{n}_q^{j}, ~ \forall j ,q ,
\end{equation}
where $\bm{n}_q^{j}$ is the corresponding symmetric complex additive Gaussian white noise (AWGN) stream at the $j$-th receive antenna, whose entries follow an independent and identical distribution with zero mean and variance $\sigma_{n}^2$. The whole transmission process is exhibited in Fig. \ref{fig2_diagram} and some key system parameters are listed in Table~\ref{tab:para}. Based on the received signals $\bm{y}_{q}^{j}$, $\forall j,q$, the receiver needs to estimate the AUS and corresponding CIR, and then demodulate the information bits.

\begin{table}[h]
	\centering
	\captionsetup{justification = raggedright,labelsep=period}
	\caption{Parameter list}	
	\label{tab:para}
	\resizebox{.6\columnwidth}{!}{
		\renewcommand\arraystretch{1.2}{%
			\begin{tabular}{c|l}
				\hline
				\textbf{Notation} & \multicolumn{1}{c}{\textbf{Defination}} \\ \hline
				$N_r$,$N_t$ & Number of antennas for satellite and UT \\ \hline
				$K$,$K_a$ & Number of potential and active terminals \\ \hline
				$\alpha_k$,$\mathcal{A}$ & Activity indicator and active terminal set \\ \hline
				$ Q $ & Number of satellites and RF chains at UT \\ \hline
				$ P $ & Number of paths \\ \hline
				$ \tilde{\beta}_{k,p,q}$ & Equivalent complex path gain \\ \hline
				$\tau_{k,p,q}$ & Relative path delay \\ \hline
				$\theta_{k,q}$, $\phi_{k,q}$ & Azimuth and elevation AoAs of satellite \\ \hline
				$\bm{a}_r(\theta_{k,q},\phi_{k,q})$ & Steering vector at the satellite \\ \hline
				$\bm{g}_{k,q}[l]$, $\bm{h}_{k,q}$ & Equivalent channel vector and channel impulse response \\ \hline
				$\bm{H}_{k,q}$  & Channel impulse response matrix \\ \hline
				$M$, $L$, $G$& Length of TS, CIR, and  non-ISI region\\ \hline
				$ N $  & Number of subcarriers \\ \hline
				$\tilde{\mathbf{s}}_{k,q,t}$, $\mathbf{c}_k$, $\mathbf{x}_{k,q,t}$ & \begin{tabular}[c]{@{}l@{}} Transmitted time domain frame, training sequence, \\ and transmitted data block\end{tabular} \\ \hline
				$\bm{Y}_{q,t}^{\rm TS}$,$\bm{Y}_{q,t}^{\rm D}$ & Received training sequence part and data part \\ \hline
				$\mathring{\bm{X}}_t$, $\mathring{\bm{X}}_f$, & Detected time, frequency domain data \\ \hline
			\end{tabular}%
	}}
\end{table}

\section{Massive Access Processing at Single Satellite}\label{S3}

In this section, we first formulate the JADCE at each satellite node as a compressed sensing problem, where the OAMP-MMV algorithm is utilized to estimate the CIR and AUS. Then the AoA of each TSL path is estimated by the 2D-ESPRIT algorithm \cite{2DESPRIT}, from which an enhanced CE performance is achieved. Finally, the DD method for a single satellite receiver is elaborated.

\begin{figure*}[tbp]
\captionsetup{font={footnotesize}, singlelinecheck = off,name={Fig.},justification=centering, labelsep=period}
\includegraphics[width=0.7\textwidth]{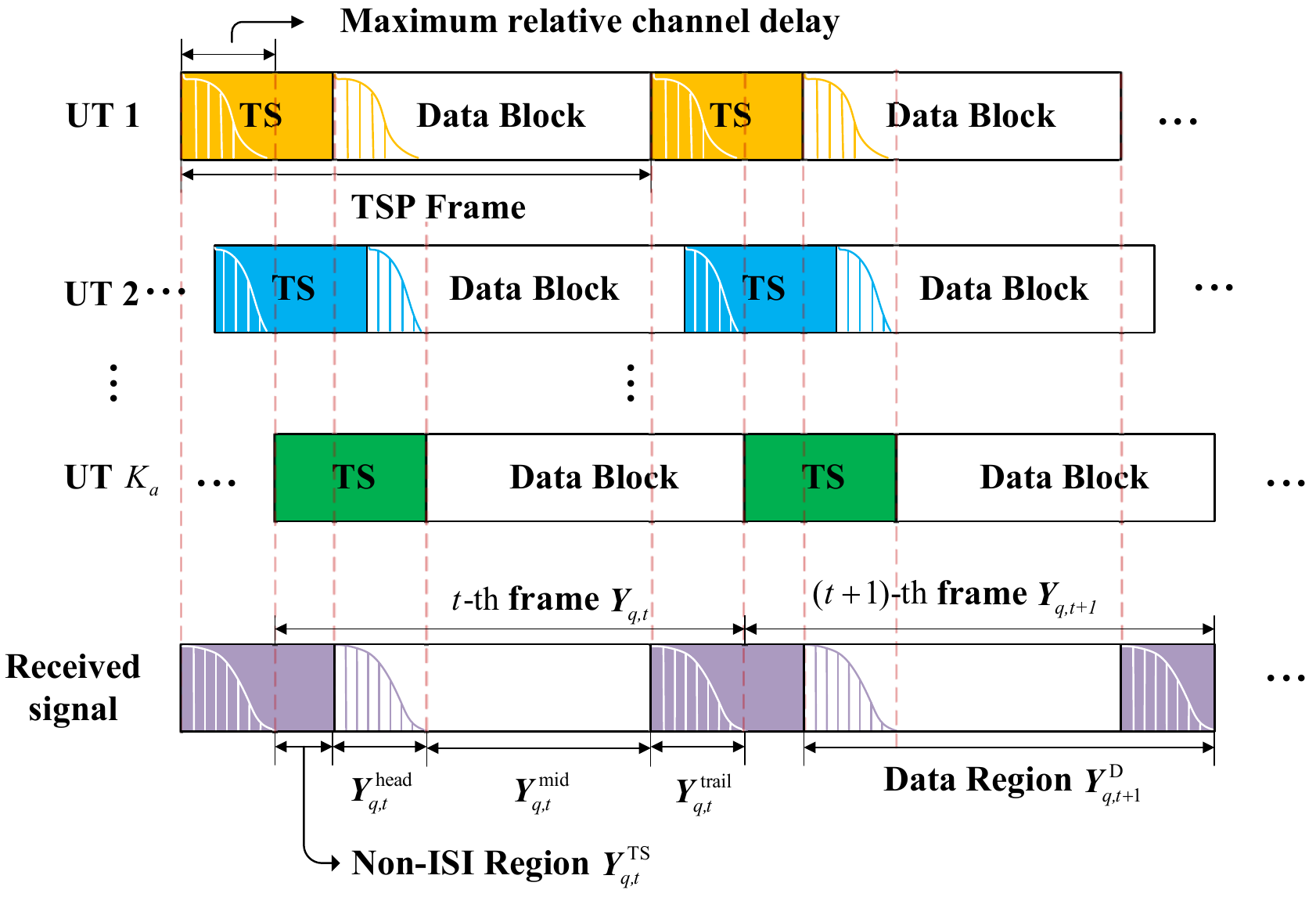}
\centering
\caption{Illustration of the proposed TSP frame structure for LEO satellite-based random access.}
\label{fig3_frame} 
\vspace{-7mm}
\end{figure*}
	
\subsection{Signal Formulation of JADCE}\label{S3.1}

The received frame structure is illustrated in Fig.~\ref{fig3_frame}. The TSP-MCS system is designed to allow a maximum RCD of $L T_s$. The TS length $M$ is designed to be longer than $L$, and a small non-ISI region of length $G = M-L+1$ at the receiver of the $q$-th satellite can be exploited for JADCE. In this non-ISI region, by converting the convolution operation into matrix multiplication, the received signal $\bm{y}_{q,\mathrm{TS}}^{j}\in \mathbb{C}^{G}$ at the $j$-th antenna from all the UTs can be rewritten as
\begin{equation}\label{RxSnon-ISI} 
 \bm{y}_{q,{\rm TS}}^j = \sum_{k = 1}^K \bm{\Psi}_k \left[\bm{H}_{k,q}\right]_{[:,j]} + \bm{n}_{q,{\rm TS}}^j ,
\end{equation}
where for notation simplicity the frame index $t$ is dropped, and $\bm{n}_{q,\mathrm{TS}}^{j}\in \mathbb{C}^{G}$ is the AWGN vector in the non-ISI region, while $\bm{\Psi}_k\in \mathcal{S}^{G\times L}$ is a Toeplitz matrix with each of its entries generated from a feasible TS symbol set $\mathcal{S}$, which is given by 
\begin{equation}\label{TxSnon-ISI} 
 \bm{\Psi}_{k} = \left[ \begin{matrix}
			c_{k,L-1} & c_{k,L-2} & c_{k,L-3} & \cdots & c_{k,0} \\
			c_{k,L}   & c_{k,L-1} & c_{k,L-2} & \cdots & c_{k,1} \\
			\vdots    & \vdots    & \vdots    & \vdots & \vdots \\
			c_{k,M-1} & c_{k,M-2} & c_{k,M-3} & \cdots & c_{k,M-L}
 \end{matrix} \right] . 
\end{equation}

By stacking $\bm{\Psi} = \left[\bm{\Psi}_{1} ~ \bm{\Psi}_{2}\cdots \bm{\Psi}_{K}\right]\in\mathbb{C}^{G\times K L}$ and $\tilde{\bm{H}}_{q} = \left[\bm{H}_{1,q}^{\rm T} ~ \bm{H}_{2,q}^{\rm T}\cdots \bm{H}_{K,q}^{\rm T}\right]^{\rm T}\in \mathbb{C}^{K L\times N_r}$, the received signal $\bm{Y}_{q}^\mathrm{TS} = \left[\bm{y}_{q,\mathrm{TS}}^{1} ~ \bm{y}_{q,\mathrm{TS}}^{2}\cdots \bm{y}_{q,\mathrm{TS}}^{N_r}\right]\in \mathbb{C}^{G\times N_r}$ at the $q$-th satellite can be written as
\begin{equation}\label{equ.csmodel} 
 \bm{Y}_{q}^\mathrm{TS} = \bm{\Psi}\tilde{\bm{H}}_{q} + \bm{N}_{q}^{\mathrm{TS}}, ~ \forall q , 
\end{equation}
where $\bm{N}_{q}^{\mathrm{TS}} = \left[\bm{n}_{q,\mathrm{TS}}^{1} \cdots \bm{n}_{q,\mathrm{TS}}^{N_r}\right]\in \mathbb{C}^{G\times N_r}$. The overall channel matrix $\tilde{\bm{H}}_{q}$ in (\ref{equ.csmodel}) exhibits the structured sparsity. Specifically, considering the sparsity of UTs' activity and delay-domain CIR, all the columns of $\tilde{\bm{H}}_{q}$ share the common support set. In other words, the channel vectors at all the receive antennas show the same sparsity pattern, i.e.,
\begin{equation}\label{CommonSupport} 
 \mathrm{supp}\left\{\left[\tilde{\bm{H}}_{q}\right]_{[:,1]}\right\} = \cdots = \mathrm{supp}\left\{\left[\tilde{\bm{H}}_{q}\right]_{[:,N_r]}\right\}.
\end{equation}
	
\subsection{OAMP-Based JADCE at Edge Satellite Node}\label{sec.oamp} 

OAMP is an efficient iterative algorithm for compressed sensing \cite{OAMP}. The conventional OAMP algorithm recursively proceeds between a linear estimation (LE) module and a non-linear (NLE) module. By restricting a de-correlated estimator in the LE module and a divergence-free estimator in the NLE module, the input and output errors become orthogonal. For notation simplicity, in this subsection, we simplify the $j$-th column of (\ref{equ.csmodel}), $\bm{y}_{q,\mathrm{TS}}^{j} = \bm{\Psi}\left[\tilde{\bm{H}}_q\right]_{[:,j]} + \bm{n}_{q,\mathrm{TS}}^{j}$, to the form $\bm{y}_{j} = \bm{\Psi}\bm{h}_{j}+\bm{n}_j$. Starting with $\bm{d}_{j}^{0} = \bm{0}_{KL\times 1}$, the optimal OAMP structure with single measurement vector in the $\iota$-th iteration for such a problem is given by \cite{OAMP}
\begin{subequations}
\begin{align}
 \label{oamp_le} 
 \text{LE:}\; \bm{r}_{j}^{\iota} &= \eta_{\iota}\left(\bm{d}_{j}^{\iota}\right) = \bm{d}_{j}^{\iota} + \bm{W}_{\iota}\left (\bm{y}_{j} - \bm{\Psi} \bm{d}_{j}^{\iota} \right), \\
 \label{oamp_nle} 
 \text{NLE:}\; \bm{d}_{j}^{\iota + 1} &= \phi_{\iota}\left(\bm{r}_{j}^{\iota}\right) = C_{\iota}\left(\hat{\phi}_{\iota}\left(\bm{r}_{j}^{\iota}\right) - \mathrm{E}\left\{\hat{\phi_{\iota}^{\prime}}\right\} \bm{r}_{j}^{\iota}\right) .
\end{align}
\end{subequations}

For the LE, $\bm{W}_{\iota}\! =\! \frac{KL}{\mathrm{tr}\left(\hat{\bm{W}}_{\iota}\bm{\Psi}\right)}\hat{\bm{W}}_{\iota}$ is a de-correlated estimator with an optimal linear minimum mean square estimation (LMMSE) $\hat{\bm{W}}_{\iota}\! =\! v_{\iota}\bm{\Psi}^{\rm H}\left(v_{\iota}\bm{\Psi}\bm{\Psi}^{\rm H}\! +\! \sigma_n^2\bm{I}_G\right)^{-1}$, where $v_{\iota}\! =\! \frac{1}{KL}\mathrm{E}\left\{\|\bm{d}_{j}^{\iota}\! -\! \bm{h}_{j}\|_2^{2}\right\}$ is the error measure between the output of the NLE and the true signal $\bm{h}_{j}$. For the NLE, $\phi_{\iota}\left(\cdot\right)$ is an element-wise divergence-free function, and the minimum mean square estimation (MMSE) $\hat{\phi_{\iota}}$ is given by
\begin{equation}\label{mmse_mean} 
 \xi_{i,j}^{\iota} = \hat{\phi}_{\iota}\left(r_{i,j}^{\iota}\right) = \mathrm{E}\left\{h_{i,j}|r_{i,j}^{\iota} = h_{i,j}+\sqrt{\tau_{\iota}} Z\right\}, ~ \forall i,
\end{equation}
where $\tau_{\iota} = \frac{1}{KL}\mathrm{E}\left\{\|\bm{r}_{j}^{\iota}-\bm{h}_{j}\|_2^{2}\right\}$ is the error measure between the output of the LE and the true signal $\bm{h}_{j}$, and $Z\sim\mathcal{N}_c\left(Z;0,1\right)$, while $r_{i,j}^{\iota}$ and $h_{i,j}$ are the $i$-th elements of $\bm{r}_{j}^{\iota}$ and $\bm{h}_{j}$, respectively. The posterior variance of $h_{i,j}$ is then given by 
\begin{equation}\label{mmse_var} 
 \zeta_{i,j}^{\iota} = \mathrm{Var}\left\{h_{i,j}|r_{i,j}^{\iota}\right\}  =  \mathrm{E}\Big\{\big(\hat{\phi}_{\iota}\big(r_{i,j}^{\iota}\big)-h_{i,j}\big)^{2}\Big\} .
\end{equation}
Besides, by defining $\bar{\zeta}_{j}^{\iota} = \frac{1}{KL}\sum_{i=1}^{KL}\zeta_{i,j}^{\iota}$, the optimal scalar $C_{\iota}$ is given by $C_{\iota} = \frac{\tau_{\iota}}{\tau_{\iota}-\bar{\zeta}_{j}^{\iota}}$ and $\mathrm{E}\left\{ {\hat{\phi}}_{\iota}^{\prime}\right\} = \frac{{\bar{\zeta}}_{j}^{\iota}}{\tau_{\iota}}$. To track the errors of the LE and NLE outputs, an auxiliary matrix is defined as $\bm{B}_{\iota} = \bm{I}_{K L}-\bm{W}_{\iota}\bm{\Psi}$, and the error measures of the LE and NLE are respectively iterated as 
\begin{subequations}
\begin{align}\label{le_err_tau} 
 \tau_{\iota} &= \frac{1}{N}\mathrm{tr}\left(\bm{B}_{\iota}\bm{B}_{\iota}^{\rm H}\right)v_{\iota} \!+\! \frac{1}{N}\mathrm{tr}\left(\bm{W}_{\iota}\bm{W}_{\iota}^{\rm H}\right)\sigma_n^2 , \\
 \label{nle_err_v} 
 v_{\iota} &= \frac{\|\bm{y}_j-\bm{\Psi}\bm{d}_j^{\iota}\|^2-G\sigma_n^2}{\mathrm{tr}\left(\bm{\Psi}^{\rm H}\bm{\Psi}\right)} .
\end{align}
\end{subequations}
To avoid the matrix inversion in $\hat{\bm{W}}_{\iota}$, a singular value decomposition (SVD) is applied in advance. Let $\bm{\Psi} = \bm{U}\bm{\Sigma}\bm{V}^{\rm H}$ and $\left\{\sigma_{g}\right\}_{g=1}^{G}$ be the singular values. The calculation of $\hat{\bm{W}}_{\iota}$ is simplified as
\begin{equation}\label{simple_LMMSE} 
 \hat{\bm{W}}_{\iota} = \bm{V}\bm{\Sigma}^{\rm H}\tilde{\bm{\Sigma}}^{\iota} \bm{U}^{\rm H} ,
\end{equation}
where 
\begin{align}\label{SVD-1} 
 \tilde{\bm{\Sigma}}^{\iota} =& \left(v_{\iota}\bm{\Sigma}\bm{\Sigma}^{\rm H}+\sigma_{n}^2\bm{I}_G\right)^{-1} = \mathrm{diag}\bigg(\left[\frac{1}{v_{\iota}\sigma_{1}^2+\sigma_{n}^2} ~ \frac{1}{v_{\iota}\sigma_{2}^2+\sigma_{n}^2}\cdots \frac{1}{v_{\iota}\sigma_{G}^2+\sigma_{n}^2}\right]^{\rm T}\bigg) .
\end{align}
Furthermore, to calculate (\ref{mmse_mean}) and (\ref{mmse_var}), the prior distribution of each element in $\bm{h}_j$ is required. Owing to the sparsity of ${\bm{h}}_{j}$, the $i$-th element of $\bm{h}_j$ follows the Bernoulli-Gaussian distribution
\begin{equation}\label{prior_bg} 
 p(h_{i,j}) = \big(1-\rho_{i,j}\big)\delta\big(h_{i,j}\big) + \rho_{i,j}\mathcal{N}_c\left(h_{i,j};\mu_{i,j}, \gamma_{i,j}\right), 
\end{equation}
where $\rho_{i,j}\in \left[0, ~ 1\right]$ is the probability of $h_{i,j}$ and $\delta (h)$ is the Dirac delta function, while $\mu_{i,j}$ and $\gamma_{i,j}$ are the prior mean and variance of Gaussian distribution, respectively. By combining (\ref{prior_bg}) with (\ref{mmse_mean}), the posterior PDF of $\hat{h}_{i,j}$ can be acquired according to Bayesian rule as
\begin{equation}\label{post_bg} 
 p\left(\hat{h}_{i,j}|r_{i,j}^{\iota}\right) = \big(1-\lambda_{i,j}^{\iota}\big)\delta\big(\hat{h}_{i,j}\big) + \lambda_{i,j}^{\iota}\mathcal{N}_c\left(\hat{h}_{i,j};a_{i,j}^{\iota}, b_{i,j}^{\iota}\right),
\end{equation}
where
\begin{subequations}
\begin{align}\label{post_ab} 
 a_{i,j}^{\iota} &= \frac{\mu_{i,j}\tau_{\iota}+r_{i,j}^{\iota}\gamma_{i,j}}{\tau_{\iota}+\gamma_{i,j}}, \quad
b_{i,j}^{\iota} = \frac{\tau_{\iota}\gamma_{i,j}}{\tau_{\iota} +\gamma_{i,j}} , \\
 \label{post_lambda} 
 \lambda_{i,j}^{\iota} &= \frac{\rho_{i,j}}{\rho_{i,j}+(1-\rho_{i,j}) e^{-\mathcal{L}^{\iota}}} , \\
 \label{post_L} 
 \mathcal{L}^{\iota} &= \ln\frac{\tau_{\iota}}{\tau_{\iota}+\gamma_{i,j}} + \frac{|r_{i,j}^{\iota}|^2}{\tau_{\iota}} - \frac{|r_{i,j}^{\iota}-\mu_{i,j}|^2}{\tau_{\iota}+\gamma_{i,j}} .
\end{align}	
\end{subequations}
Then the posterior mean and variance can be calculated as
\begin{subequations}
\begin{align}\label{post_mean} 
 \xi_{i,j}^{\iota} &= \mathrm{E}\left\{h_{i,j}|r_{i,j}^{\iota}\right\}  =\lambda_{i,j}^{\iota}a_{i,j} ^{\iota}, \\
 \label{post_var} 
 \zeta_{i,j}^{\iota} &= \mathrm{Var}\left\{h_{i,j}|r_{i,j}^{\iota} \right\}= \lambda_{i,j}^{\iota}\big(|a_{i,j}^{\iota}|^2+b_{i,j}^{\iota}\big) - |\xi_{i,j}^{\iota}|^2 .
\end{align}
\end{subequations}
	
Substituting $\bm{h}_{j}$ by each column of $\tilde{\bm{H}}_q$, the CE problem (\ref{equ.csmodel}) can be solved column by column. However, two problems occur when applying the OAMP algorithm to a practical system. First, the common support structure of (\ref{CommonSupport}) is not utilized, which neglects the spatial correlation among different receive antennas. Second, the hyper-parameters in the prior assumption are unknown when executing the MMSE in (\ref{mmse_mean}) and (\ref{mmse_var}). For the first concern, an OAMP-MMV algorithm was proposed in \cite{OAMP_MMV}, where the prior distribution of each row in $\tilde{\bm{H}_q}$ is modeled as a multi-dimensional Bernoulli-Gaussian distribution. By introducing a group Gram-Schmidt orthogonalization process, the orthogonality of the input and output errors in the LE and NLE can be ensured. However, due to the introduction of a prior covariance matrix for each row of $\tilde{\bm{H}_q}$, high complexity matrix inversion is inevitable, which makes its implementation impractical. For the second concern, a hyper-parameter learning method, such as the expectation maximization (EM) algorithm, can be utilized for prior parameter estimation \cite{YK_Mei}. 
	
To overcome the aforementioned problems, we partially utilize the common support property when updating the hyper parameters using the EM algorithm. Denote the unknown prior parameters by $\bm{\theta} = \{\rho_{i,j},\mu_{i,j},\gamma_{i,j},\forall i,j\}$. The EM algorithm constitutes of two key steps:
\begin{align} 
 \label{em_e}
 \mathcal{Q}\left(\bm{\theta},\bm{\theta}^{\iota}\right) &= \mathrm{E}\left\{\ln p(\tilde{\bm{H}}_q,\bm{Y}_{q}^{\mathrm{TS}})|\bm{Y}_{q}^{\mathrm{TS}},\bm{\theta}^{\iota}\right\} , \\
 \label{em_m}
 \bm{\theta}^{{\iota}+1} &= \arg \max\limits_{\bm{\theta}} \mathcal{Q}\left(\bm{\theta},\bm{\theta}^{\iota}\right) ,
\end{align}
where the expectation (\ref{em_e}) is taken over the conditional distribution $p\left(\tilde{\bm{H}}_q|\bm{Y}_q^{TS};\bm{\theta}^{\iota}\right)$, which can be approximated by the product of (\ref{post_bg}) $\forall i,j$. Similar to the deviations of Appendix~A in \cite{YK_Mei}, we obtain the parameter update procedure as follows
\begin{subequations}
\begin{align}
 \label{equ:prior_sparsity} 
 \rho_{i,j}^{{\iota}+1} &= \lambda_{i,j}^{\iota}, \forall i,j , \\
 \label{equ:prior_mean} 
 \mu_{i,j}^{{\iota}+1} &=  \frac{\sum_{i=1}^{KL}\lambda_{i,j}^{\iota}a_{i,j}^{\iota}}{\sum_{i=1}^{KL}\lambda_{i,j}^{\iota}},\forall i,j , \\
 \label{equ:prior_variance} 
 \gamma_{i,j}^{{\iota}+1} &=  \frac{\sum_{i=1}^{KL}\lambda_{i,j}^{\iota}\left(|\mu_{i,j}^{\iota}-a_{i,j}^{\iota}|^2+b_{i,j}^{\iota}\right)}{\sum_{i=1}^{KL}\lambda_{i,j}^{\iota}},\forall i,j.
\end{align}
\end{subequations}
In addition, a proper initialization can be found in \cite{Meng_AMP_init} to avoid being trapped in local extrema. 
	
So far, the common support set property (\ref{CommonSupport}) has not been utilized. Considering that all the antennas receive signal simultaneously, it is reasonable to assign a common activity indicator for each row in $\tilde{\bm{H}}_q$, which indicates that a refined $\rho_{i,j}$ can be given by
\begin{equation}\label{ai_refine} 
 \rho_{i,1}^{{\iota}+1} = \cdots = \rho_{i,N_r}^{{\iota}+1} = \frac{1}{N_r}\sum_{j=1}^{N_r}\lambda_{i,j}^{\iota}.
\end{equation}
Denote the overall estimated channel matrix by $\hat{\bm{H}}_{q}$. Then the support set of $\hat{\bm{H}}_{q}$ provides the activity information of potential UTs and the RCD information for each path. To acquire an estimation of the AUS, we first identify the support set of $\hat{\bm{H}}_{q}$. Let $\kappa_{i}$ be the support indicator for the $i$-th row of $\hat{\bm{H}}_{q}$, which means that the $i$-th row of $\hat{\bm{H}}_{q}$ is non-zero if $\kappa_{i} = 1$ and zero otherwise. A simple but efficient energy based activity detector can be constructed as \cite{Ke_TSP} 
\begin{equation}\label{equ:AD} 
 \kappa_i = \mathbb{I}\left\{\left(\frac{1}{N_r}\sum_{j=1}^{N_r}\mathbb{I}\left\{\big|\hat{h}_{i,j}\big|^2>\epsilon\right\}\right)>\eta_{th}\right\}, \\
\end{equation}
where $\mathbb{I}\left\{x\right\}$ is an indicator function whose value is $1$ if $x$ is true and $0$ otherwise, while $\hat{h}_{i,j} = \left[\hat{\bm{H}}_{q}\right]_{i,j}$ In addition, $\epsilon = 0.02 \max\left\{|h_{i,j}|, \forall i,j\right\}$ and $\eta_{th} = 0.5$ are empirically chosen.  
	
By defining the support subset for each UT as $\Omega_{k} = \left\{i\;|\;(k-1)L+1\le i\le k L \right\}$, an energy based activity detector can be constructed as
\begin{equation}\label{equ:AD1} 
 \hat{\alpha}_k =
		\begin{cases}
			1 & \sum_{i\in \Omega_{k}}\kappa_{i} > 0 , \\
			0 & \mathrm{otherwise} .
		\end{cases}
\end{equation}
Then the estimated AUS is given by $\hat{\mathcal{A}}_{q} = \left\{\hat{\alpha}_k\;|\; \hat{\alpha}_k = 1, \forall k\right\}$. The overall procedure of the JADCE algorithm at the $q$-th satellite is summarized in Algorithm~\ref{alg:JADCE}.
	
\begin{algorithm}[htbp]
  \KwIn{Observation at $q$-th satellite ${\bm{Y}}_{q}^{\mathrm{TS}}$, TS sensing matrix $\bm{\Psi}$, noise variance $\sigma_{n}^2$, and maximum number of iterations $N_{\mathrm{iter}}$.} 
  \KwOut{Channel matrix $\hat{\bm{H}}_{q}$ and AUS $\hat{\mathcal{A}}_{q}$.}
  Calculate SVD of $\bm{\Psi}$ offline as $\bm{\Psi} = \bm{U}\bm{\Sigma}\bm{V}^{\rm H}$; \\
	Initialize $\bm{d}^0_{j} = \bm{0}_{KL\times 1}$, $\forall j$, and $v_0 = 1$; Initialize prior parameters $\rho_{i,j}, \mu_{i,j}, \gamma_{i,j}$, $\forall i,j$, as recommended in \cite{Meng_AMP_init}; Set iteration number to ${\iota}=1$;\\   
  \For{${\iota} \leq N_{\mathrm{iter}}$} { 
    Calculate LMMSE estimator (\ref{simple_LMMSE}); \\	
	  Perform LE (\ref{oamp_le}), $\forall j$; \\
		Calculate $\tau_{\iota}$ (\ref{le_err_tau}); \\
		Acquire posterior estimation of NLE $\hat{\phi}_{\iota}$ (\ref{post_mean}) and (\ref{post_var}); \\
		Perform NLE (\ref{oamp_nle}), $\forall j$; \\
		Calculate $v_{\iota}$ (\ref{nle_err_v}); \\ 	
		Update prior parameters according to (\ref{equ:prior_sparsity})-(\ref{equ:prior_variance}); \\
		Refine acticity indicator (\ref{ai_refine});
  }
	Acquire final estimated channel matrix $\hat{\bm{H}}_{q}$ via (\ref{post_mean}) and AUS $\hat{\mathcal{A}}_{q}$ via (\ref{equ:AD1}); \\
	\Return $\hat{\bm{H}}_{q}$ and $\hat{\mathcal{A}}_{q}$ for satellite $q$.
\caption{OAMP-MMV based JADCE}
\label{alg:JADCE} 
\LinesNumbered
\end{algorithm}

\vspace*{-2mm}	
\subsection{Enhanced AoA Estimation for Channel Reconstruction}\label{S3.3}

The estimated equivalent channel matrix $\hat{\bm{H}}_q$ and AUS $\hat{\mathcal{A}}_{q}$ provide the pairing of the active UTs and corresponding delay-spatial domain channels. However, the OAMP-MMV in Algorithm~\ref{alg:JADCE} neglects the spatial structure of receive array, and it cannot reveal information about the user direction. To acquire a high-resolution estimation of the AoAs, the 2D-ESPRIT algorithm \cite{2DESPRIT,Liao_Tcom} is adopted for enhanced parametrized CE. For notation simplicity, we drop the index $q$.

\subsubsection{2D-ESPRIT for AoA Estimation}

The steering vector for each AoA in the UPA is formed by the Kronecker product of $\bm{v}_k^{x}$ and $\bm{v}_k^{y}$.  Due to the angular domain sparsity of TSL, all the multipath components from the same UT share the same AoA, which can be regarded as multiple snapshots. To facilitate 2D angle estimation, let $\mathcal{S}_{k}\! =\! \left\{ i \;|\; \kappa_{i} = 1,(k-1)L+1\le i \le kL \right\}$ be the effective channel support set for the $k$-th UT. Then the equivalent observation for the $k$-th UT's channel can be denoted as $\tilde{\bm{X}}_{k}\! =\! \left[\hat{\bm{H}}_{q}\right]_{[\mathcal{S}_k,:]}^{\rm T}\! \in\! \mathbb{C}^{N_{r}\times |\mathcal{S}_{k}|}$. Typically, the number of multipath components $|\mathcal{S}_{k}|$ in TSL is limited, which may restrict the performance of the ESPRIT algorithm. We utilize a spatial smoothing preprocessing technique \cite{Liao_Tcom} for enhancing robustness. First, we define two spatial smoothing parameters $\left\{G_{s}\right\}, $ with $1\le G_{s}\le N_{s}, s\in \left\{x,y\right\}$, and obtain the two sub-dimensions as $M_{s}^{\mathrm{sub}}\! =\! N_s\! -\! G_s\! +\! 1$. The size of the total sub-dimensions is $M_{\mathrm{sub}}\! =\! M_{x}^{\mathrm{sub}}M_{y}^{\mathrm{sub}}$. Then we define the selection matrix $\bm{J}^{\left(g_s\right)}$ for each dimension with $\bm{J}^{\left(g_s\right)}\! =\! \left[\bm{0}_{M_{s}^{\mathrm{sub}}\times \left(g_{s}-1\right)} ~ \bm{I}_{M_s^{\mathrm{sub}}} ~ \bm{0}_{M_s^{\mathrm{sub}}\times \left(G_s-g_s\right)}\right]\! \in\! \mathbb{R}^{M_s^{\mathrm{sub}}\times N_{s}}$ for $1\le g_s\le G_s$. Therefore, a total of $G\! =\! G_x G_y$ 2D selection matrices $\bm{J}_{g_x,g_y}\! =\! \bm{J}^{\left(g_x\right)}\otimes\bm{J}^{\left(g_y\right)}\! \in\! \mathbb{R}^{M_{\mathrm{sub}}\times N_r}$ can be obtained. By applying these 2D selection matrices to $\tilde{\bm{X}}_{k}$,  the smoothed observation $\bar{\bm{X}}_{k}\! \in\! \mathbb{C}^{M_{\mathrm{sub}}\times G|\mathcal{S}|_{k}}$ is formed as
\begin{equation}\label{SSP} 
 \bar{\bm{X}}_{k} = \left[\bm{J}_{1,1}\tilde{\bm{X}}_{k} \cdots \bm{J}_{g_x,g_y}\tilde{\bm{X}}_{k} \cdots \bm{J}_{G_x,G_y}\tilde{\bm{X}}_{k}\right].
\end{equation}
The classical 2D-ESPRIT algorithm \cite{2DESPRIT} is applied to the smoothed observation $\bar{\bm{X}}_{k}$ of (\ref{SSP}) to estimate the corresponding azimuth angle $\hat{\theta}_{k}$ and elevation angle $\hat{\phi}_{k}$ directly. The corresponding details of 2D-ESPRIT algorithm can be found in Line 5 - Line 9, and the definitions of mentioned auxiliary matrices can be found in Appendix \ref{apd1}. 

\begin{algorithm}[!t]	
	\KwIn{Estimated channel matrix $\hat{\bm{H}}_q$ and AUS $\hat{\mathcal{A}}_{q}$.}
	\KwOut{Refined channel $\bar{\bm{H}}_{q}$.}
	Acquire $\kappa_{i}$, $\forall i$, according to (\ref{equ:AD}) and set
	$\mathcal{S}_{k}\! =\! \left\{ i | \kappa_{i} = 1,(k-1)L+1\le i \le kL \right\}$, $\forall k\! \in\! \hat{\mathcal{A}}_q$; \\
	\For{$\forall k \in \hat{\mathcal{A}}_q$} {
		\% \textbf{Spatial smoothing preprocessing} \\
		$\tilde{\bm{X}}_{k} = \left[\hat{\bm{H}}_{q}\right]_{[\mathcal{S}_k,:]}^{\rm T}\in\mathbb{C}^{N_{r}\times |\mathcal{S}_{k}|}$; 
		Construct observation $\bar{\bm{X}}_{k}$ according to (\ref{SSP}); \\
			\% \textbf{2D-ESPRIT-based AoA estimation \cite{2DESPRIT}:}\\
			Compute $\bm{e}_s$ via the largest left singular vector of $\left[\mathcal{R}e\left\{\bar{\bm{Y}}_{k}\right\}, \mathcal{I}m\left\{\bar{\bm{Y}}_{k}\right\}\right]$, where $\bar{\bm{Y}}_{k}\!\! = \!\!\left(\bm{Q}_{M_{y}^{\mathrm{sub}}}^{H}\!\otimes\! \bm{Q}_{M_{x}^{\mathrm{sub}}}\right)\!\bm{\bar{X}}_{k}$;\\
			Compute $\omega_{\mu_{x}} $  and $\omega_{\mu_{y}}$ as the solution to equations $\bm{K}_{\mu1}\bm{e}_s\omega_{\mu_{x}} = \bm{K}_{\mu2}\bm{e}_s$ and $\bm{K}_{\nu1}\bm{e}_s\omega_{\mu_{y}} = \bm{K}_{\nu2}\bm{e}_s$, respectively;\\
			Compute the spatial frequency estimates as $\mu_x = 2\tan^{-1}\left(\omega_{\mu_{x}}\right)$ and $\mu_{y} = 2\tan^{-1}\left(\omega_{\mu_{y}}\right)$;\\
			Acquire the AoAs estimation as $\hat{\theta}_{k} = \tan^{-1}\left(\frac{\mu_y}{\mu_x}\right)-\pi$
			and	$\hat{\phi}_{k} = \sin^{-1}\left(\frac{1}{\pi}\sqrt{\mu_x^2+\mu_{y}^2}\right)$;\\
		\% \textbf{Channel reconstruction}\\
		Estimate channel gain according to (\ref{CG_LS2}); \\
		Reconstruct the complete channel according to (\ref{C_Re}); \\		
	}     	
	\Return $\bar{\bm{H}}_{q}$.
	\caption{Channel estimation refinement}
	\label{alg:CER} 
	\LinesNumbered
\end{algorithm}

\subsubsection{Channel Gain Estimation and Channel Reconstruction}

So far, the channel delay taps and corresponding AoAs have been estimated by the OAMP-MMV and 2D-ESPRIT algorithms, respectively. To reconstruct the complete channel, the channel gain for each tap has to be estimated. Note that we have already acquired an initial estimation for each path, thus, the channel gain for each path can be further estimated according to the following equation
\begin{equation}\label{CG_LS} 
 \left[\hat{\bm{H}}_{q}\right]_{[\mathcal{S}_{k}\left\{i\right\},:]}^{\rm T} = \beta_{k,i} \hat{\bm{a}}_{r,k} ,
\end{equation}
where $\mathcal{S}_{k}\{i\}$ denotes the $i$-th element of $\mathcal{S}_{k}$, $\hat{\bm{a}}_{r,k} = \bm{v}_{k}^{y}\left(\hat{\theta}_{k},\hat{\phi}_{k}\right) \otimes \bm{v}_{k}^{x}\left(\hat{\theta}_{k},\hat{\phi}_{k}\right)$, and $\beta_{k,i}$ can be estimated by the least squares (LS) method as
\begin{equation}\label{CG_LS2} 
 \hat{\beta}_{k,i} = \left(\hat{\bm{a}}_{r,k}^{\rm H}\hat{\bm{a}}_{r,k}\right)^{-1}\hat{\bm{a}}_{r,k}^{\rm H}\left[\hat{\bm{H}}_{q}\right]_{[\mathcal{S}_{k}\left\{i\right\},:]}^{\rm T}.
\end{equation}
Finally, the spatial channel of the $k$-th active UT can be reconstructed as
\begin{equation}\label{C_Re} 
 \left[\bm{\bar{H}}_{q}\right]_{[\mathcal{S}_{k}\left\{i\right\},:]} = \hat{\beta}_{k,i}  \left( \bm{v}_{k}^{y}\left(\hat{\theta}_{k},\hat{\phi}_{k}\right)  \otimes \bm{v}_{k}^{x}\left(\hat{\theta}_{k},\hat{\phi}_{k}\right)\right)^{\rm T},  \forall k\in \hat{\mathcal{A}_{q}}, \forall i\in \mathcal{S}_{k}.
\end{equation}
The overall AoA estimation and channel reconstruction process are listed in Algorithm~\ref{alg:CER}.
	
\subsection{Single Satellite DD Algorithm}\label{S3.4}

As illustrated in Fig.~\ref{fig3_frame}, the multipath components from different active UTs inevitably contaminate the received data. Fortunately, based on the estimated channel and the known TS, the contaminated data can be recovered by subtracting the known contamination from the received data. Specifically, the $t$-th received frame $\bm{Y}_{q,t} = \left[\left(\bm{Y}_{q,t}^{\mathrm{TS}}\right)^{\rm T} ~ \left(\bm{Y}_{q,t}^{\mathrm{D}}\right)^{\rm T} \right]^{\rm T}\in \mathbb{C}^{\left(M+N\right)\times N_r}$ consists of a non-ISI TS part $\bm{Y}_{q,t}^{\mathrm{TS}}\in \mathbb{C}^{G\times N_r}$ and a data part $\bm{Y}_{q,t}^{\mathrm{D}}\in \mathbb{C}^{\left(M+N-G\right)\times N_r}$. The received data part contains three subparts $\bm{Y}_{q,t}^{\rm D} = \left[\left(\bm{Y}_{q,t}^{\mathrm{head}}\right)^{\rm T} ~ \left(\bm{Y}_{q,t}^{\mathrm{mid}}\right)^{\rm T} ~ \left(\bm{Y}_{q,t}^{\mathrm{trail}}\right)^{\rm T}\right]^{\rm T}$, where $\bm{Y}_{q,t}^{\mathrm{head}}\in \mathbb{C}^{\left(L-1\right)\times N_r}$ is polluted by the trail of the TS, $\bm{Y}_{q,t}^{\mathrm{trail}}\in\mathbb{C}^{\left(L-1\right)\times N_r}$ is polluted by the head of the next TSP-MCS frame, and $\bm{Y}_{q,t}^{\mathrm{mid}}\in \mathbb{C}^{\left(M+N-G-2L+2\right)\times N_r}$ is a clean subpart without any contamination. 
	
To facilitate subcarrier-wise data detection, we first subtract the ISI at the contaminated head part and trail part according to the known TS and estimated channel, and then move the clean trailing data part to the clean head part by a cyclic shift and addition operation. Consequently, a clean receive frame is reconstructed. Specifically, we first create a local frame that contains the TS part and empty data part for the $k$-th UT, denoted by $\bm{s}_{k}^{o} = \left[\bm{c}_{k}^{\rm T} ~ \bm{0}_{N\times 1}^{\rm T} ~ \bm{c}_{k}^{\rm T} \cdots \right]^{\rm T}$. Then, by simulating the transmission progress at the receiver, the received frame is obtained as the convolution of $\bm{s}_{k}^{o}$ and the estimated CIR $\left[\bar{\bm{H}}_{k,q}\right]_{[:,j]}$, $\forall j$, where $\bar{\bm{H}}_{k,q} = \left[\bar{\bm{H}}_{q}\right]_{[\mathcal{I}_0:,]}$ with $\mathcal{I}_0=\{(k-1)L+1\le i\le k L\}$ is the estimated CIRM between the $k$-th UT and the $q$-th satellite. By summing over all the estimated active UTs, the received frame at $j$-th antenna can be calculated as 
\begin{equation}\label{local_ts} 
 \hat{\bm{y}}_{q}^{j} = \sum_{k\in \hat{\mathcal{A}}_q} \bm{s}_{k}^{o} \star \left[\bar{\bm{H}}_{k,q}\right]_{[:,j]}, \forall j.
\end{equation}
Since $\hat{\bm{y}}_{q}^{j}$, $\forall j$, are generated at the satellite and contain the estimated ISI, we refer to them as the pseudo observations. We denote the pseudo observation in the $t$-th frame by $\hat{\bm{Y}}_{q,t} = \left[\hat{\bm{y}}_{q,t}^{1} ~ \hat{\bm{y}}_{q,t}^{2}\cdots \hat{\bm{y}}_{q,t}^{N_r}\right]\in\mathbb{C}^{(M+N)\times N_r}$, and the TS contamination to the head part and trail part can be respectively indexed by $\hat{\bm{Y}}_{q,t}^{\mathrm{head}} = \left[\hat{\bm{Y}}_{q,t}\right]_{[\mathcal{I}_1,:]}$ and $\hat{\bm{Y}}_{q,t}^{\mathrm{trail}} = \left[\hat{\bm{Y}}_{q,t}\right]_{[\mathcal{I}_2,:]}$, where $\mathcal{I}_1 = \{G+1\le i\le G+L-1\}$ and $\mathcal{I}_2 = \{G+N+1\le i\le M+N\}$. Based on the pseudo observation $\hat{\bm{Y}}_{q,t}^{\mathrm{head}}$, $\hat{\bm{Y}}_{q,t}^{\mathrm{trail}}$ and the true observation ${\bm{Y}}_{q,t}^{\mathrm{D}}$, the clean data $\tilde{\bm{Y}}_{q,t}^{\rm D} \in \mathbb{C}^{N\times N_r}$ can be reconstructed as
\begin{equation}\label{non-MI} 
 \tilde{\bm{Y}}_{q,t}^{\rm D} = \left[\bm{Y}_{q,t}^{\mathrm{head}}-\hat{\bm{Y}}_{q,t}^{\mathrm{head}} + \bm{Y}_{q,t}^{\mathrm{trail}}-\hat{\bm{Y}}_{q,t}^{\mathrm{trail}} ~ \bm{Y}_{q,t}^{\mathrm{mid}}\right].
\end{equation}
Based on the cleaned data $\tilde{\bm{Y}}_{q,t}^{\rm D}$, the frequency-domain data $\mathring{\bm{Y}}_{q,t}^{\rm D}\in\mathbb{C}^{N\times N_r}$ can be acquired by
\begin{equation}\label{fre_data} 
 \mathring{\bm{Y}}_{q,t}^{\rm D} = \bm{F}_{N}\tilde{\bm{Y}}_{q,t}^{\rm D},
\end{equation}
where $\bm{F}_{N}$ is the $N\times N$ DFT matrix with normalized power, whose $\left(m,n\right)$-th element is given by $\left[\bm{F}_{N}\right]_{m,n} = \frac{1}{\sqrt{N}} e^{\frac{-\textsf{j}2\pi(m-1)(n-1)}{N}}$. The delay-domain channel between the $k$-th UT and the $n_r$-th receive antenna at the $q$-th satellite can be acquired as $\bm{h}_{k,q}^{n_r}\in\mathbb{C}^L$ with $\big[\bm{h}_{k,q}^{n_r}\big]_i  = \left[\bar{\bm{H}}_{q}\right]_{(i+(k-1)L),n_r}$, $1\le i\le L$, where $\bar{\bm{H}}_{q}$ is provided by Algorithm~\ref{alg:CER}. By defining $\tilde{\bm{h}}_{k,q}^{n_r}\! =\! \left[\left(\bm{h}_{k,q}^{n_r}\right)^{\rm T} ~  \bm{0}_{(N-L)\times 1}^{\rm T} \right]^{\rm T}\! \in\! \mathbb{C}^{N}$, the frequency-domain channel $\mathring{\bm{h}}_{k,q}^{n_r}\! \in\! \mathbb{C}^{N}$ is obtained as 
\begin{equation}\label{fre_channel} 
 \mathring{\bm{h}}_{k,q}^{n_r} =  \bm{F}_{N}\tilde{\bm{h}}_{k,q}^{n_r}. 
\end{equation}
Based on (\ref{fre_data}) and (\ref{fre_channel}), the data for all the active UTs on the $n$-th subcarrier in the $t$-th frame can be solved from the following equation (with the subscript $t$ omitted for notation simplicity)
\begin{equation}\label{distributed_dd} 
 \mathring{\bm{y}}_{q}[n] = \mathring{\bm{H}}_{q}[n]\mathring{\bm{x}}[n] + \bm{n}[n],\forall n, 
\end{equation}
where $\mathring{\bm{y}}_{q}[n] = \left[\mathring{\bm{Y}}_{q,t}^{\rm D}\right]_{[n,:]}^{\rm T}$, $\mathring{\bm{x}}[n] = \left[\mathring{x}_{1}[n] ~ \mathring{x}_{2}[n] \cdots \mathring{x}_{|\bar{\mathcal{A}}_q|}[n] \right]^{\rm T} \in \mathbb{C}^{|\bar{\mathcal{A}}_q|}$ is the source data on the $n$-th subcarrier, and the frequency-channel matrix $\mathring{\bm{H}}_{q}[n] \in \mathbb{C}^{N_r\times |\bar{\mathcal{A}}_q| }$ is given by
\begin{equation}\label{equ_ls} 
 \mathring{\bm{H}}_{q}[n] =
		\left[ \begin{matrix}
			\mathring{h}_{1,q}^{1}[n] & \mathring{h}_{2,q}^{1}[n] & \cdots & \mathring{h}_{|\bar{\mathcal{A}}_q|,q}^{1}[n]  \\
			\mathring{h}_{1,q}^{2}[n] & \mathring{h}_{2,q}^{2}[n] & \cdots & \mathring{h}_{|\bar{\mathcal{A}}_q|,q}^{2}[n]  \\
			\vdots & \vdots & \ddots & \vdots  \\
			\mathring{h}_{1,q}^{N_r}[n] & \mathring{h}_{2,q}^{N_r}[n] & \cdots & \mathring{h}_{|\bar{\mathcal{A}}_q|,q}^{N_r}[n]  \\
		\end{matrix} \right] . 
\end{equation}
Equation (\ref{distributed_dd}) provides a DD formulation for each satellite independently, which can be effectively solved by the LS method (zero-forcing detection).

Similar to \cite{dft_s_ofdm}, in DFT-s-OFDM, only $M_s \le N$ subcarriers are used for information transmission while the remaining subcarriers are padded by zero. After solving $\mathring{\bm{x}}[n]$ for all non-empty subcarriers $n$, the original information symbol in the $t$-th frame $\mathring{\bm{X}_{t}} \in \mathbb{C}^{M_s\times|\bar{\mathcal{A}}_q|}$ for all the active UTs can be recovered by subcarrier demapping and $M_{s}$-point IDFT, which is acquired by
\begin{equation}\label{equ.idft} 
 \mathring{\bm{X}}_{t} = \bm{F}_{M_s}^{\rm H}\bm{\varXi}^{\rm H}\mathring{\bm{X}}_{f} ,
\end{equation}
where $\mathring{\bm{X}_{f}} = \left[\mathring{\bm{x}}[1]\cdots \mathring{\bm{x}}[N]\right]^{T} \in \mathbb{C}^{N\times|\bar{\mathcal{A}}_q|}$ is the stacked-frequency domain equalization result, which can be acquired from (\ref{distributed_dd}), while $\bm{\varXi}\in \mathbb{C}^{N\times M_s}$ is a predefined permutation matrix whose $M_s$ columns are extracted from $\bm{I}_{N}$, and it defines the subcarrier mapping relationship. Besides, $\mathring{\bm{X}}_{t} \in \mathcal{M}^{M_s\times |\bar{\mathcal{A}}_q|}$ represents the information symbol for all the active UTs, where $\mathcal{M}$ is the symbol constellation set, e.g., QPSK constellation.	Based on $\mathring{\bm{X}_{t}}$, the original data bits for each UT can be recovered after channel decoding.
	
\section{Massive Access Processing Based on Multi-Satellite Cooperation}\label{S4}

Section~\ref{S3} has provided a solution for the JADCE and DD problem in the single satellite-enabled massive access system. However, the AD and DD performance are constrained by the quality of TSLs and the channel correlations among active UTs, respectively. In this section, we propose a majority voting scheme to acquire a more reliable estimation of AUS. Furthermore, to mitigate the DD performance degradation caused by high channel correlation, we propose the cooperative DD scheme, where both perfect backhaul and quantized backhaul of received information are discussed. 
		
\subsection{AD Enhancement Based on Majority Voting}\label{S4.1}

\begin{figure*}[!t]
\vspace*{-6mm}
\begin{center}
	\hspace{-2mm}
	\begin{minipage}[t]{0.49\linewidth}
	\captionsetup{font={footnotesize}, singlelinecheck = off,name={Fig.},justification=centering, labelsep=period}
	\includegraphics[scale=0.49]{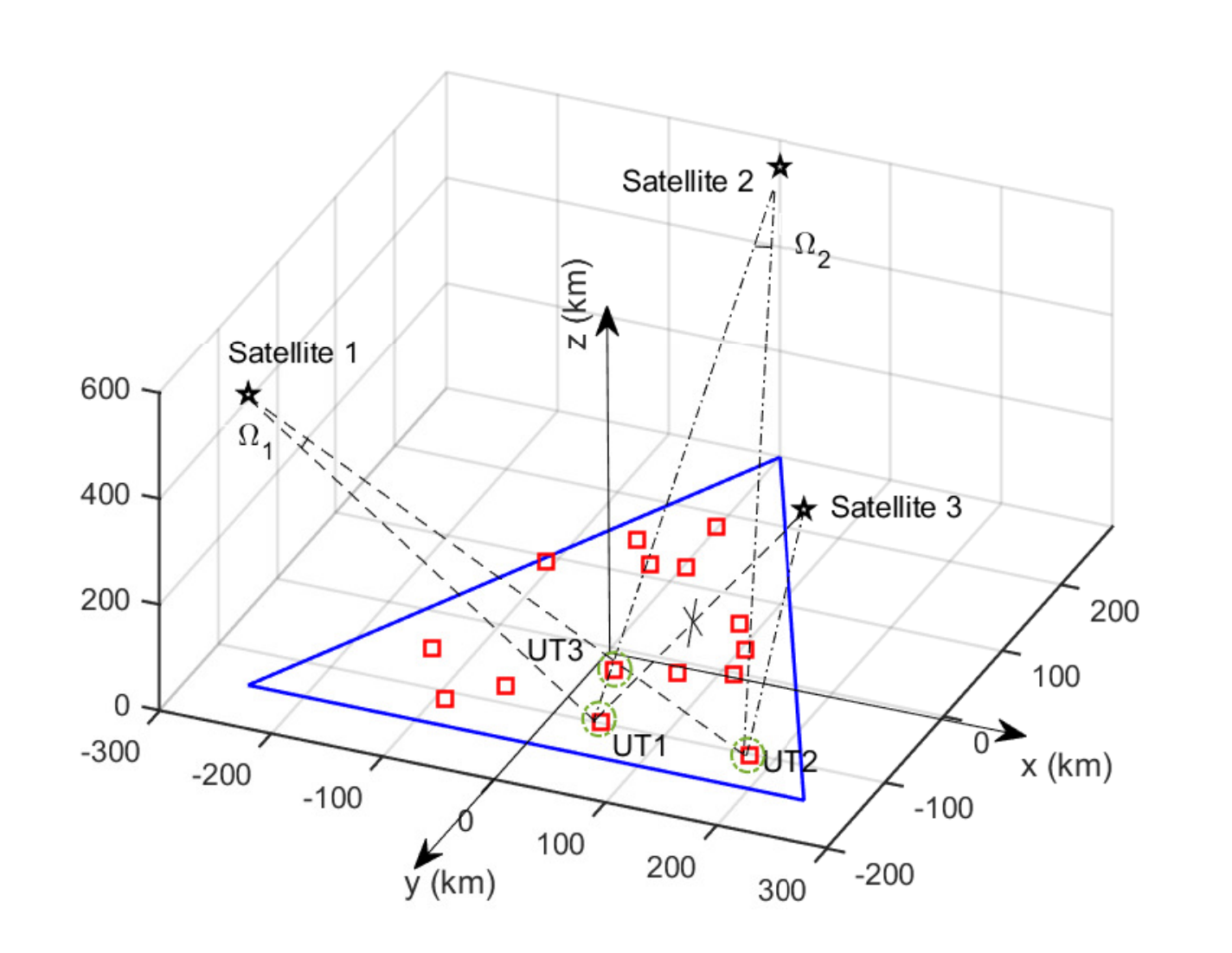}
  	\caption{Example of geographic distribution of UTs and satellites. Three satellites nodes serve as BSs to provide the access service for terrestrial UTs.}
	\label{fig4_access} 
	\end{minipage}
	\begin{minipage}[t]{0.49\linewidth}
	\captionsetup{font={footnotesize}, singlelinecheck = off,name={Fig.},justification=centering, labelsep=period}
	\includegraphics[scale=0.49]{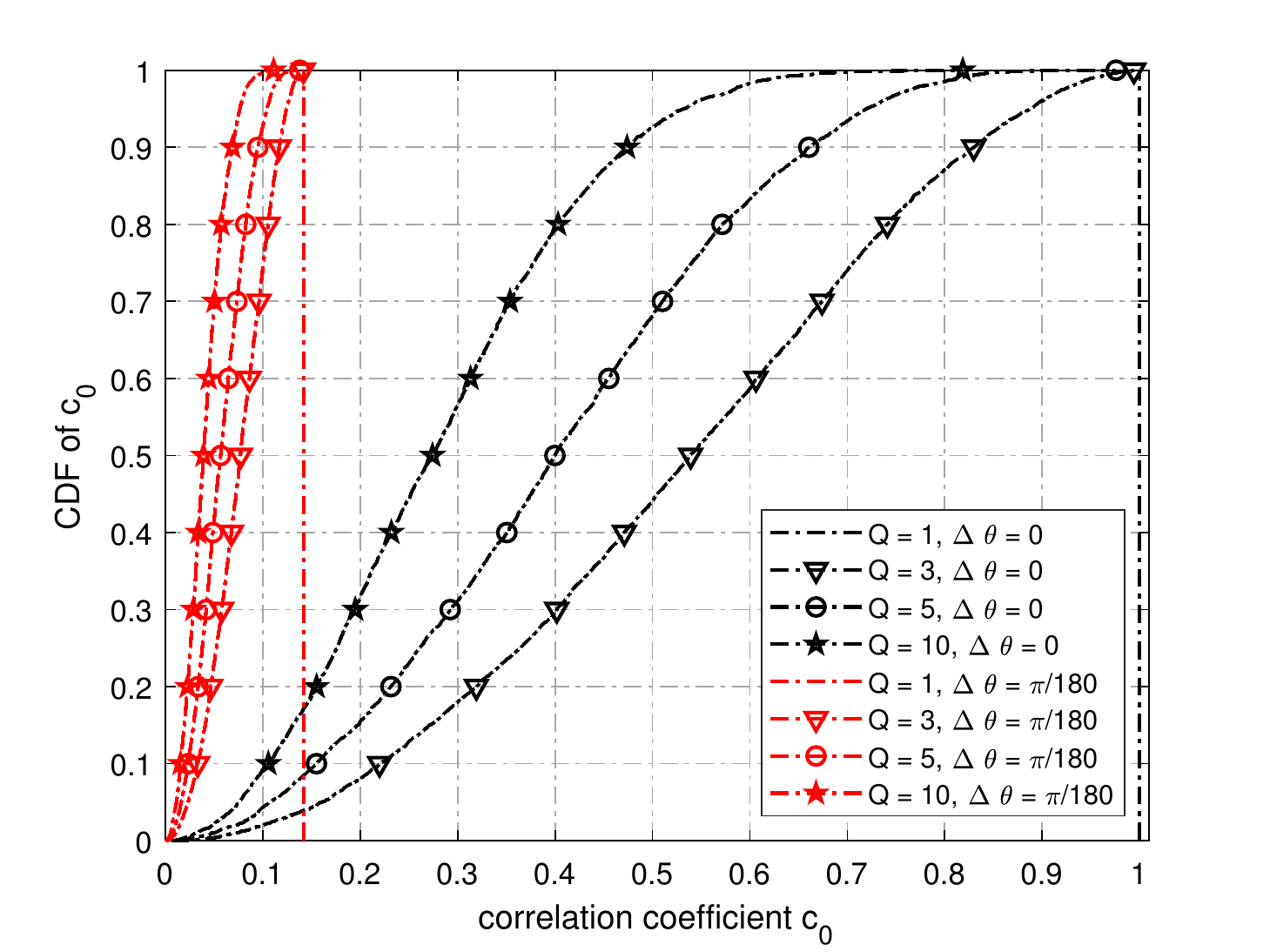}
	\caption{CDF curve of correlation coefficient $c_{0}$.}
	\label{fig5_coeff} 
	\end{minipage}
\end{center}
\vspace{-10mm}
\end{figure*}

Estimation of the AUS may suffer from missed detection or false alarm. Missed detection of the AUS at a satellite  may occur due to the deterioration of the corresponding TSL, caused by the blockage of obstacles, rain attenuation, or handover at the cell edge (e.g., the TSL between UT 1 and satellite 3 in Fig.~\ref{fig4_access}). False alarms may happen when some burst noise or interference occurs. The false alarm case can be effectively distinguished by increasing the detection threshold or adding an additional cyclic redundancy check to the frame at the cost of increased computational complexity and decreased spectral efficiency. To increase the reliability of AD, we propose a majority voting scheme by fusing the decision results from multiple edge nodes. 
	
The algorithms of Section~\ref{S3} provide an initial estimate of the AUS at each satellite independently. In our diversity transmission scheme, active UTs transmit the same data packets to the target satellites simultaneously, and the detected AUS should be the same for all the received satellites. To overcome the error detection of an individual satellite, a majority voting method is used to determine a reliable AUS by fusing the data at the central node. Let $\hat{\alpha}_k^{q}$ be the estimated active status of UT $k$ by satellite $q$. A refined activity estimation can be acquired as
\begin{equation}\label{AUS_vote} 
 \mathring{\alpha}_k =
		\begin{cases}
			& 1, \quad \frac{1}{Q}\sum_{q=1}^{Q}\hat{\alpha}_k^{q} \ge 0.5, \\
			& 0, \quad \mathrm{otherwise} .
		\end{cases}
\end{equation}
A refined AUS is then obtained as $\bar{\mathcal{A}} = \left\{\mathring{\alpha_{k}}|\mathring{\alpha_{k}} = 1, \forall k\right\}$.

\subsection{Multi-satellite Cooperative Linear DD Algorithm}\label{S4.2}

When the number of UTs increases, the possibility of high spatial correlation between adjacent UTs' channel vectors also increases. This high spatial correlation may lead to the rank deficient random access channel matrix $\mathring{\bm{H}}_{q}[n]$, $\forall n$, which degrades the bit error rate (BER) performance for multi-user decoding. The spatial correlation caused by geographical similarity can be solved from two aspects. The first one is to introduce a sufficiently large angle difference in the space. For example, as shown in Fig.~\ref{fig4_access}, UT~1 and UT~2 share similar AoAs when transmitting information to satellite~1. By adding extra satellite~2, a greater angular difference can be observed, which helps to suppress the spatial domain inter-user interference. The second method is to utilize the diversity of channel gain. For example, UT~1 and UT~3 share similar AoAs for all the satellites. In this case, the diversity gain can help to reduce the spatial correlation. Existing literature \cite{XY_Zhou} adopts a user grouping strategy to promise a sufficiently large spacing in elevation angle and azimuth angle, such that the ill-conditioned channel matrix can be avoided. An alternative solution is to introduce frequency domain or time domain differences between different UTs. These strategies will need extra grouping mechanisms or consume more time-frequency resources. We use a more flexible spatial diversity approach. By transmitting the same data source to separate target LEO satellites, multiple observations can be acquired from different spatial angles. 
	
To provide an insightful explanation, let us assume a simple two-UT case, where two geographically neighboring UTs, each equipped with a uniform linear array, are simultaneously served by $Q$ satellites. The overall channel matrix is given by 
\begin{equation}\label{H-2ut} 
 \bm{H}_{all} = 
		\left[ \begin{matrix}
			{\beta}_{1,1}\bm{a}\left(\theta_{1,1}\right) & {\beta}_{1,2}\bm{a}\left(\theta_{1,2}\right)  \\
			\vdots & \vdots  \\
			{\beta}_{Q,1}\bm{a}\left(\theta_{Q,1}\right) & {\beta}_{Q,2}\bm{a}\left(\theta_{Q,2}\right)  \\
		\end{matrix}
		\right] \in \mathbb{C}^{QN_r\times 2},
\end{equation}
where $\beta_{q,k}$ and $\bm{a}\left(\theta_{q,k}\right)$ are the channel coefficient and steering vector between the $q$-th satellite and the $k$-th UT, respectively. Assuming that $\bm{a}\left(\theta\right) = \left[1 ~ e^{\textsf{j}\theta}\cdots e^{\textsf{j}(N_{r}-1)\theta}\right]^{\rm T}\in\mathbb{C}^{N_{r}}$ and $\beta_{q,k}$ follows $\mathcal{N}_c\left(\beta_{q,k};0,1\right)$, the correlation coefficient between the two columns of $\bm{H}_{all}$ is given by 
\begin{equation}\label{equ.coeff} 
 c = \frac{1}{N_{r}}\frac{\sum_{q=1}^{Q} \beta_{q,1}^{*} \beta_{q,2}\bm{a}^{\rm H}\left(\theta_{q,1}\right)\bm{a}\left(\theta_{q,2}\right)}{\sqrt{\left(\sum_{q=1}^{Q}|\beta_{q,1}|^2\right)\left(\sum_{q=1}^{Q}|\beta_{q,2}|^2\right)}} .
\end{equation}
For simplicity, we assume $\theta_{q,2}-\theta_{q,1} = \Delta{\theta}$, $\forall q$, then the modulus of $c$ can be further simplified as 
\begin{equation}\label{equ.coeff2} 
 c_{0} = \frac{|\sum_{q=1}^{Q}\beta_{q,1}^{*}\beta_{q,2}|}{\sqrt{\left(\sum_{q=1}^{Q}|\beta_{q,1}|^2\right)\left(\sum_{q=1}^{Q}|\beta_{q,2}|^2\right)}}\cdot \frac{1}{N_{r}} \frac{\sin \frac{N_{r}\Delta\theta}{2}}{\sin\frac{\Delta\theta}{2}}.
\end{equation}
Fig.~\ref{fig5_coeff} plots the cumulative distribution function (CDF) of $c_0$ by 5000 Monte Carlo simulations. It can be seen that as the number of satellites increases, the distribution of the correlation coefficient $c_{0}$ becomes closer to zero. Furthermore, as $Q\to \infty$, the columns of $\bm{H}_{all}$ become asymptotically orthogonal, which shows that the randomness of the channel coefficients promote the diversity gain, even for $\Delta \theta = 0$. Because multiple satellites promote angle difference, a lower correlation coefficient can be achieved for UTs in the same geographical area.
		
By aggregating the multi-angle observations of the UT's data, an improved DD performance can be achieved. Specifically, the central server node aggregates the data from multiple edge satellite nodes and stacks them to yield the following cooperative DD problem, which can be also solved by LS method. 
\begin{equation}\label{LS_equalization} 
 \left[\mathring{\bm{y}}_{1}^{\rm T}[n]\cdots \mathring{\bm{y}}_{Q}^{\rm T}[n]\right]^{\rm T} = \left[\mathring{\bm{H}}_{1}^{\rm T}[n]\cdots \mathring{\bm{H}}_{Q}^{\rm T}[n]\right]^{\rm T} \mathring{\bm{x}}[n] + \bm{n}[n] , \forall n .
\end{equation}
The overall multi-satellite cooperative linear DD algorithm is summarized in Algorithm~\ref{alg:UIC-CDD}.

\begin{algorithm}[!h]
  \KwIn{Data observation $\bm{Y}_{q,t}^{\rm D}$, estimated CIRM $\bar{\bm{H}}_{q}$, estimated AUS $\hat{\mathcal{A}}_{q}, \forall q$.}
  \KwOut{Source data bits for active UTs.}
  Generate local pseudo observation according to (\ref{local_ts}); \% \textbf{For edge satellite nodes} \\
	Construct non-multipath interference data $\tilde{\bm{Y}}_{q,t}^{\rm D}$ according to (\ref{non-MI}); \\
  Feed observations $\tilde{\bm{Y}}_{q,t}^{\rm D}$, AUS $\hat{\mathcal{A}}_{q}$, delay-domain channels $\bar{\bm{H}}_{q}$, $\forall q$, back to central node; \\
  Determine improved AUS estimation $\bar{\mathcal{A}}$ according to (\ref{AUS_vote}); \% \textbf{For central server node} \\
  Construct frequency-domain observations and channel matrices according to (\ref{fre_data})-(\ref{fre_channel}); \\
  Solve cooperative DD problem (\ref{LS_equalization})/(\ref{mixed_resolution}) in perfect/quantized backhaul case;\\
  Acquire information symbol according to (\ref{equ.idft}); \\
  Recover the source data bits by channel decoding; \\
  \Return Decoded source bits for estimated AUS.
\caption{Multi-satellite cooperative linear DD algorithm}
\label{alg:UIC-CDD} 
\LinesNumbered
\end{algorithm}

\vspace*{-2mm}
\subsection{Processing with Quantized Backhaul}\label{S4.3}

The multi-satellite cooperative linear DD algorithm (Algorithm 3) assumes a perfect backhaul of observations $\tilde{\bm{Y}}_{q,t}^{\rm D}$ from edge nodes to the central node, which requires high-capacity links. A practical solution to reduce this high-capacity demand is to feed back the quantized observations and recover the original data utilizing a suitable algorithm. For MSCTP, a terrestrial server acts as the central node, and all the LEO satellites serve as edge nodes, which feed back their observations to the central server. For MSCBP, a satellite serves as the central node, and it needs to reserve its own observation and receives the results from edge nodes through inter-satellite links. We discuss the case of MSCBP, but the results can be extended to the case of MSCTP. We assume that the CIR from edge nodes can be perfectly fed back to the satellite central node since the sparsity makes CIR easy to be compressed and recovered using the existing schemes \cite{CSI_feedback1,CSI_feedback2}.

\begin{figure*}[!b]
\vspace*{-8mm}
\centering
\includegraphics[width=0.8\textwidth]{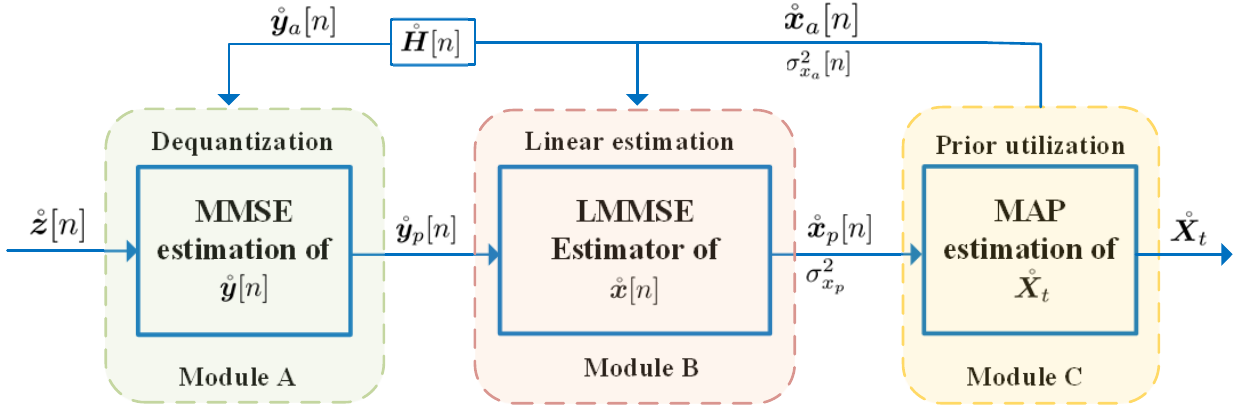}
\captionsetup{font={footnotesize}, singlelinecheck = off,name={Fig.},justification=centering, labelsep=period}
\caption{The block diagram of the proposed multi-satellite cooperative Bayesian dequantization DD algorithm.}
\label{fig6_vi}
\end{figure*}
	
We consider one central satellite node simultaneously receives the quantized observations from the other $Q\! -\! 1$ edge nodes. By defining $\mathring{\bm{H}}[n]\! =\! \left[\mathring{\bm{H}}_{1}^{\rm T}[n]\cdots \mathring{\bm{H}}_{Q}^{\rm T}[n]\right]^{\rm T}\! \in\! \mathbb{C}^{Q N_r\times |\bar{\mathcal{A}}|}$, $\mathring{\bm{y}}[n] \! =\! \left[\mathring{\bm{y}}_{1}^{\rm T}[n]\cdots \mathring{\bm{y}}_{Q}^{\rm T}[n]\right]^{\rm T}\! \in\! \mathbb{C}^{Q N_r}$, (\ref{LS_equalization}) is changed into the following mixed-resolution DD problem
\begin{align}\label{mixed_resolution} 
 \mathring{\bm{z}}[n] = \psi\left(\mathring{\bm{y}}[n]\right) = \psi \left( \mathring{\bm{H}}[n]\mathring{\bm{x}}[n] + \bm{n}[n]\right), \forall n,
\end{align}
where $\mathring{\bm{z}}[n] = \left[\mathring{\bm{z}}_{1}^{T}[n]\cdots \mathring{\bm{z}}_{Q}^{T}[n]\right]^{T} \in \mathbb{C}^{Q N_r}$. Without loss of generality, we further assume that $q=1$ is the central satellite node and the other satellites are edge nodes. In (\ref{mixed_resolution}), the scalar transform function $\psi\left(\cdot\right)$ defines the quantization effect and can be expressed as
\begin{equation}\label{quantized_observation} 
 \left[\mathring{\bm{z}}[n]\right]_{r} = \psi\left(\left[\mathring{\bm{y}}[n]\right]_{r} \right) = \left\{
  \begin{array}{cl}
	 \left[\mathring{\bm{y}}[n]\right]_{r}, & r\in \mathcal{R}_{C}, \\
	 \mathcal{Q}_{\mathcal{B}}\left(\left[\mathring{\bm{y}}[n]\right]_{r}\right), & r\in \mathcal{R}_{E} ,
	\end{array}	\right.
\end{equation}
where $\mathcal{R}_{C} = \left\{1,\cdots,N_{r}\right\}$ denotes the set of the row indexes corresponding to the central node, and $\mathcal{R}_{E} = \left\{N_{r}+1,\cdots,QN_{r}\right\}$ is the set of the row indexes corresponding to edge nodes, while $\mathcal{Q}_{\mathcal{B}}\left(\cdot\right)$ is a $\mathcal{B}$-bit complex quantizer that can quantize the real part and imaginary part of the input signal respectively and element-wisely. The output of $\mathcal{Q}_{\mathcal{B}}\left(\cdot\right)$ is decided by the resolution of the feedback bits $\mathcal{B}$ and a set of thresholds $\left\{e_{0}, e_{1},\cdots, e_{2^{\mathcal{B}-1}}\right\}$ such that $-\infty=e_{0}<e_{1}<\cdots<e_{2^{\mathcal{B}-1}}<e_{2^{\mathcal{B}} } = +\infty$. The quantizer assigns a fixed value $\varpi_{b}$ for the output when the input falls into the interval $\left(e_{b-1}, ~ e_{b}\right]$, e.g., $\varpi_{b} = \left(e_{b-1}+e_{b}\right)/2$.
	
To fully utilize the discrete constellation prior of the symbols and the quantization constraints for the observations from edge nodes, the proposed multi-satellite cooperative Bayes dequantization DD algorithm proceeds iteratively, as shown in its block diagram of Fig.~\ref{fig6_vi}. 

\subsubsection{Module~A}

It performs the posterior estimation of the unquantized observation $\mathring{\bm{y}}[n]$ based on the prior information $\mathring{\bm{y}}_{a}[n]$ and the quantized result $\mathring{\bm{z}}[n]$. Similar to the Bayesian inference process \cite{VI,Bayes}, the posterior mean of the unquantized observation $\mathring{\bm{y}}_{p}[n]$ is given by 
\begin{equation}\label{equ.dequant} 
 \left[\mathring{\bm{y}}_{p}[n]\right]_{r} = \left\{ \begin{array}{cl} 
	 \left[\mathring{\bm {y}}_{a}[n]\right]_{r} + \dfrac{ \phi\left( \frac{\left[\mathring{\bm{z}}[n]\right]_{r}^{(\text{lo})} - \left[\mathring{\bm{y}}_{a}[n]\right]_{r}}{\sigma_{n}/\sqrt{2}}\right) - \phi\left( \frac{\left[\mathring{\bm{z}}[n]\right]_{r}^{(\text{hi})} - \left[\mathring{\bm{y}}_{a}[n]\right]_{r}}{\sigma_{n}/\sqrt{2}}\right)}{\Phi\left( \frac{\left[\mathring{\bm{z}}[n]\right]_{r}^{(\text{hi})} - \left[\mathring{\bm {y}}_{a}[n]\right]_{r}}{\sigma_{n}/\sqrt{2}}\right) - \Phi\left( \frac{\left[\mathring{\bm{z}}[n]\right]_{r}^{(\text{lo})} - \left[\mathring{\bm{y}}_{a}[n]\right]_{r}}{\sigma_{n}/\sqrt{2}}\right)} \frac{\sigma_{n}}{\sqrt{2}} , & r\in \mathcal{R}_{E}, \\
	 \left[\mathring{\bm{z}}_{a}[n]\right]_{r} , & r\in \mathcal{R}_{C} ,
 \end{array}\right.
\end{equation}
where $\mathring{\bm{y}}_{a}[n]\! =\! \mathring{\bm{H}}[n]\mathring{\bm{x}}_{a}[n]\! \in\! \mathbb{C}^{Q N_r}$, $\mathring{\bm{x}}_{a}[n]\! \in\! \mathbb{C}^{|\bar{\mathcal{A}}|}$ is the prior frequency-domain signal provided by Module C, and $\mathring{\bm{z}}[n]^{(\text{lo})}$ and $\mathring{\bm{z}}[n]^{(\text{hi})}$ are the lower and upper bounds of the quantization output corresponding to observation $\mathring{\bm{y}}[n]$, while $\phi\left(\cdot\right)$ and $\Phi\left(\cdot\right)$ are the PDF and CDF of a standard normal random variable evaluated at a given value, respectively. The output of Module~A, $\mathring{\bm{y}}_{p}[n]$, $\forall n$, are passed to Module~B.

\subsubsection{Module~B} 

Given $\mathring{\bm{y}}_{p}[n]$ and $\mathring{\bm{x}}_{a}[n]$, $\forall n$, an LMMSE estimator is constructed to acquire the posterior estimate of the frequency-domain information $\mathring{\bm{x}}_p[n]$ as
\begin{equation}\label{equ.lmmse} 
 \mathring{\bm{x}}_p[n] = \mathring{\bm{x}}_a[n] + \left(\mathring{\bm{H}}^{\rm H}[n] \mathring{\bm{H}}[n] + \frac{\sigma_{n}^2}{\sigma_{x_{a}}^{2}[n]} \bm{I}_{|\bar{\mathcal{A}|}}\right)^{-1} \mathring{\bm{H}}^{\rm H}[n] \left(\mathring{\bm{y}}_{p}[n]-\mathring{\bm{H}}[n]\mathring{\bm{x}}_a[n]\right) ,
\end{equation}
where $\sigma_{x_{a}}^{2}[n]$ is the prior variance of the frequency-domain information provided by Module~C. In addition, the posterior covariance matrix of $\mathring{\bm{x}}_{p}[n]$ is approximated by $\sigma_{x_{p}}^{2}\bm{I}_{|\bar{\mathcal{A}}|}$, where $\sigma_{x_{p}}^{2}$ is the posterior variance of the frequency-domain information given by
\begin{align}\label{PostVar} 
 \sigma_{x_{p}}^{2} =& \frac{\sigma_{n}^2}{N} \sum_{n=1}^{N} \left\langle\left(\mathring{\bm{H}}^{\rm H}[n]\mathring{\bm{H}}[n] + \frac{\sigma_{n}^{2}}{\sigma_{x_{a}}^{2}[n]} \bm{I}_{|\bar{\mathcal{A}|}}\right)^{-1}\right\rangle .
\end{align}
The posterior estimates, $\mathring{\bm{x}}_p[n]$, $\forall n$, and $\sigma_{x_{p}}^{2}$ are then passed to module~C.

\subsubsection{Module~C} 

To acquire the original information symbol for each active UT and utilize the discrete prior of constellation modulation, the frequency-domain information is first transformed back into the time domain before performing UT-wise maximum a posterior estimation. Specifically, by stacking  $\mathring{\bm{X}_{f}} = \left[\mathring{\bm{x}}_{p}[1] \cdots \mathring{\bm{x}}_{p}[N]\right]^{\rm T} \in \mathbb{C}^{N\times|\bar{\mathcal{A}}|}$ from Module B and defining $\mathring{\bm{X}_{p}} = \bm{\varXi}^{\rm H}\mathring{\bm{X}_{f}} \in \mathbb{C}^{M_s\times|\bar{\mathcal{A}}|}$, the symbol information corrupted by noise can be expressed as
\begin{equation}\label{equ.idft0}	 
 \mathring{\bm{X}_{p}}  = \bm{F}_{M_s}\mathring{\bm{X}_{t}} + \bm{N}_t,
\end{equation}
where $\bm{F}_{M_s}$ is the $M_s\times M_s$ DFT matrix, $\mathring{\bm{X}}_{t} = \left[\dot{\bm{x}}_{1} \cdots \dot{\bm{x}}_{|\bar{\mathcal{A}}|}\right]\in \mathcal{M}^{M_s\times |\bar{\mathcal{A}}|}$, and each column of $\mathring{\bm{X}_{t}}$ carriers the original symbol information for the corresponding active UT, while $\bm{N}_t$ is the observation noise with each of its elements following $\mathcal{N}_{c}\left(N_{t};0,\sigma_{x_{p}}^{2}\right)$. Multiplying the both sides of (\ref{equ.idft0}) by $\bm{F}_{M_s}^{\rm H}$ can lead to an estimate of $\mathring{\bm{X}_{t}}$ directly. However, such a method neglects the discrete constellation prior of symbols. We resort to variational Bayesian inference \cite{VI} for more accurate estimation. Since different UTs' information can be decoded column-by-column, we focus on the following estimation problem for the $u$-th column:
\begin{equation}\label{equ.dft}	
 \dot{\bm{x}}_{p} = \bm{F}_{M_s}\dot{\bm{x}}_{t} + \dot{\bm{n}}_{t}, 
\end{equation}
where $\dot{\bm{x}}_{p}\! =\! \left[\mathring{\bm{X}_{p}}\right]_{[:,u]}\! \in\! \mathbb{C}^{M_s}$, $\dot{\bm{x}}_{t}\! =\! \left[\mathring{\bm{X}}_{t}\right]_{[:,u]}\! \in\! \mathbb{C}^{M_s}$ and $\dot{\bm{n}}_{t}\! =\! \left[\bm{N}_{t}\right]_{[:,u]}$, with $1\le u \le |\bar{\mathcal{A}}|$. In (\ref{equ.dft}), for notation simplicity, we drop the subscript $u$, e.g., $\dot{\bm{x}}_{p,u}$ is simplified to $\dot{\bm{x}}_{p}$, since the same procedure can be applied for all the active UTs in parallel. The posterior distribution of $\dot{{x}}_{t,m} = \left[\dot{\bm{x}}_{t}\right]_m$ can be acquired from the joint distribution $p(\dot{\bm{x}}_{p},\dot{\bm{x}}_{t})$ as
\begin{equation}\label{equ.qx}	
 \ln q\left(\dot{{x}}_{t,m}\right) = \mathrm{E}_{\backslash \dot{{x}}_{t,m}}\left\{\ln p\left(\dot{\bm{x}}_{p}|\dot{\bm{x}}_{t};\bm{F}_{M_s},\sigma_{x_{p}}^{2}\right) + \ln p\left(\dot{\bm{x}}_{t}\right)\right\},
\end{equation}
where $\mathrm{E}_{\backslash \dot{{x}}_{t,m}}\{\cdot\}$ means that the expectation is taken with respect to all the variables except for $\dot{{x}}_{t,m}$. From (\ref{equ.qx}), we have 
\begin{equation} \label{equ.gm}
	\begin{aligned}
	\ln q\left(\dot{{x}}_{t,m} = s_{i}\right) & \propto \mathrm{E}_{\backslash \dot{{x}}_{t,m}}\left\{
	-\frac{1}{\sigma_{x_{p}}^{2}}{\|\bm{\dot{x}}_{p}-\mathbf{F}_{M_s}\bm{\dot{x}}_{t}\|}^2 + \ln p\left(\dot{\bm{x}}_{t}\right) \right\} \\
	& \propto -\frac{1}{\sigma_{x_{p}}^{2}} \|\bm{\dot{x}}_{p}-\bm{f}_{m}s_{i}-\sum_{m^{\prime}\neq m} \bm{f}_{m^{\prime}}\dot{x}_{t,m^{\prime}}\|^2 + \ln p\left(\dot{{x}}_{t,m} = s_{i}\right), 				
	\end{aligned}
\end{equation}
where $\bm{f}_{m} = \left[\bm{F}_{M_s}\right]_{[:,m]}$ and $\dot{{x}}_{t,m^{\prime}}$ is the corresponding expectation from last iteration. By further expanding the right-hand side of (\ref{equ.gm}) and discarding the irrelevant constant terms, we can obtain the normalized posterior distribution of $\dot{{x}}_{t,m}$, as given by  

\begin{equation}\label{equ.post_symbol}	
	q\left(\dot{{x}}_{t,m} = s_{i}\right) = \dfrac{e^{g_m \left(s_i\right)}}{\sum_{s^{\prime}\in\mathcal{M}} e^{g_m\left(s^{\prime}\right)}}, i = 1,\cdots,|\mathcal{M}|.
\end{equation}
 where $g_{m}\left(\cdot\right)$ in (\ref{equ.post_symbol}) is given by 
\begin{equation}\label{equ.fkt}	
 g_{m}(s_{i}) =  -\frac{1}{\sigma_{x_{p}}^{2}}\left(\|\bm{f}_{m}\|^2|s_{i}|^2 +  2\mathcal{R}e\left[\left(\sum_{m^{\prime}\neq m}\bm{f}_{m^{\prime}}^{\rm H}\bm{f}_{m}\dot{{x}}_{t,m^{\prime}}^{*}-\dot{\bm{x}}_{p}^{\rm H}\bm{f}_{m}\right)s_{i}\right]\right) + \ln p\left(\dot{{x}}_{t,m} = s_{i}\right).
\end{equation} 
Then the posterior mean and variance of $\dot{x}_{t,m}$ can be acquired by 
\begin{subequations}
 \begin{align}\label{equ.xt_postM} 
	\mathrm{E}\left\{\dot{x}_{t,m}\right\} &= \sum_{i=1}^{|\mathcal{M}|}s_{i}q(\dot{{x}}_{t,m} = s_{i}), \\
	\label{equ.xt_postV} 
	\mathrm{Var}\left\{\dot{x}_{t,m}\right\} &= \sum_{i=1}^{|\mathcal{M}|}|s_{i}|^2q(\dot{{x}}_{t,m} = s_{i}) - \left(\mathrm{E}\left\{\dot{x}_{t,m}\right\}\right)^2 .
 \end{align}
\end{subequations}

\begin{algorithm}[!bt]
	\KwIn{Quantized observation data ${\mathring{\bm{z}}}[n]^{(\text {lo})},{\mathring{\bm{z}}}[n]^{(\text {hi})}$, channel matrix $\mathring{\bm{H}}[n]$, AUS $\bar{\mathcal{A}}$, noise variance $\sigma_{n}^2$, maximum iteration number $N_{\mathrm{iter}}$.} 
	\KwOut{Source data bits for active UTs.}
	Initialize $\mathring{\bm{x}}_{a}[n] = \bm{0}_{|\bar{\mathcal{A}}|\times 1}$, $\mathring{\bm{y}}_{a}[n] = \bm{0}_{Q N_r\times 1}$, $\sigma_{x_a}^{2}[n] = 1$, $\forall n$, and $p\left(\dot{x}_{t,m,u}\right) = \frac{1}{|\mathcal{M}|}$, $\forall m,u$; Set iteration number $\iota=1$;\\   
	\For{$ \iota \leq N_{\mathrm{iter}}$} { 
		Calculate $\mathring{\bm{y}}_{p}[n]$, $\forall \text{ non-empty subcarrier } n$, by (\ref{equ.dequant}); \% \textbf{Module A} \\	
		Calculate $\mathring{\bm{x}}_{p}[n]$, $\forall \text{ non-empty subcarrier } n$, by (\ref{equ.lmmse}); \% \textbf{Module B} \\
		Calculate	$\sigma_{x_{p}}^{2}$ as in (\ref{PostVar}); \\
		Estimate posterior mean and variance of $\mathring{\bm{X}}_{t}$ by (\ref{equ.xt_postM}) and (\ref{equ.xt_postV}); \% \textbf{Module C} \\
		Update prior distribution of $p(\dot{{x}}_{t,m,u})$ by (\ref{equ.post_symbol}), i.e, $p(\dot{{x}}_{t,m,u}) = q(\dot{{x}}_{t,m,u}), \forall m,u$; \\ 
		Update  $\mathring{\bm{x}}_{a}[n]$ and $\sigma_{x_{a}}^{2}[n]$, $\forall n$, by (\ref{priorM}) and (\ref{priorV}); \\
		$\mathring{\bm{y}}_{a}[n] = \mathring{\bm{H}}[n]\mathring{\bm{x}}_{a}[n]$, $\forall \text{ non-empty subcarrier } n$;
	}
	Apply channel decoding to each column of $\mathring{\bm{X}}_{t}$; \\
	\Return Decoded source bits for estimated AUS.
	\caption{Multi-satellite cooperative Bayesian dequantization DD algorithm}
	\label{alg:quant_DD} 
	\LinesNumbered
\end{algorithm} 

After acquiring the posterior mean and variance of $\mathring{\bm{X}}_{t}$, the prior information $\mathring{\bm{x}}_{a}[n]$ and $\sigma_{x_{a}}^{2}[n]$ are updated. First we reintroduce the subscript $u$, e.g., $\dot{\bm{x}}_{t}$ is back to the full notation $\dot{\bm{x}}_{t,u}$ and $\dot{{x}}_{t,m}$ is back to $\dot{{x}}_{t,m,u}$. Next we define the intermediate variable $\mathring{\bm{X}}_a \in \mathbb{C}^{N\times |\mathcal{A}|}$ as 
\begin{align}\label{priorM} 
 \mathring{\bm{X}}_{a} = \bm{\varXi} \bm{F}_{M_s} \left[\dot{\bm{x}}_{t,1} \cdots \dot{\bm{x}}_{t,|\mathcal{A}|} \right] . 
\end{align}
Then $\mathring{\bm{x}}_{a}[n] = \left[\mathring{\bm{X}}_{a}\right]_{\left[n,:\right]}^{\rm T}$. Similarly, by defining the intermediate variable $\bm{v}_{a} \in \mathbb{C}^{N}$ as
\begin{align}\label{priorV} 
 \bm{v}_{a} = \bm{\varXi} \left[\frac{1}{\left|\bar{\mathcal{A}}\right|} \sum_{u=1}^{|\bar{\mathcal{A}}|} \mathrm{Var}\left\{\dot{{x}}_{t,1,u}\right\} \cdots \frac{1}{\left|\bar{\mathcal{A}}\right|} \sum_{u=1}^{|\bar{\mathcal{A}}|} \mathrm{Var}\left\{\dot{{x}}_{t,M_s,u}\right\}\right]^{\rm T} ,
\end{align}
we have $\sigma_{x_{a}}^{2}[n] = \left[{\bm{v}}_{a}\right]_{n}$. $\mathring{\bm{x}}_{a}[n]$ and $\sigma_{x_{a}}^{2}[n]$, $\forall n$, are fed back to Module~A and Module~B. Actually, with the known subcarrier mapping $\bm{\varXi}\in \mathbb{C}^{N\times M_s}$, $\sigma_{x_{a}}^{2}[n]=0$ for the empty subcarriers, and Module~A and Module~B do not need to perform inference on these empty subcarriers. 
	
The proposed multi-satellite cooperative Bayesian dequantization DD algorithm for the case of MSCBP is summarized in Algorithm~\ref{alg:quant_DD}. 
For the MSCTP case, no high-resolution observation data can be acquired. In this case, the above algorithm can still apply by simply modifying the index sets as $\mathcal{R}_{C} = \emptyset$, and $\mathcal{R}_{E} = \left\{1,\cdots,QN_r\right\}$. However, due to the loss of accurate high-resolution observations, some degradation in the system performance is inevitable.

\section{Computational Complexity}\label{S_add}
In this section, we discuss the computational complexity of the edge satellite nodes and the central server node.
For edge satellite nodes, algorithm 1 and algorithm 2 are performed. For the central serve node, algorithm 3 is performed, and LS/Algorithm 4 may serve as one of the steps in algorithm 3 in perfect/quantized backhaul cases, respectively. Therefore, we discuss the computational complexity at edge satellite nodes and the central server node separately.

Table II exhibits the computational complexity of Alg. 1 in detail. For steps in Alg. 1, the complexity of Line 1 is omitted since it can be calculated offline. During the calculation of the number of multiplications, we utilized the results from SVD for reduced complexity. For example, in Line 4 and the calculation of traces of matrices, by applying the results from SVD, lots of computations of full-dimensional matrix multiplications can be reduced to the multiplication of diagonal matrices. By neglecting the terms with much fewer computations, the overall complexity in one iteration is approximate $\mathcal{O}\left(G^2KL + 3G KLN_r\right)$. If the iteration times is $N_{\text{iter}}$, then the overall complexity of the Alg. 1 is given by $\mathcal{O}\left(G^2KLN_{\text{iter}} + 3G KLN_rN_{\text{iter}}\right)$.

The complexity of the proposed channel estimation refinement algorithm (Alg. 2) is also exhibited in Table.~\ref{TAB1}. Since Alg. 2 is non-iterative and the refinement process can be implemented for each active UT parallelly, it can be computed much faster than Alg. 1. During the main process of Alg. 2, the complexities of Line 1, and Lines 3-4 are omitted since they are far fewer than other steps. The complexity of Line 11 and Line 12 are both $\mathcal{O}\left(2N_rP \right)$ since only simple vector multiplications are involved. The 2D-ESPRIT algorithm in Lines 6-9 contributes most to the overall computations. Luckily, for each active UT, only one set of angles $\{\hat{\theta}_{k}, \hat{\phi}_{k}\}$ needs to be estimated, therefore, we only need to compute the largest singular value vector, which can be effectively solved by the existing methods such as power iteration method. The complexity for power iteration is $\mathcal{O}\left(M_{\text{sub}}G^{\prime}PI_{\text{iter}} \right)$,  where 
$M_{\text{sub}} = M_x^{\text{sub}}M_y^{\text{sub}}$ is the sub-dimensions during spatial smooth preprocessing, $G^{\prime} = G_xG_y$ is the number of selection matrices. $P$ is the maximum number of multipath component, $I_{\text{iter}}$ is the number of iteration of power iteration method. Therefore, the overall complexity of Alg. 2 is approximately $\mathcal{O}\left( K_a M_{\text{sub}}G^{\prime}PI_{\text{iter}} + 4K_aN_rP \right)$.

\begin{table}[!t]
	\centering
	\captionsetup{ justification = raggedright,labelsep=period}
	\caption{Computational Complexity at edge satellite nodes}	
	\label{TAB1}
	\resizebox{.6\columnwidth}{!}{
		\renewcommand\arraystretch{1.2}{\begin{tabular}{|cc|c|}
				\hline
				\multicolumn{1}{|c|}{\multirow{8}{*}{Algorithm 1}} & \textbf{Operation} & \textbf{Number of multiplication per iteration}                                      \\ \cline{2-3} 
				\multicolumn{1}{|c|}{} & Line 4             & $\mathcal{O}\left(G^2(KL+1) + 2G \right)$ \\ \cline{2-3} 
				\multicolumn{1}{|c|}{} & Line 5             & $\mathcal{O}\left(3GKLN_r \right)$        \\ \cline{2-3} 
				\multicolumn{1}{|c|}{} & Line 6             & $\mathcal{O}\left(2G \right)$             \\ \cline{2-3} 
				\multicolumn{1}{|c|}{} & Line 7             & $\mathcal{O}\left(16KLN_r \right)$        \\ \cline{2-3} 
				\multicolumn{1}{|c|}{} & Line 8             & $\mathcal{O}\left(KLN_r \right)$          \\ \cline{2-3} 
				\multicolumn{1}{|c|}{} & Line 9             & $\mathcal{O}\left(KLN_r \right)$          \\ \cline{2-3} 
				\multicolumn{1}{|c|}{} & Line 10 \& Line 11 & $\mathcal{O}\left(5KLN_r \right)$         \\ \hline
				\multicolumn{2}{|c|}{\textbf{Overall Complexity}}                       & $\mathcal{O}\left(G^2KLN_{\text{iter}} + 3G KLN_rN_{\text{iter}}\right)$    \\ \hline \hline 
				\multicolumn{1}{|c|}{\multirow{4}{*}{Algorithm 2}} & \textbf{Operation} & \textbf{Number of multiplication}                                                    \\ \cline{2-3} 
				\multicolumn{1}{|c|}{}                             & Line 6-9    & $\mathcal{O}\left(M_{\text{sub}}G^{\prime}PI_{\text{iter}} \right)$         \\ \cline{2-3} 
				\multicolumn{1}{|c|}{} & Line 11             & $\mathcal{O}\left(2N_rP \right)$          \\ \cline{2-3} 
				\multicolumn{1}{|c|}{} & Line 12             & $\mathcal{O}\left(2N_rP \right)$          \\ \hline
				\multicolumn{2}{|c|}{\textbf{Overall Complexity}}                       & $\mathcal{O}\left(M_{\text{sub}}G^{\prime}PI_{\text{iter}} + 4N_r P\right)$ \\ \hline
	\end{tabular}}}
	 \vspace{-5mm}
\end{table}

According to TABLE~\ref{TAB1}, the process that contributes most to the overall computational complexity of JADCE is mainly matrices multiplications. Therefore, it is reasonable to assume the edge LEO satellites to have such computation power. In the next, we discuss the complexity of Algorithm 3 and Algorithm 4. For MSCTP, this process is performed by the terrestrial server node, while for MSCBP, this process is performed by the central satellite node with super computation power.  

The computational complexities of the central server node (Alg. 3 and Alg. 4) are exhibited in TABLE~\ref{TAB2}. In Alg. 3, the complexity of the server node only involves Lines 3 - 8. The complexity of Line 4 is neglected since it is much less than other steps. Line 5 and Line 7 mainly involve the matrices multiplication with the DFT matrix, which can be efficiently solved by the fast Fourier transform (FFT) algorithm. When the dimension of the matrix is the power of 2, the complexity of Line 5 and Line 7 are respectively given by $\mathcal{O}\left(QK_aN_rN\log N + QN_rN\log N\right) = \mathcal{O}\left(QK_aN_rN\log N \right)$ and $\mathcal{O}\left(K_aM_s\log M_s \right)$, respectively. Therefore, the overall complexity of Alg. 3 is mainly dominated by Line 6.

In Line 6 of Alg. 3, according to whether perfect backhaul or quantized backhaul is performed, we use the LS algorithm/Alg. 4 to solve the problems Equ. (41)/Equ. (42), respectively. The overall computational complexity of DD at the central server node is also shown in TABLE~\ref{TAB2}, where the specific complexity of Alg. 4 is further elaborated. By further assuming $M_s = N$, and setting the iteration number of Alg. 4 to $N_\text{iter}$, the overall complexity of Algorithm is 
$\mathcal{O}\left(N_\text{iter}NK_a^3 + 2N_\text{iter}NQN_rK_a^2 + N_\text{iter}N^2|\mathcal{M}|\right)$, which is also the approximate complexity for DD in MSCTP and MSCBP schemes. Therefore, a relatively higher computation ability is required at the central server node. 

\begin{table}[!t]
	\centering
	\captionsetup{ justification = raggedright,labelsep=period}
	\caption{Computational Complexity at the central server node}	
	\label{TAB2}
	\resizebox{.6\columnwidth}{!}{
		\renewcommand\arraystretch{1.2}{\begin{tabular}{|cc|c|}
				\hline
				\multicolumn{1}{|c|}{\multirow{5}{*}{Algorithm 3}} &
				\textbf{Operation} &
				\textbf{Number of multiplication} \\ \cline{2-3} 
				\multicolumn{1}{|c|}{}          & Line 5          & $\mathcal{O}\left(QK_aN_rN\log N\right)$     \\ \cline{2-3} 
				\multicolumn{1}{|c|}{} &
				\begin{tabular}[c]{@{}c@{}}Line 6\\ Problem (41)\end{tabular} &
				$\mathcal{O}\left(NK_a^3 + 2NQN_rK_a^2 + NQN_rK_a\right)$ \\ \cline{2-3} 
				\multicolumn{1}{|c|}{} &
				\begin{tabular}[c]{@{}c@{}}Line 6\\ Problem (42)\end{tabular} &
				$ \text{See Algorithm 4} $ \\ \cline{2-3} 
				\multicolumn{1}{|c|}{}          & Line 7          & $\mathcal{O}\left(16KLN_r \right)$           \\ \hline 
				\multicolumn{2}{|c|}{\textbf{Overall Complexity}} & Dominated by Line 6                          \\ \hline \hline
				\multicolumn{1}{|c|}{\multirow{6}{*}{Algorithm 4}} &
				\textbf{Operation} &
				\textbf{Number of multiplication per iteration} \\ \cline{2-3} 
				\multicolumn{1}{|c|}{}          & Line 3          & $\mathcal{O}\left(6QNN_r\right)$             \\ \cline{2-3} 
				\multicolumn{1}{|c|}{} &
				Line 4 &
				$\mathcal{O}\left(NK_a^3 + 2NQN_rK_a^2 + 2NQN_rK_a\right)$ \\ \cline{2-3} 
				\multicolumn{1}{|c|}{}          & Line 6          & $\mathcal{O}\left(M_s^2|\mathcal{M}|\right)$ \\ \cline{2-3} 
				\multicolumn{1}{|c|}{}          & Line 8          & $\mathcal{O}\left(K_aM_s\log M_s \right)$    \\ \cline{2-3} 
				\multicolumn{1}{|c|}{}          & Line 9          & $\mathcal{O}\left(NQN_rK_a \right)$          \\ \hline
				\multicolumn{2}{|c|}{\textbf{Overall Complexity}} &
				$\mathcal{O}\left(N_\text{iter}NK_a^3 + 2N_\text{iter}NQN_rK_a^2 + N_\text{iter}N^2|\mathcal{M}|\right)$ \\ \hline
	\end{tabular}}}
\vspace{-5mm}
\end{table}

\section{Simulation Results}\label{S5}

This section conducts simulations to validate the superiority of our proposed schemes. We consider a typical massive access scenario with $Q = 3$ LEO satellites and $K = 100$ potential UTs which are uniformly distributed in a triangular area formed by the satellites' projection onto the ground, as shown in Fig.~\ref{fig4_access}. The sides of the triangle and the satellite-to-satellite distance are both $500$\,km and the altitude of the satellites is $550$\,km.\footnote{Outside this triangular area, the interference from neighboring cells can be conveniently handled in our proposed framework. For the JADCE problem, the proposed methods can still apply by enlarging the problem size. Besides, the multi-satellite cooperative DD algorithm can be flexibly configured by allocating different frequency resources (subcarriers) to different cells. Another method to eliminate interference is to dynamically configure the time-frequency domain resources, e.g. \cite{RKinf}. Hence, for performance evaluation convenience, the interference from outside the cell is ignored.} In view of the large-scale fading difference caused by the vast geometric range of all the potential UTs and high-speed relative movement between UTs and LEO satellites, we assume that power control and Doppler compensation are performed at the UT side, and this can be achieved coarsely by leveraging the prior extrinsic information of deterministic LEO satellites' trajectory and known UTs' geographical position \cite{Doppler_Comp1,Doppler_Comp2}. Other simulation parameters are set as follows. The carrier frequency is set to $f_c = 14.5$\,GHz \cite{SpaceX}, and the subcarrier spacing and system bandwidth are set to $480$\,kHz and $259.2$\,MHz, respectively. We adopt the QPSK modulation with DFT-s-OFDM waveform, and the data size of the DFT-s-OFDM is set to $\left(M_s, N\right) = \left(540, 540\right)$. The training sequence $\bm{c}_k$ for each UT is randomly generated from a standard normal distribution. The size of each satellite antenna array is set to $N_x=N_y=10$ unless otherwise specified. The power distribution factor among multipath and the pulse shaping filter are set to $K_f = 10 $ dB and $r\left(\tau\right) = \delta\left(\tau\right)$, respectively. In addition, the maximum  RCD is set to $L = 17$. The performance evaluation metrics for AD, CE, and DD are defined as user activity error probability (AEP), normalized mean square error (NMSE), and BER, given respectively as
\begin{align}\label{AEP} 
 \mathrm{AEP} = \frac{1}{K}\sum_{k=1}^{K}|{\mathring{\alpha}}_k-\alpha_k|,  
 \mathrm{NMSE} = \frac{\parallel\bar{\bm{H}}-\tilde{\bm{H}}\parallel^2_{\mathrm{F}}}{\parallel\tilde{\bm{H}}\parallel^2_{\mathrm{F}}},
 \mathrm{BER} = \frac{E_{{\mathring{\mathcal{A}}}\cap \mathcal{A}} + P_sM_sM_\mathrm{ord}|\mathcal{A}\textbackslash{\mathring{\mathcal{A}}}|}{P_sM_{s}M_{\mathrm{ord}}{|\mathcal{A}|}},
\end{align}
where $\bar{\bm{H}} = \left[\bar{\bm{H}}_{1} ~ \bar{\bm{H}}_{2}\cdots \bar{\bm{H}}_{Q}\right]$ is the estimated channel, and $\tilde{\bm{H}} = \left[\tilde{\bm{H}}_{1} ~ \tilde{\bm{H}}_{2}\cdots \tilde{\bm{H}}_{Q}\right]$ is the true channel, while for $\mathrm{BER}$, $E_{{\mathring{\mathcal{A}}}\cap \mathcal{A}}$ is the total number of error bits for correctly estimated AUS, $|\mathcal{A}\textbackslash{\mathring{\mathcal{A}}}|$ is the number of missed detected UTs, $P_s$ is the number of the total transmit frames, and $M_{\mathrm{ord}}$ is the modulation order.

\subsection{Performance Evaluation with Perfect Backhaul}\label{S5.1}

\subsubsection{Comparison with Existing Schemes}

\begin{figure*}[!tb]
	\vspace{-5mm}
	\centering
	\captionsetup{font={footnotesize}, singlelinecheck = off,name={Fig.},justification=centering, labelsep=period}
	\hspace{-4mm}
	\subfigure[CE performance]{
		\includegraphics[width=.345\columnwidth,keepaspectratio]{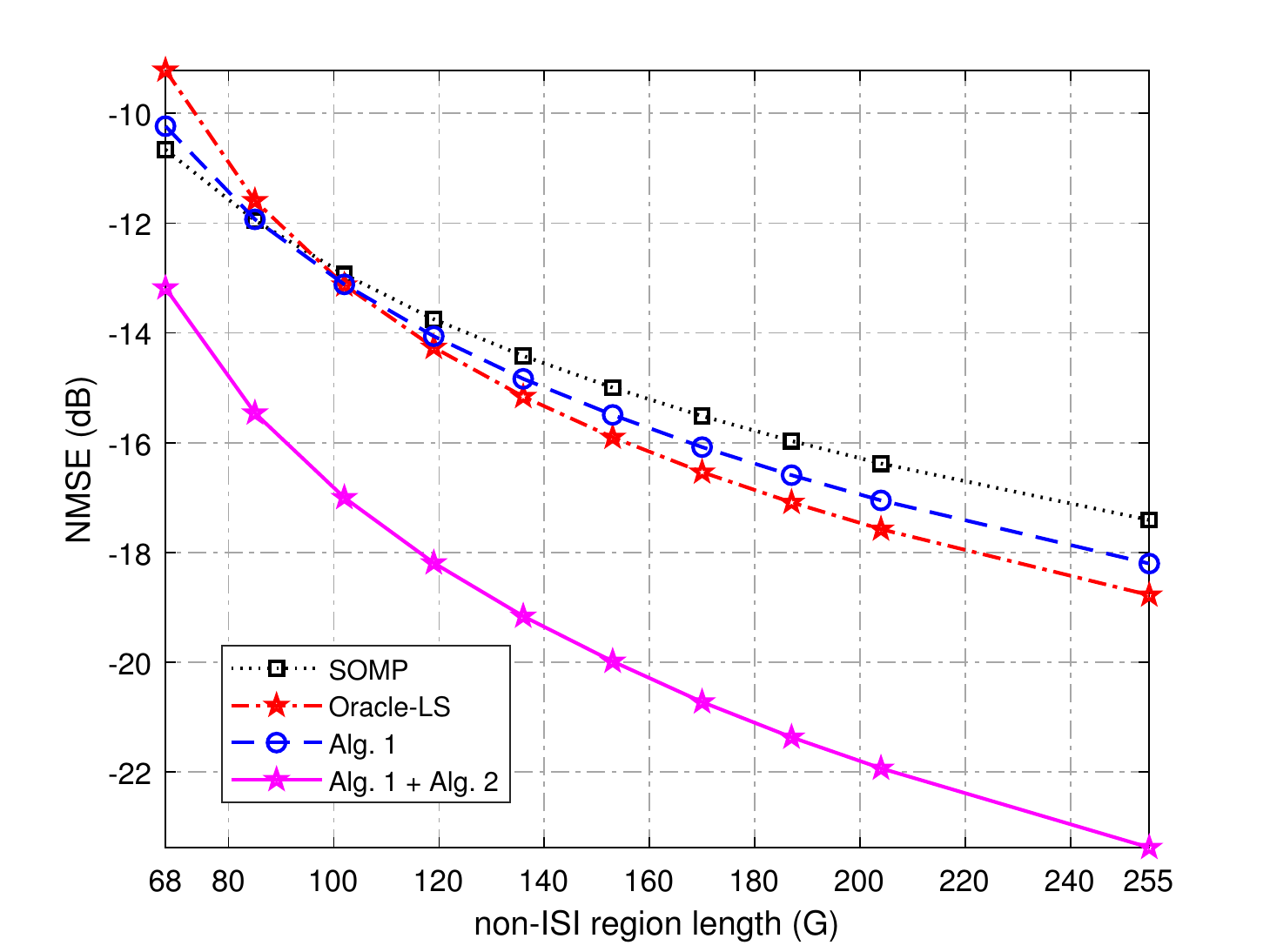}
	}
	\hspace{-9mm}
	\subfigure[AD performance]{
		\includegraphics[width=.345\columnwidth,keepaspectratio]{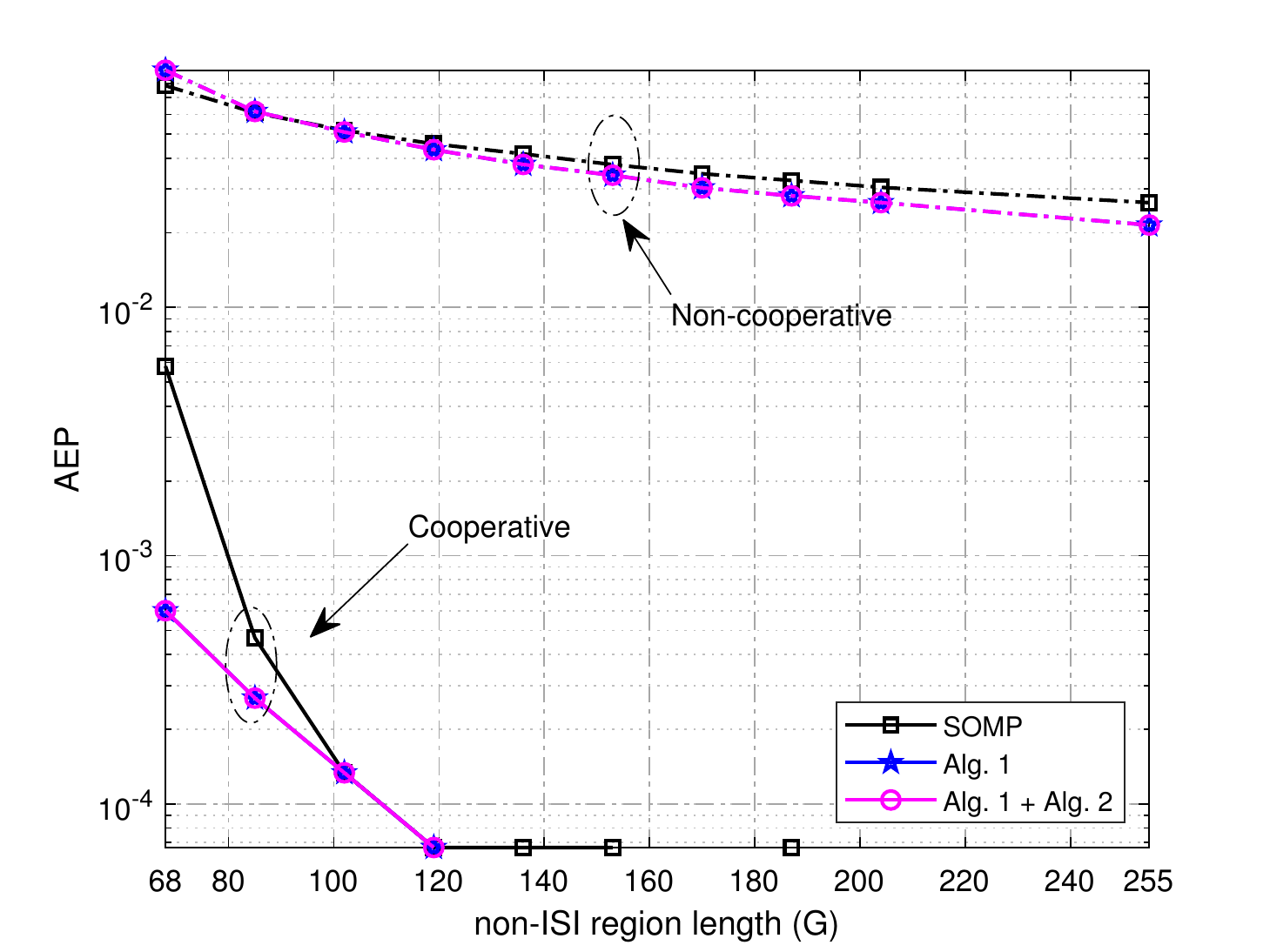}
	}
	\hspace{-9mm}
	\subfigure[DD performance]{
		\includegraphics[width=.345\columnwidth,keepaspectratio]{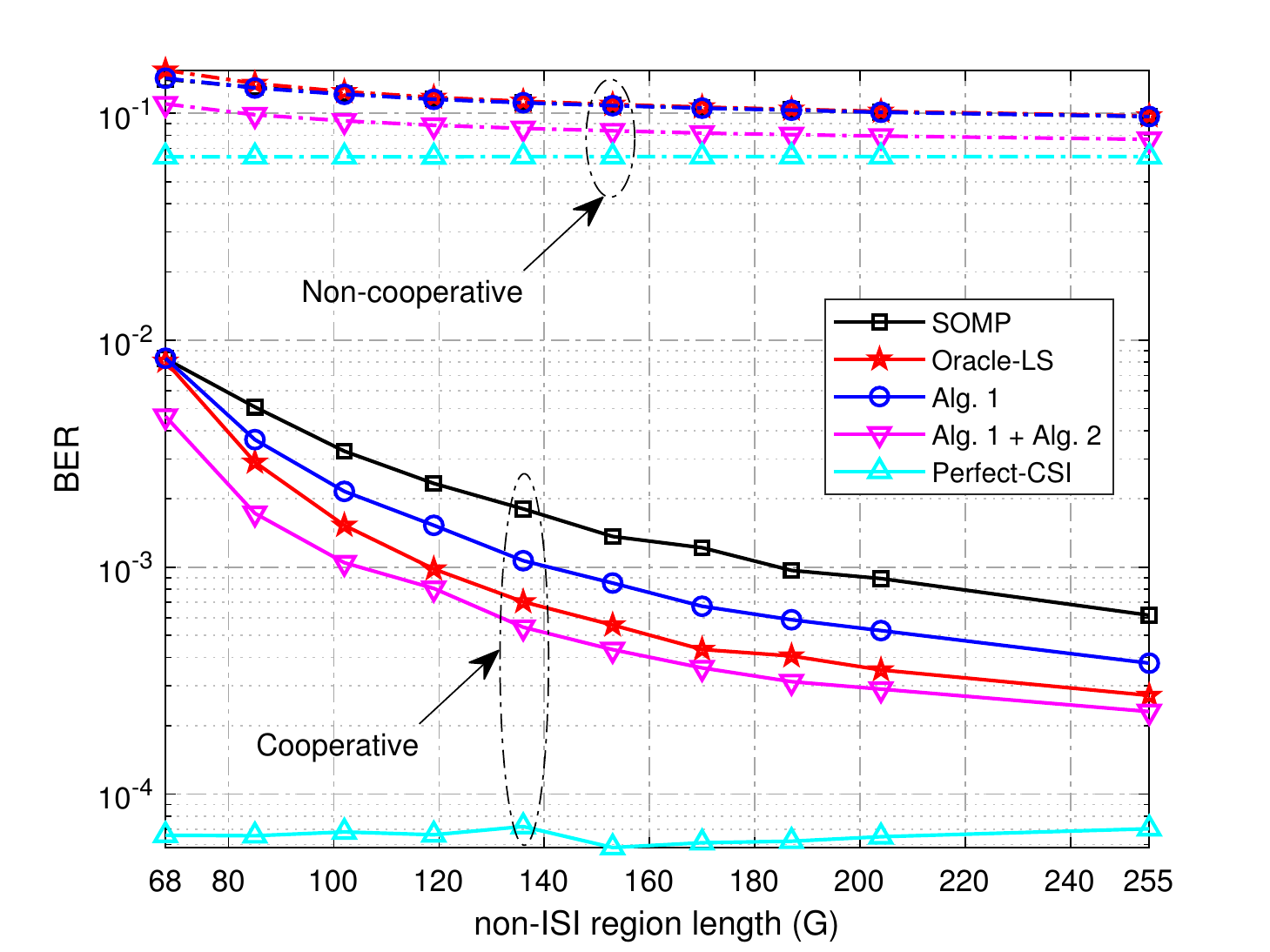}
	}
	\caption{Comparison of different schemes given different non-ISI region length, with $ K_a = 15$, $P = 3$ and $\text{SNR}=12$\,dB.}
	\label{Fig7}
	\vspace{-8mm}
\end{figure*}

We first compare our schemes with the classical simultaneous orthogonal matching pursuit (SOMP) algorithm \cite{SOMP} and Oracle-LS method \cite{Oracle_LS}. Fig.~\ref{Fig7} shows the NMSE, AEP, and BER performance as the functions of the non-ISI region length achieved by various algorithms. In this simulation, the signal-to-noise ratio (SNR) is set to $12$\,dB, the number of active UTs is $K_a\! =\! 15$, and each TSL has $P\! =\! 3$ paths. In addition, `Alg.\,1' means that only Algorithm~\ref{alg:JADCE} is executed at the satellite node when performing CE, and `Alg.\,1~+~Alg.\,2' stands for both Algorithms~\ref{alg:JADCE} and \ref{alg:CER} are serially executed at the satellite node. For the array manifold information of the receiver, Algorithm~\ref{alg:JADCE} only utilizes the common sparsity property among different antennas, which achieves good performance when the non-ISI region is sufficiently large. Because the high-resolution 2D-ESPRIT algorithm captures the angle information of the channel accurately, a significant performance improvement can be further obtained by Algorithm~\ref{alg:CER}. Oracle-LS \cite{Oracle_LS} assumes that the support sets of $\left\{\tilde{\bm{H}}_{q}\right\}_{q=1}^{Q}$ are known, and it conducts LS estimation of the non-zero rows in the CIRM. Since the true AUS can be acquired from the known support sets of $\left\{\tilde{\bm{H}}_{q}\right\}_{q=1}^{Q}$, Oracle-LS does not perform AD.
It can be seen from Fig.~\ref{Fig7}\,(a) that our Alg.\,1 outperforms SOMP, and Oracle-LS with the idealized perfectly known support set is better than Alg.\,1 when $G > 100$, while our combined Alg.\,1~+~Alg.\,2 dramatically outperforms the other schemes, in terms of CE performance. In particular, over the range of the non-ISI region lengths tested, our Alg.\,1~+~Alg.\,2 exhibits an NMSE performance gain of $3$ to $5$\,dB over our Alg.\,1. Clearly, benefiting from the high-resolution estimation of AoAs, Algorithm~\ref{alg:CER} can significantly enhance CE performance.

Fig.~\ref{Fig7}\,(b) compares the AEP performance of the three algorithms with two different processing schemes. Specifically, based on the estimated CIRM by different CE algorithms, we use the same energy-based detector of (\ref{equ:AD}) to acquire the support set of CIRM, and we compare the non-cooperative processing with cooperative processing. Here, `non-cooperative' means that the detection of AUS, (\ref{equ:AD1}), is executed independently at each satellite, and the AEP for the non-cooperative processing is defined as $\mathrm{AEP} = \frac{1}{KQ}\sum_{q=1}^{Q}\sum_{k=1}^{K}|{\hat{\alpha}}_{k,q}-\alpha_k|$. By contrast, `cooperative' means that the majority voting scheme of (\ref{AUS_vote}) is further adopted to refine the AD, and the AEP is defined in (\ref{AEP}). It can be seen from Fig.~\ref{Fig7}\,(b) that the two proposed methods achieve slightly better AEP performance than SOMP in non-cooperative processing. Also, our two methods outperform SOMP considerably in cooperative processing. It can be observed that cooperative processing dramatically improves the AEP performance over non-cooperative processing. This can be explained intuitively as follows. When one LEO satellite misjudges the active state of one UT due to unfavorable channel conditions between this UT and the LEO, observations from other LEO satellites can correct this mistake to a large extent. In addition, Alg.\,1~+~Alg.\,2 produces the same AEP performance as Alg.\,1, since no further AD refinement is executed in Algorithm~\ref{alg:CER}.

In the BER performance comparison depicted in	Fig.~\ref{Fig7}\,(c), we also include the idealized Perfect-CSI case, which uses the true CIRM $\left\{\tilde{\bm{H}}_{q}\right\}_{q=1}^{Q}$ and the true AUS $\mathcal{A}$ to perform DD. In this figure, non-cooperative processing means that (\ref{distributed_dd}) is utilized to perform equalization at each satellite independently, and the final BER is defined by the average result of all satellites, while cooperative processing means that (\ref{LS_equalization}) is utilized to perform equalization, as proceeded in Algorithm \ref{alg:UIC-CDD}, and its BER is defined in (\ref{AEP}). In the non-cooperative case, SOMP, Oracle-LS, and Alg.\,1 have similar BER performance, while our Alg.\,1~+~Alg.\,2 clearly achieves a better BER performance because it has a more accurate CE result. For the cooperative case, our Alg.\,1 outperforms SOMP, and Oracle-LS with the perfectly known support set is better than Alg.\,1, while our Alg.\,1~+~Alg.\,2 outperforms Oracle-LS, in terms of DD. Observe that the cooperative processing outperforms the non-cooperative processing by 2 to 3 orders of magnitude. This is because cooperative processing can suppress the high channel correlation effect, while non-cooperative processing is prone to the ill-conditioned channel condition. 

\begin{figure*}[!h]
	\centering
	\captionsetup{font={footnotesize}, singlelinecheck = off,name={Fig.},justification=centering, labelsep=period}
	\hspace{-4mm}
	\subfigure[CE performance]{
	  \vspace{-6mm}
		\includegraphics[width=.345\columnwidth,keepaspectratio]{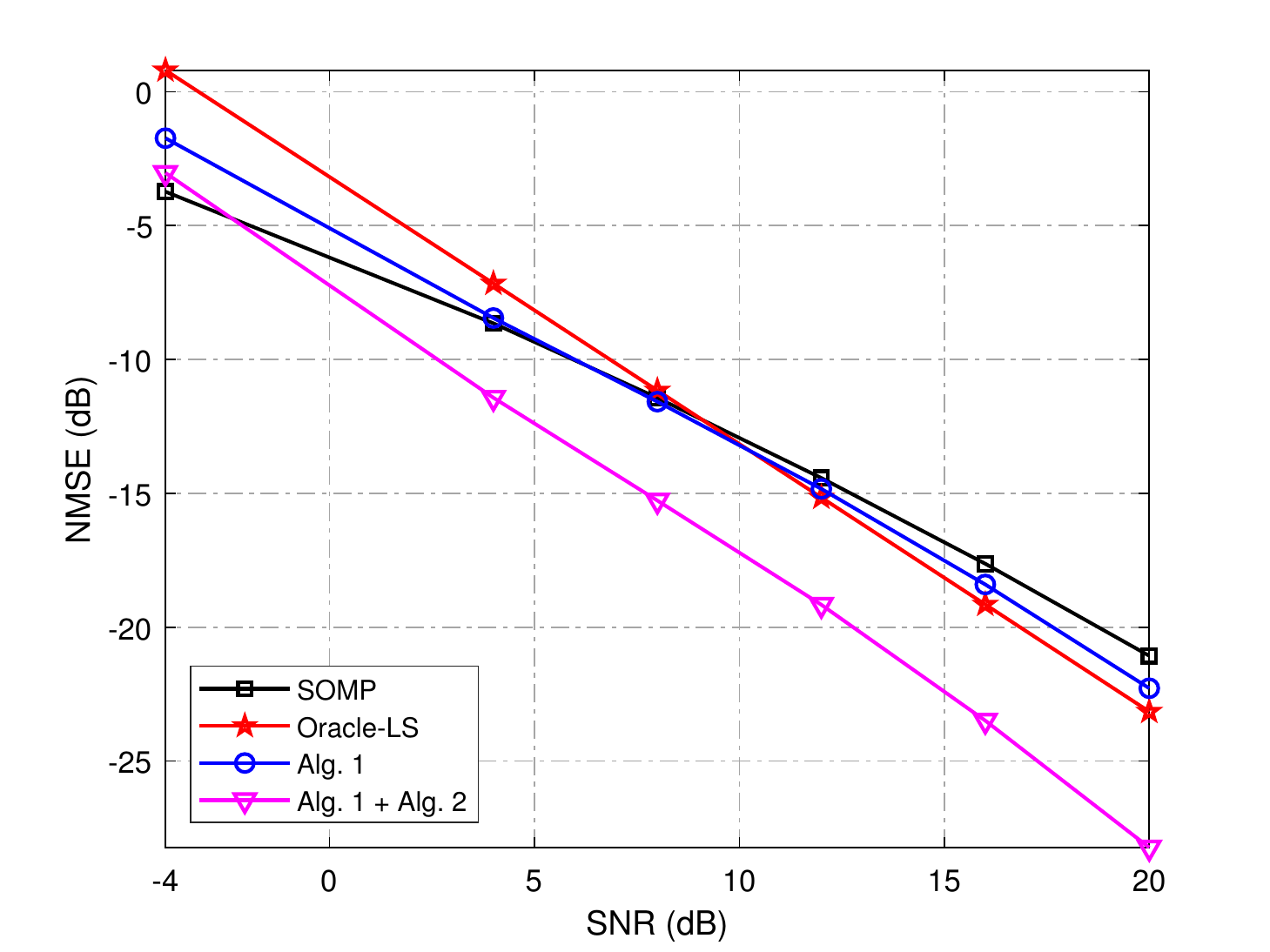}
	}
	\hspace{-9mm}
	\subfigure[AD performance]{
		\vspace{-6mm}
		\includegraphics[width=.345\columnwidth,keepaspectratio]{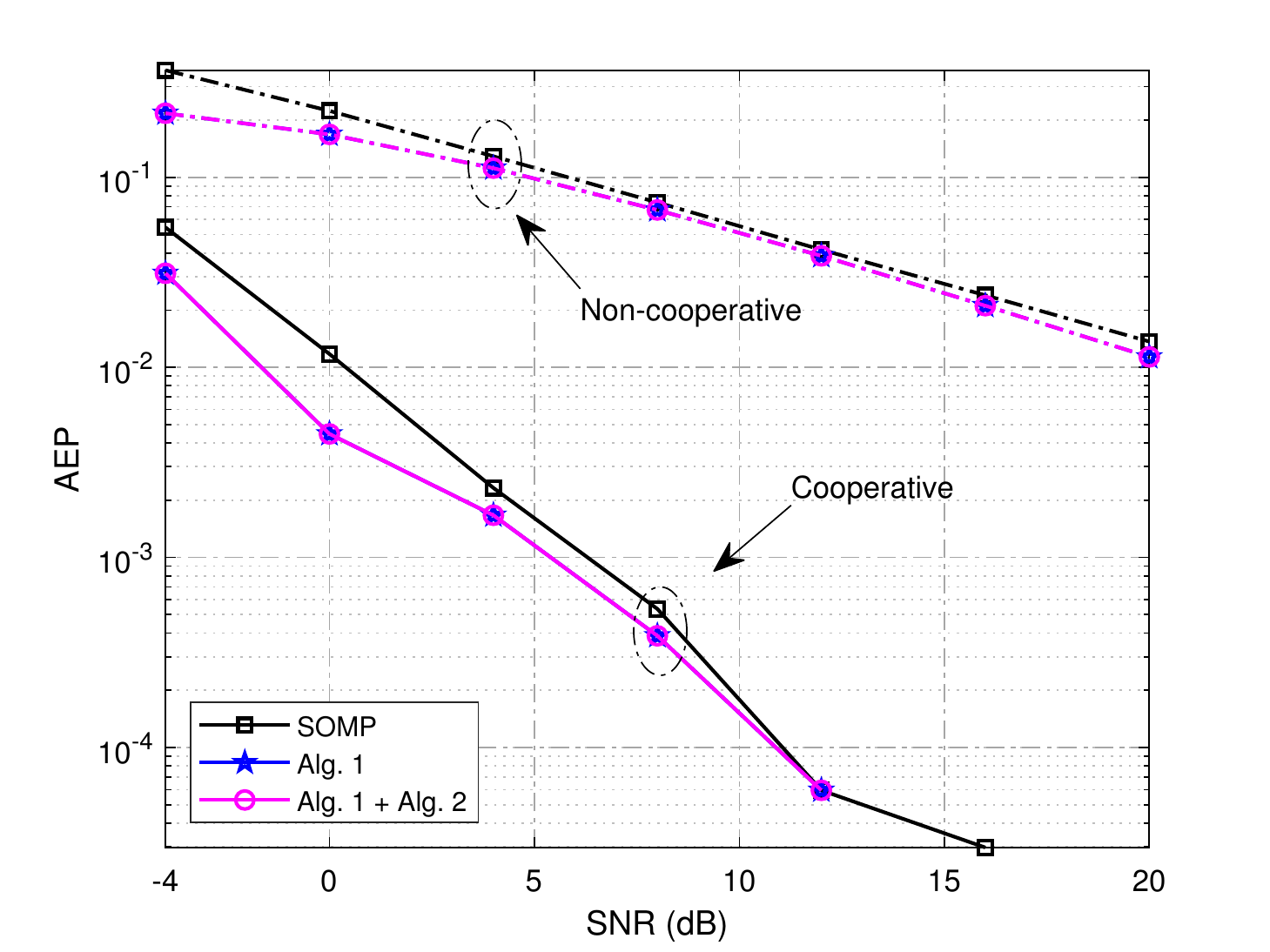}
	}
	\hspace{-9mm}
	\subfigure[DD performance]{
		\vspace{-6mm}
		\includegraphics[width=.345\columnwidth,keepaspectratio]{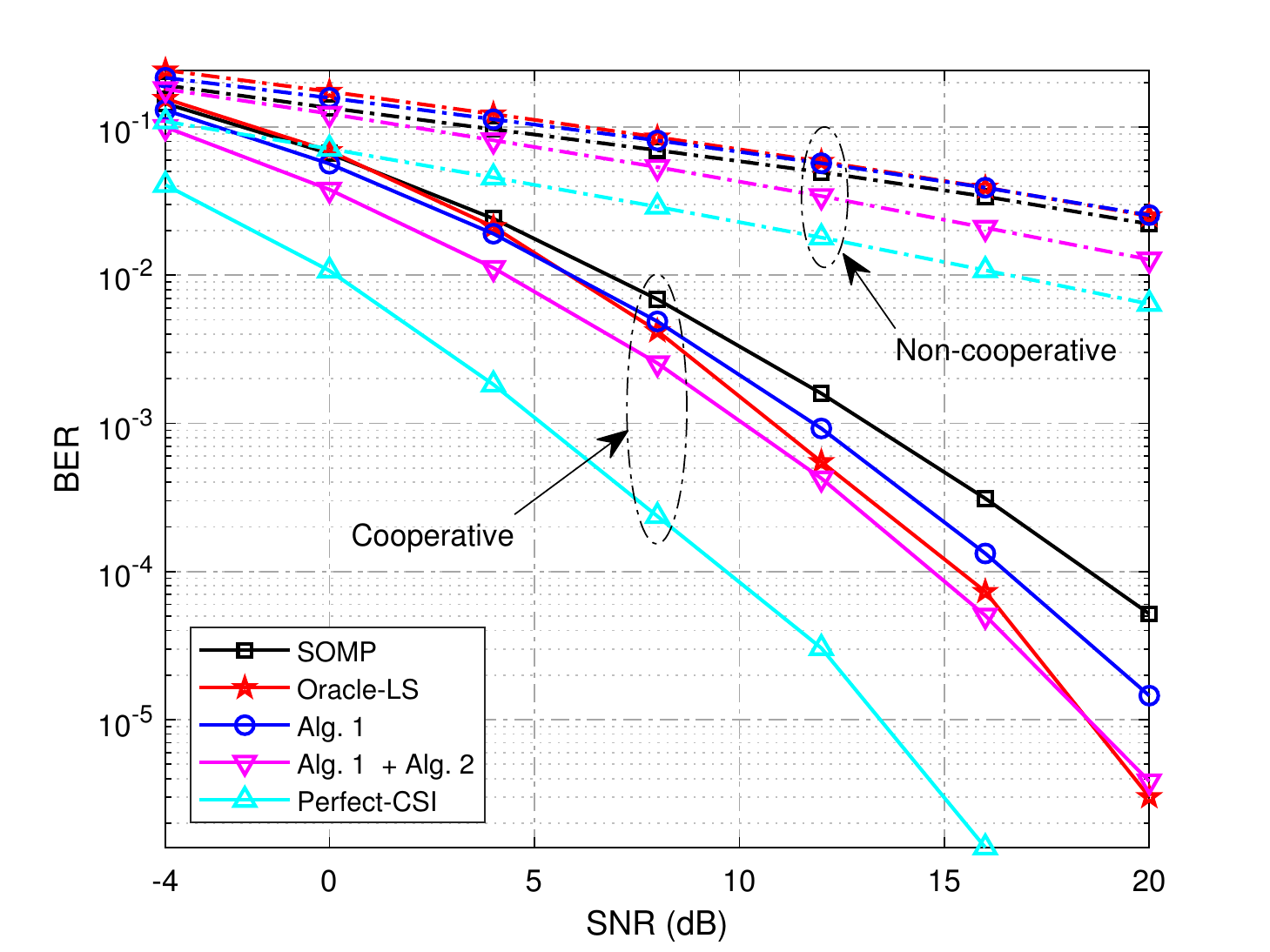}
	}
\caption{Comparison of different schemes given different SNR values, with $G = 136$, $ K_a = 15$ and $P = 3$.}
\label{Fig8}
\end{figure*}

Fig.~\ref{Fig8} depicts the NMSE, AEP, and BER performance as the functions of SNR achieved by various algorithms, given the non-ISI region length $G = 136$, the number of active UTs $K_a = 15$ and the number of multipath components $P = 3$. It can be observed from Fig.~\ref{Fig8}\,(a) that Oracle-LS is inferior to our Alg.\,1 in the low SNR region but the former becomes better than the latter in the high SNR region. Observe that the NMSE performance gap between Alg.\,1 and Alg.\,1~+~Alg.\,2 increases with the improvement of SNR. This is because higher SNR provides a better initial estimation of the CIRM for 2D-ESPRIT, which in turn is better for the exploitation of the structure of the antenna array. Similar to Fig.~\ref{Fig7}\,(b), our Alg.\,1 and Alg.\,1~+~Alg.\,2 have the same AEP performance, and they outperform SOMP, as can be seen from Fig.~\ref{Fig8}\,(b). Also cooperative processing yields significant better AEP than non-cooperative processing. Besides, similar observations to those for Fig.~\ref{Fig7}\,(c) can be drawn regarding the BER performance of different algorithms depicted in Fig.~\ref{Fig8}\,(c). In particular, the BER of our Alg.\,1~+~Alg.\,2 is the closest to that of the perfect-CIR case.

\subsubsection{Comparison of Different Numbers of Active UTs}

In this part, we investigate the impact of the number of active UTs on the performance of the proposed schemes. 

\begin{figure*}[!t]
	\vspace{-3mm}
	\centering
	\captionsetup{font={footnotesize}, singlelinecheck = off,name={Fig.},justification=centering, labelsep=period}
	\hspace{-4mm}
	\subfigure[CE performance]{
		\includegraphics[width=.345\columnwidth,keepaspectratio]{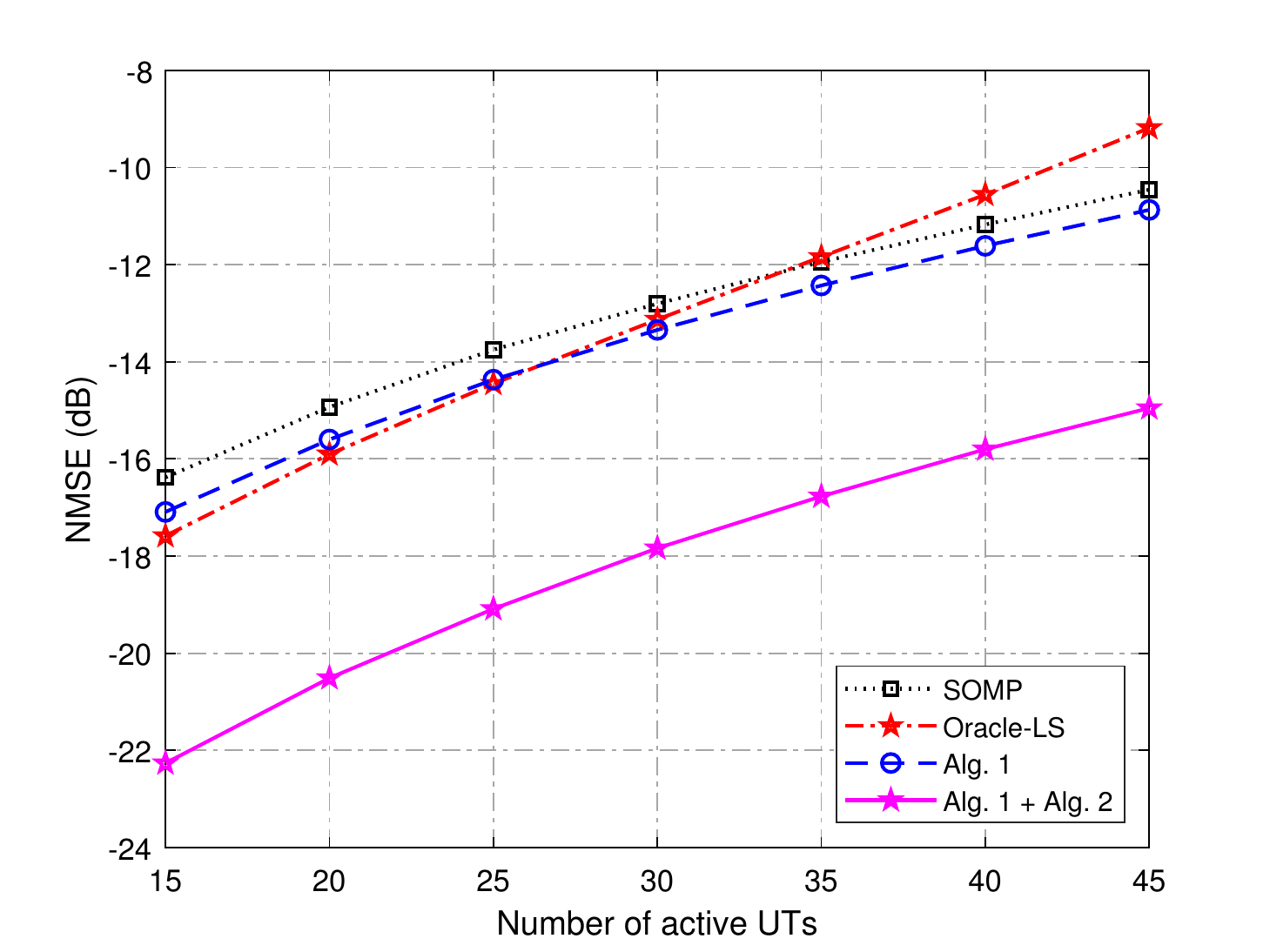}
	}
	\hspace{-9mm}
	\subfigure[AD performance]{
		\includegraphics[width=.345\columnwidth,keepaspectratio]{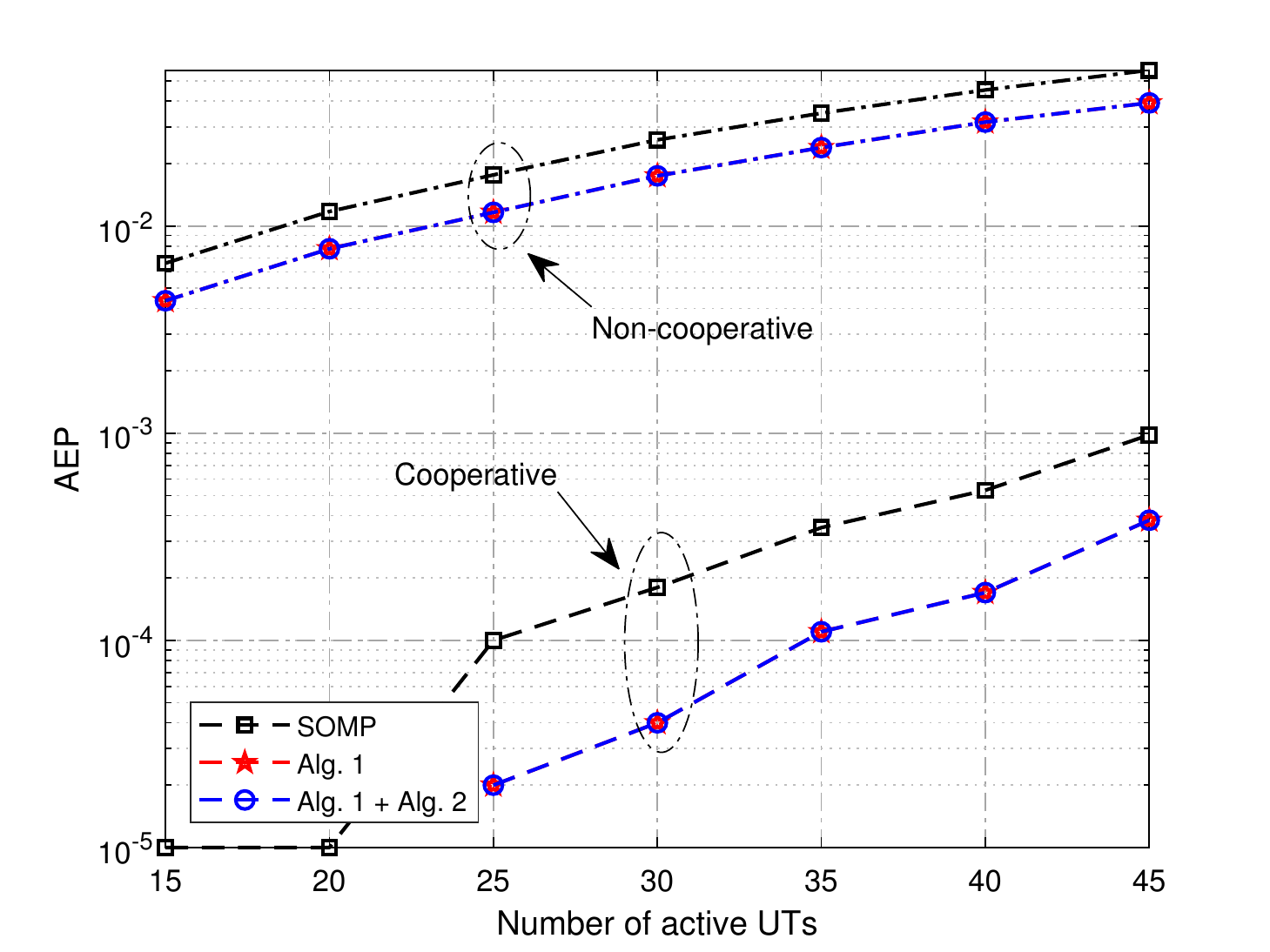}
	}
	\hspace{-9mm}
	\subfigure[DD performance]{
		\includegraphics[width=.345\columnwidth,keepaspectratio]{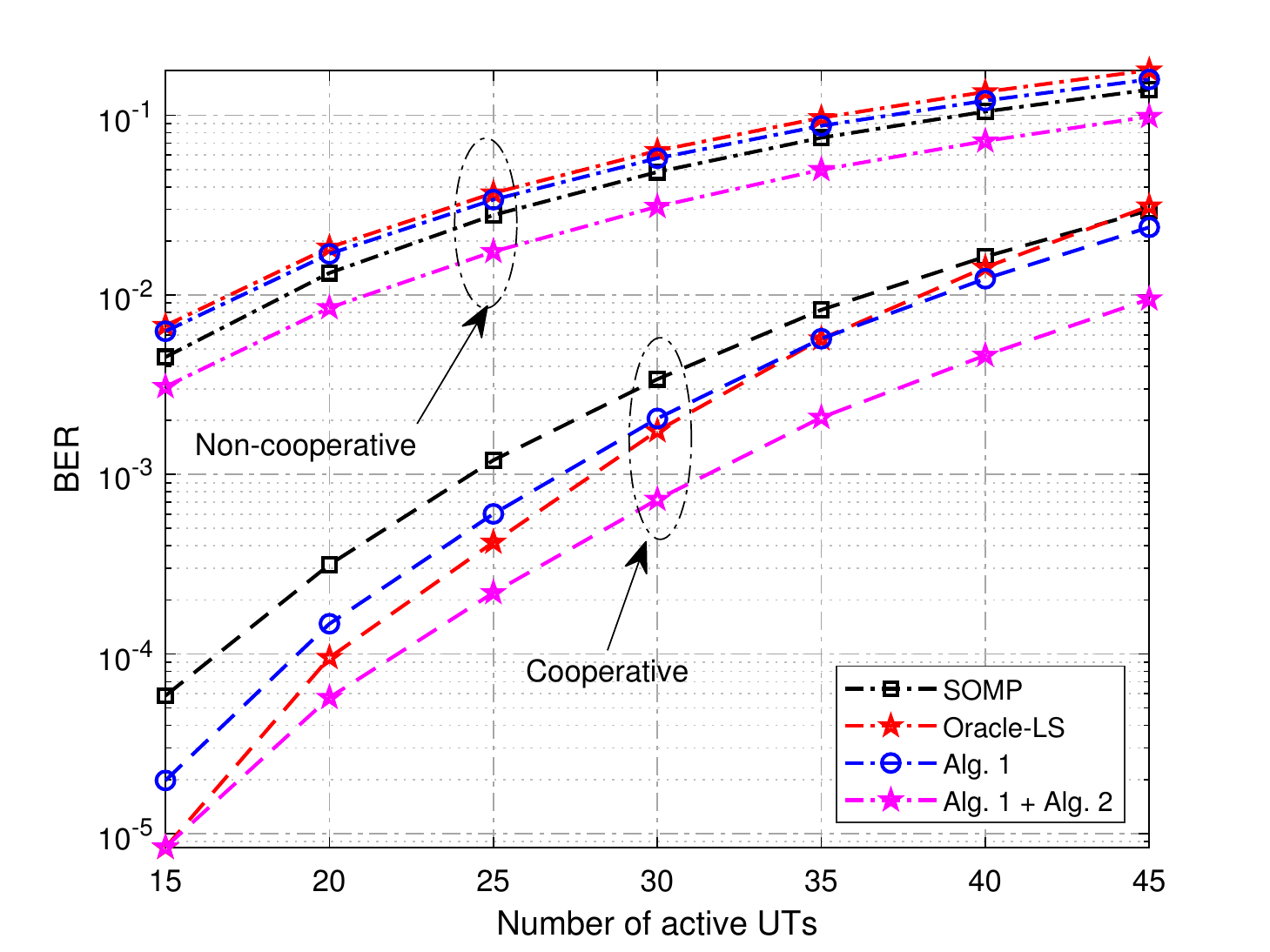}
	}
	\caption{Comparison of different schemes given different numbers of active UTs, with $G = 136$, $P = 2$ and SNR $ = 12$ dB.}
	\label{FigSim4}
	\vspace*{-3mm}
\end{figure*}

In Fig.~\ref{FigSim4}, the NMSE, AEP, and BER performance as the functions of the number of active UTs are depicted under various algorithms, given the non-ISI region length $G = 136$, the received $\text{SNR} = 12 \text{dB}$, and the number of multipath components $P = 2$. It can be observed from Fig.~\ref{FigSim4}\,(a) that the proposed Alg. 1 + Alg. 2 yields the best CE performance. As expected, increasing the number of active UTs $K_a$ degrades all the CE, AD, and DD performance. 
Similar to Fig.~\ref{Fig8}, the Alg.\,1~+~Alg.\,2 outperforms the other algorithms under  different $K_a$. Also, cooperative processing yields significantly better AEP and BER performance than non-cooperative processing. Note that the BER performance deteriorates obviously when $K_a$ is large. This is because, for the limited size of the received antenna array, when the number of active UTs is too large, the multi-user interference also becomes more severe in the spatial domain.  

\begin{figure*}[!h]
	\centering
	\captionsetup{font={footnotesize}, singlelinecheck = off,name={Fig.},justification=centering, labelsep=period}	
		\centering
		\includegraphics[scale=0.5]{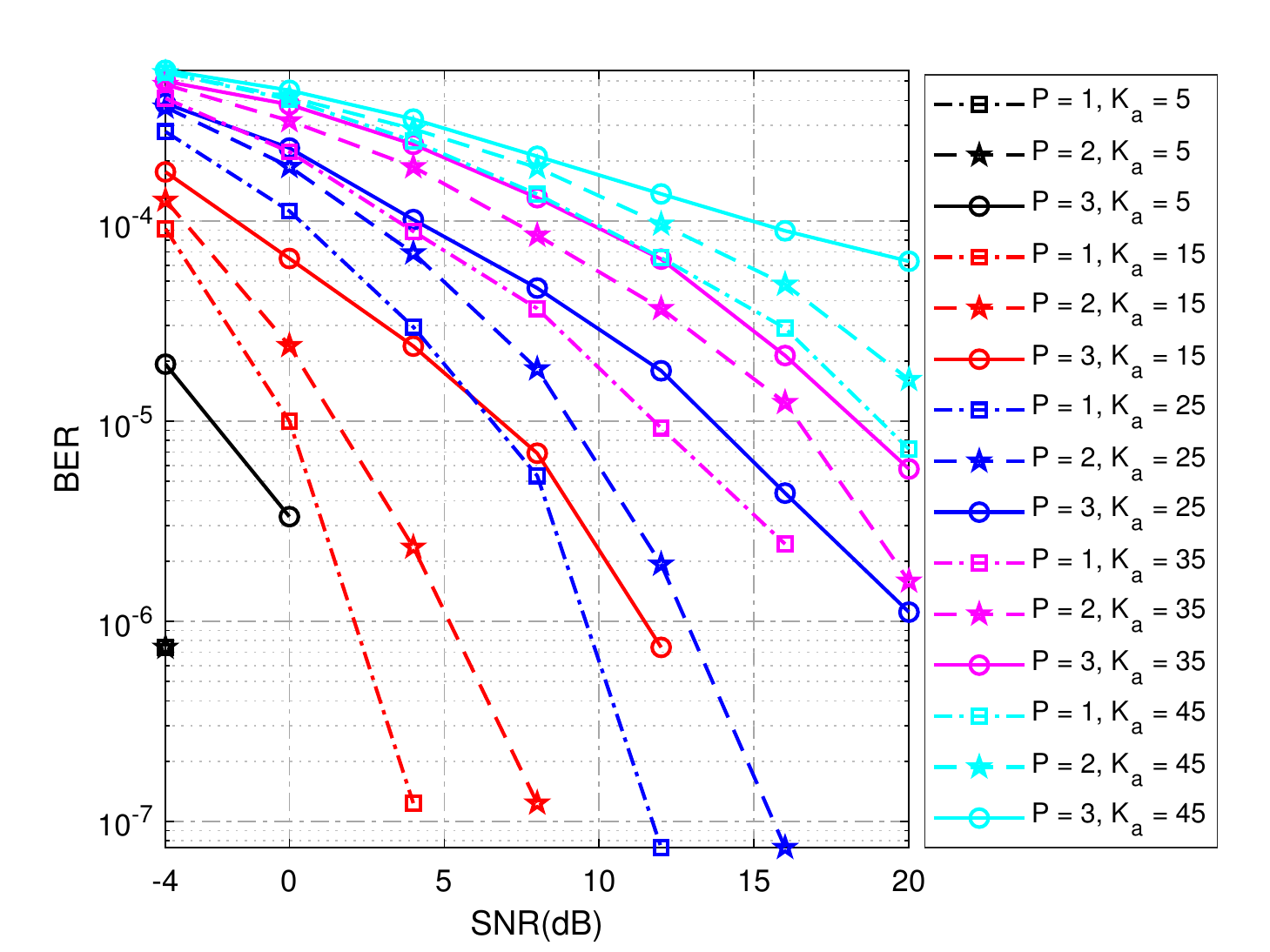}
		\caption{Cooperative DD performance as the functions of SNR given different $K_a$ and $P$, where $G = 136$.}
		\label{Fig10}
\end{figure*}	
		
In Fig.~\ref{Fig10}, we examine the BER performance of the cooperative DD method (\ref{LS_equalization}) as the functions of SNR under various combinations of the  number of multipath components $P$ and the number of active UTs $K_a$, with the non-ISI region length fixed to $G = 136$. The BER results are obtained based on the estimated CIRM from Alg.\,1~+~Alg.~2. As expected, given the same $K_a$, the BER performance degrades as $P$ increases. This indicates that in SatCom, if we can effectively perform beamforming at the UT side to concentrate more energy on the LoS path, the channel energy of NLoS paths can be suppressed to promise a better BER. Also, given the same $P$, increasing the number of active UTs $K_a$ leads to the increase of multi-user interference, which degrades the BER directly.	

\subsubsection{Comparison of the System Scalability}\label{S5.a2}
In this part, we investigate the scalability of the system, where the impacts of the numbers of cooperative satellite nodes, the numbers of potential UTs, and the size of receive array are investigated. 

\begin{figure*}[!h]
	\centering
	\captionsetup{font={footnotesize}, singlelinecheck = off,name={Fig.},justification=centering, labelsep=period}
	\hspace{-4mm}
	\subfigure[CE performance]{
		\vspace{-6mm}
		\includegraphics[width=.345\columnwidth,keepaspectratio]{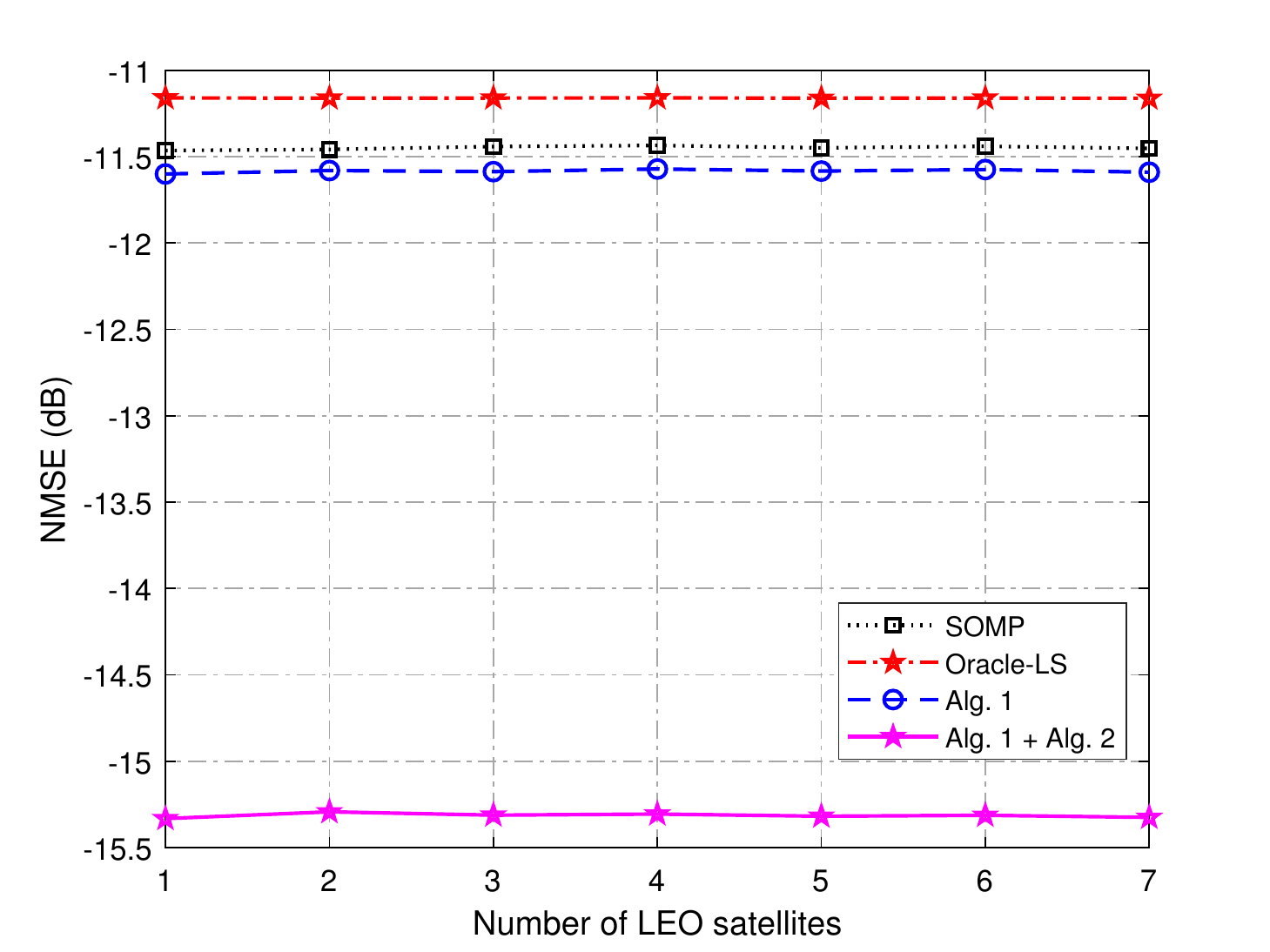}
	}
	\hspace{-9mm}
	\subfigure[AD performance]{
		\vspace{-6mm}
		\includegraphics[width=.345\columnwidth,keepaspectratio]{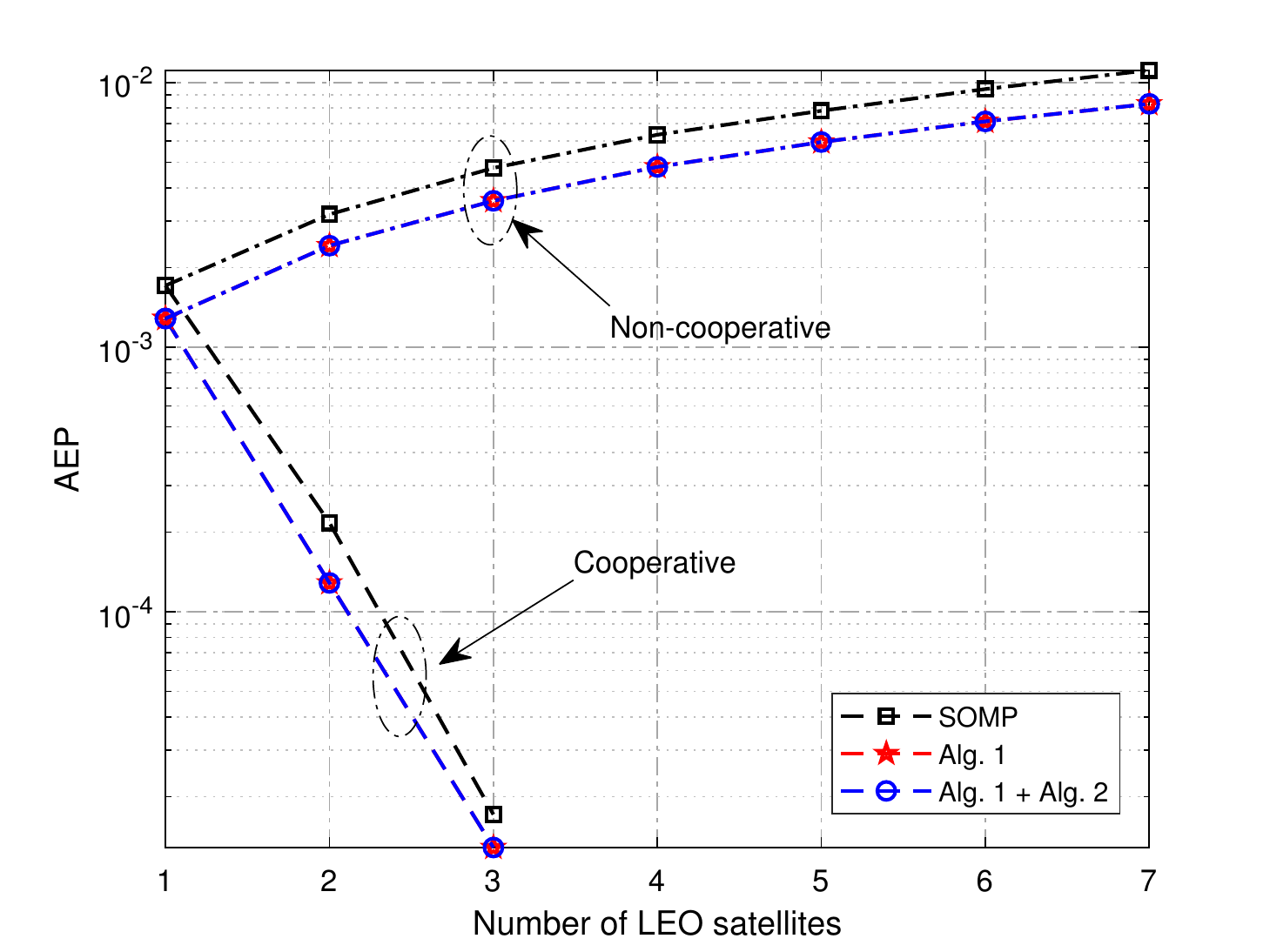}
	}
	\hspace{-9mm}
	\subfigure[DD performance]{
		\vspace{-6mm}
		\includegraphics[width=.345\columnwidth,keepaspectratio]{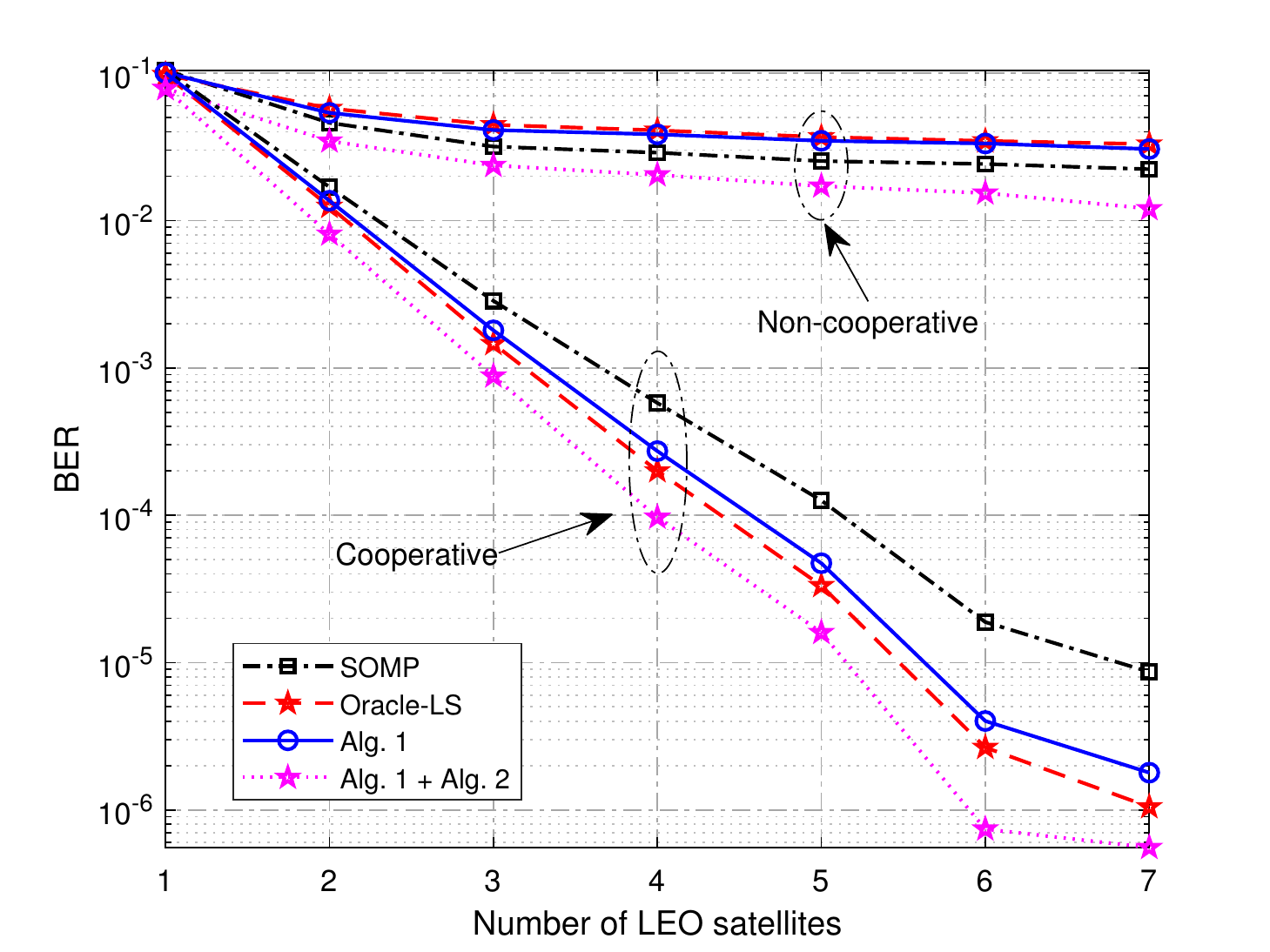}
	}
	\caption{Comparison of different schemes given different numbers of LEO satellites, with $G = 136$, $P = 3$, and SNR $ = 8$ dB.}
	\label{FigSim2}
\end{figure*}

In Fig.~\ref{FigSim2}, the NMSE, AEP, and BER performance as the functions of the numbers of LEO satellites $Q$ are depicted. As expected, increasing the number of LEO satellites does not influence the NMSE performance, since the CE is performed at each satellite independently. The non-cooperative scheme yields a larger AD error with the increased number of LEO satellites and the cooperative scheme behaves conversely. This is because in non-cooperative the AD error is accumulated with the increasing number of LEO satellites, while the cooperative scheme uses a majority scheme to eliminate the detection error. Furthermore, the BER performance becomes better with the increased number of LEO satellites, since more satellites lead to more observations. To determine the concrete number of satellites that participate in the cooperative processing, one may preset the targeted BER and AEP performance, and then the concrete number of satellites needed to achieve this threshold can be read from the figures. 

\begin{figure*}[!h]
	\vspace{-3mm}
	\centering
	\captionsetup{font={footnotesize}, singlelinecheck = off,name={Fig.},justification=centering, labelsep=period}
	\hspace{-4mm}
	\subfigure[CE performance]{
		\vspace{-6mm}
		\includegraphics[width=.345\columnwidth,keepaspectratio]{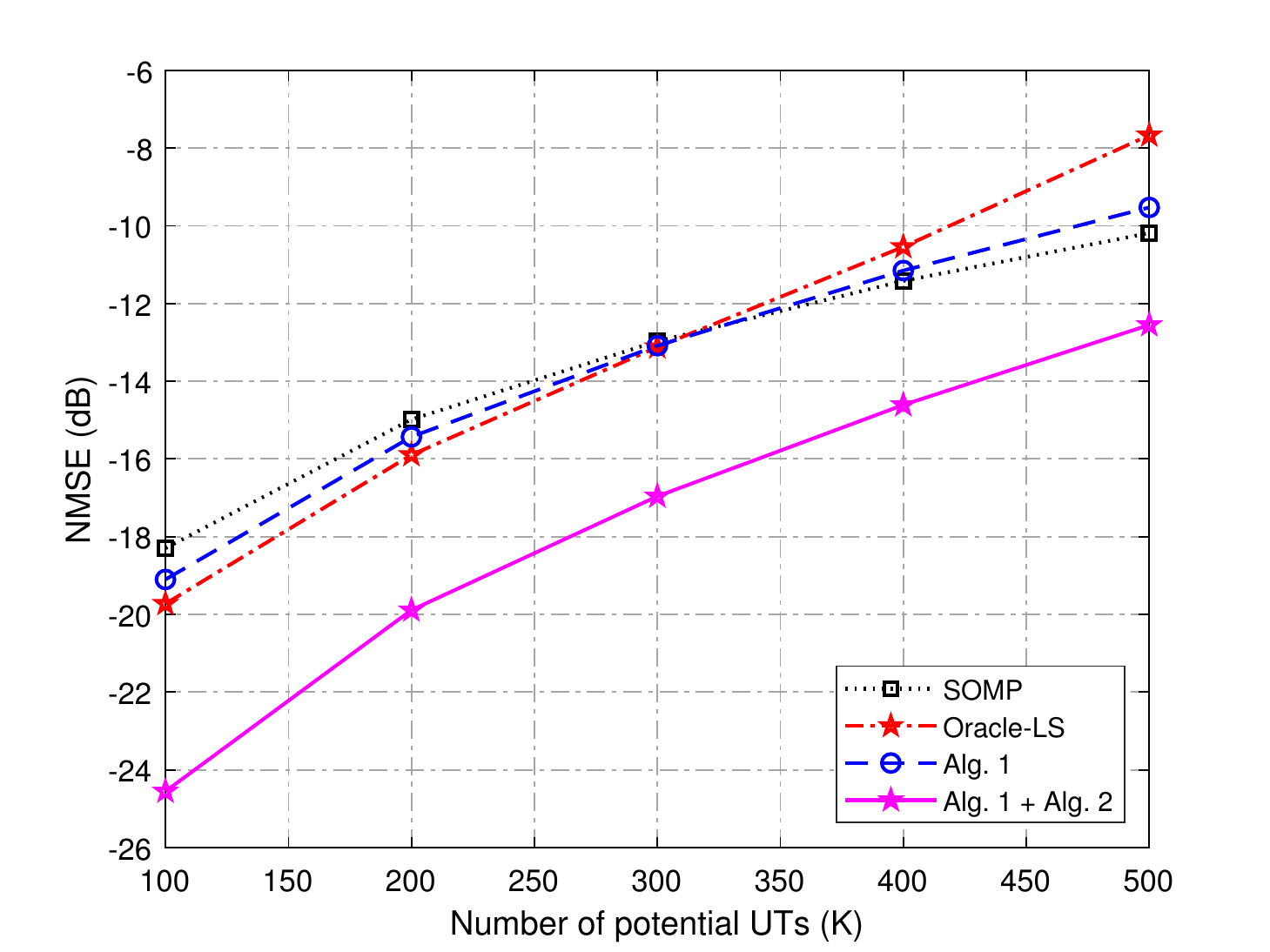}
	}
	\hspace{-9mm}
	\subfigure[AD performance]{
		\vspace{-6mm}
		\includegraphics[width=.345\columnwidth,keepaspectratio]{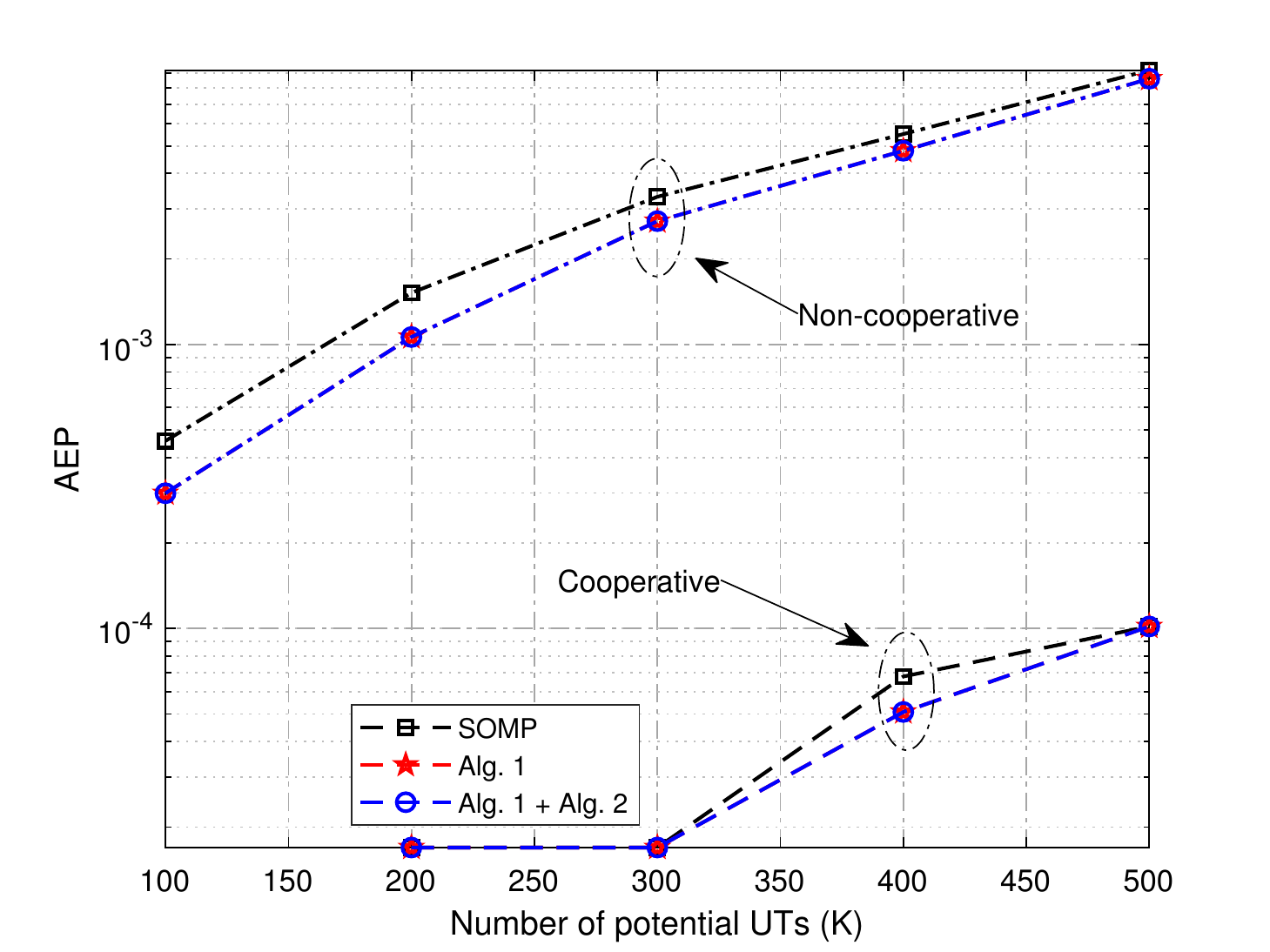}
	}
	\hspace{-9mm}
	\subfigure[DD performance]{
		\vspace{-6mm}
		\includegraphics[width=.345\columnwidth,keepaspectratio]{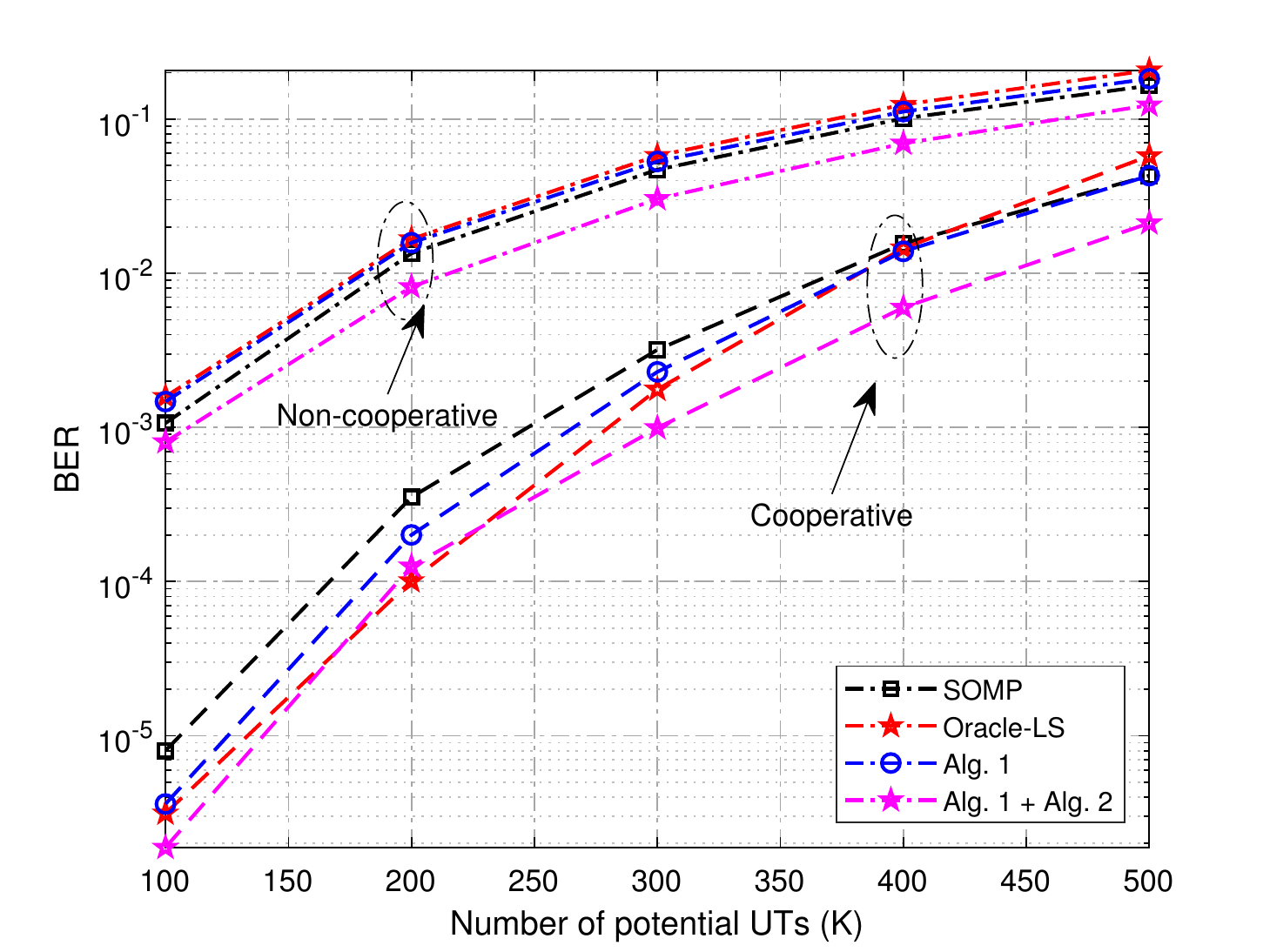}
	}
	\caption{Comparison of different schemes given different numbers of potential UTs, with $G = 136$, $P = 3$ and SNR $ = 12$ dB.}
	\label{FigSim5}
	 \vspace*{-7mm}
\end{figure*}

In Fig.~\ref{FigSim5}, we compare the NMSE, AEP, and BER performance of different schemes as the functions of the numbers of total potential UTs, with the non-ISI region length fixed to $G = 170$, $\text{SNR} = 12$ dB, and the $P = 3$. Besides, we fixed the active UTs ratio to $K_a/K = 0.1$. The proposed Alg. 1 + Alg. 2 still outperforms the other algorithms with the scaling of the $K$. The proposed cooperative AD and DD schemes also outperform non-cooperative schemes. All the schemes exhibit performance deterioration with the increment of $K$. Actually, with the increased number of $K$, we have to allocate a longer preamble to achieve satisfactory performance, which reduces the transmission efficiency unavoidably.

\begin{figure*}[!h]
	\centering
	 \vspace{-5mm}
	\captionsetup{font={footnotesize}, singlelinecheck = off,name={Fig.},justification=centering, labelsep=period}
	\centering
	\includegraphics[scale=0.5]{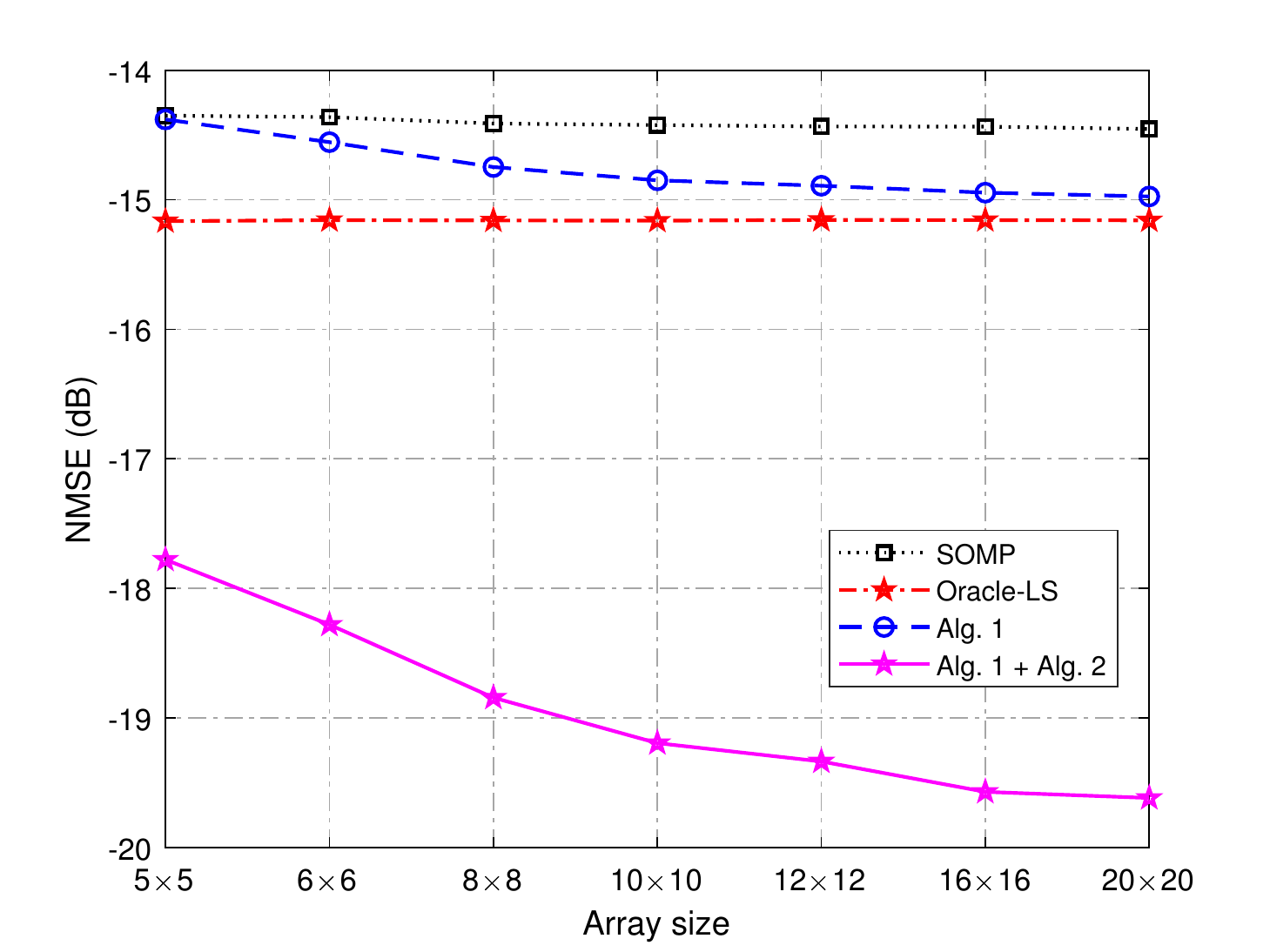}
	\caption{NMSE performance as the functions of the size of receive antenna array, where $G = 136$, $P = 3$ and $\text{SNR} = 12$\,dB.}
	\label{FigSim1}
	\vspace{-5mm}
\end{figure*}

In Fig.~\ref{FigSim1}, we examine the NMSE performance of different CE methods as the functions of the size of receive antenna array, with the non-ISI region length fixed to $G = 136$, $\text{SNR} = 12$ dB, and $P = 3$. As expected, increasing the size of the receive antenna array would lead to enhanced performance of CE in Alg.1 + Alg. 2. This is because a larger size of antenna array leads to more accurate AoA estimation at the receiver, which helps to refine the CE results. By contrast, increased antenna array size can not provide improved CE performance for the Oracle-LS algorithm and SOMP algorithm, since they didn't utilize the spatially structured information. Besides, it is observed that Alg. 1 can also acquire better performance with a larger antenna size due to the effective utilization of MMV structure.

\subsection{Performance Evaluation with Quantized Backhaul}\label{S5.2}
The previous subsection has compared the BER performance of the proposed perfect backhaul schemes, which means that the observations at each edge satellite node are perfectly fed back to the central server node. We now investigate the BER performance of the proposed quantized backhaul schemes, where the observation is quantized first and then fed back. According to whether a terrestrial server or a satellite serves as the central node, cooperative DD can be classified into the cases of MSCTP and MSCBP. The LS method is adopted in perfect backhaul case to provide the performance benchmark for other comparison schemes. 
\begin{figure*}[!h]
	\centering
	\hspace{-5mm}
	\begin{minipage}[t]{0.33\linewidth}
		\centering
		\captionsetup{font={footnotesize}, singlelinecheck = off,name={Fig.},justification=centering, labelsep=period}
		\includegraphics[scale=0.4]{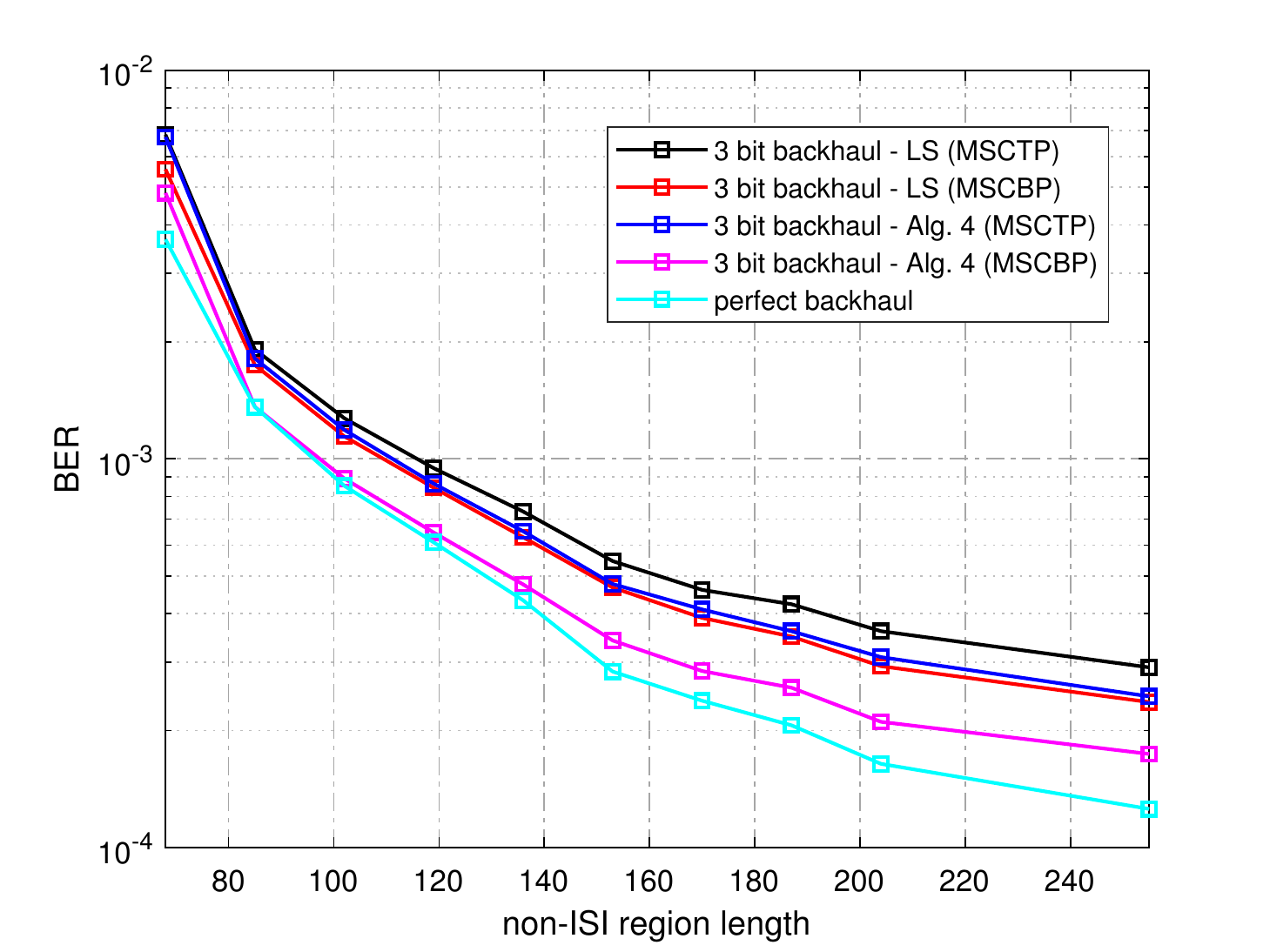}
		\caption{DD performance of different schemes with quantization resolution $\mathcal{B} = 3$ bits, given $\text{SNR} = 12$\,dB, $K_a = 15$ and $P = 3$.}
		\label{Fig11}
	\end{minipage}
	\hfill
	\begin{minipage}[t]{0.33\linewidth}
		\centering
		\captionsetup{font={footnotesize}, singlelinecheck = off,name={Fig.},justification=centering, labelsep=period}
		\includegraphics[scale=0.4]{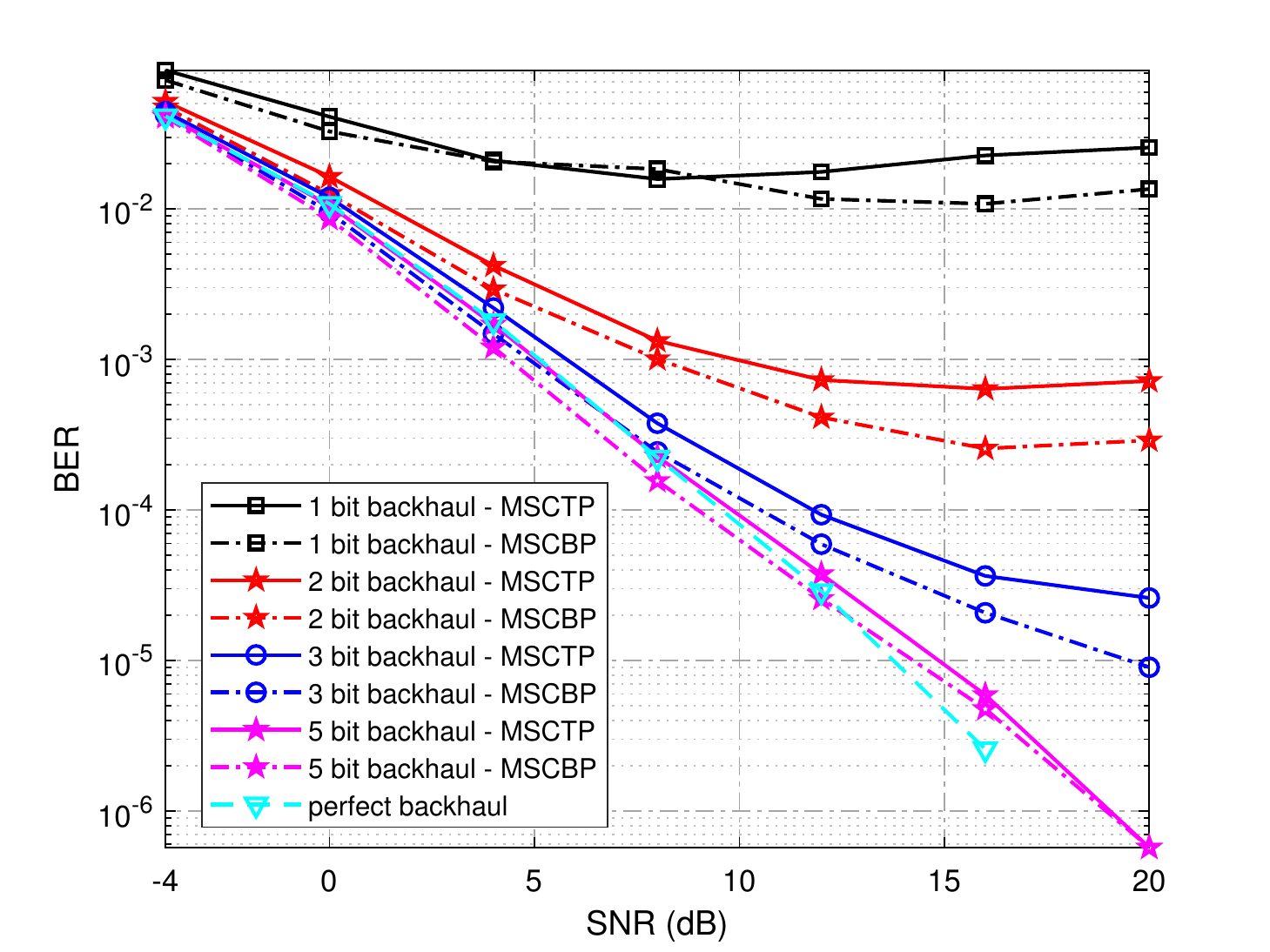}
		\caption{DD performance of Alg. 4 with various quantization resolutions, given $G = 136$, $K_a = 15$ and $P = 3$.}
		\label{Fig12}
	\end{minipage}
	\hfill
	\begin{minipage}[t]{0.33\linewidth}
		\centering
		\captionsetup{font={footnotesize}, singlelinecheck = off,name={Fig.},justification=centering, labelsep=period}
		\includegraphics[scale=0.4]{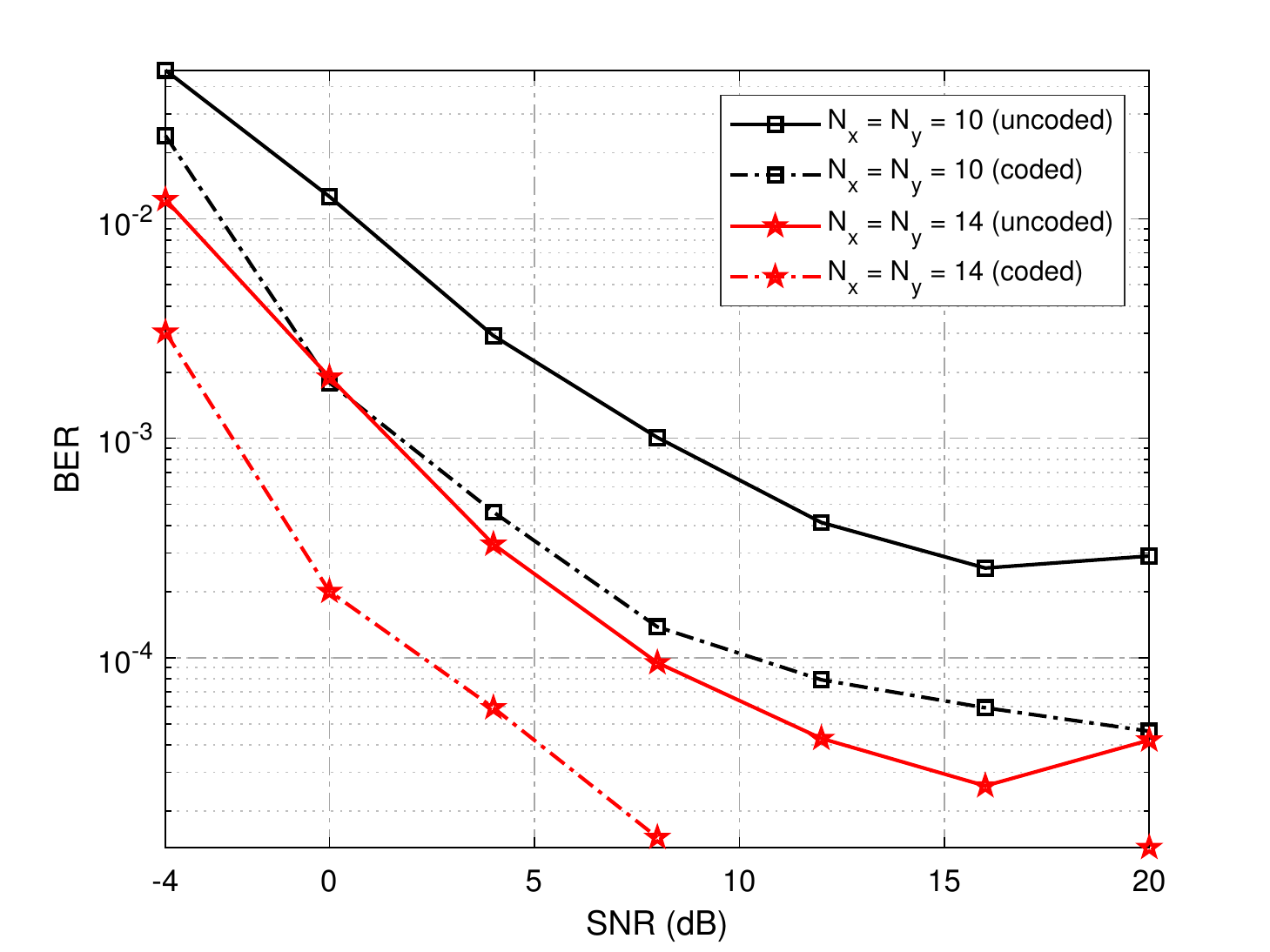}
		\caption{DD performance of Alg.~\ref{alg:quant_DD} (MSCBP), with two different array sizes, uncoded/coded schemes and $\mathcal{B} = 2$ bits, given $G = 136$, $K_a = 15$ and $P = 3$.}
		\label{Fig13}
	\end{minipage}
	 \vspace{-3.0mm}
\end{figure*}
	
Fig.~\ref{Fig11} exhibits the DD performance of different schemes, given the $\text{SNR} = 12$\,dB, the number of active UTs $K_a = 15$, the number of multipath components $P = 3$, and the quantization resolution $\mathcal{B} = 3$\,bits. To validate the effectiveness of the proposed multi-satellite cooperative Bayesian dequantization DD algorithm (Alg.~\ref{alg:quant_DD}), we compare it with the LS method in both  MSCTP and MSCBP cases. In Fig.~\ref{Fig11}, LS (MSCTP) and LS (MSCBP) mean the LS method is directly applied to solve the quantized data detection problems in MSCTP and MSCBP respectively, while Alg.~4 (MSCTP) and Alg.~4 (MSCBP) perform quantized posterior inference according to Alg.~\ref{alg:quant_DD}. In Fig.~\ref{Fig11}, a better BER performance can be observed by Alg.\,4 over LS in both cases, which is due to the better utilization of the prior symbol constellation distribution and the quantizer output. The figure also reveals that given a target BER, Alg. 4 can use much fewer non-ISI region length than LS. Besides, it is also obvious that better BER performance can be achieved in MSCBP scenario. This is because the local observations of the central satellite node in MSCBP can promise a lower information loss than the fully-quantized observations in MSCTP. Moreover, the Alg.\,4~(MSCBP) shows the smallest BER gap compared with the perfect backhaul case, which exhibits the effectiveness of the proposed scheme.
	
Fig.~\ref{Fig12} plots the DD performance of Alg.~\ref{alg:quant_DD} under both the MSCTP and MSCBP scenarios with different quantization resolutions, given $G = 136$, $K_a = 15$ and $P = 3$. It is obvious that as the quantization resolution improves, the BER gap between MSCTP/MSCBP and the perfect backhaul becomes smaller, and MSCBP outperforms MSCTP at all the quantization resolutions. It is also worth noting that in the low SNR region and with the quantization resolution $\mathcal{B} = 5$ bits, the MSCBP framework can achieve better BER than the perfect backhaul. This is because Alg.~\ref{alg:quant_DD} can better utilize the discrete constellation information during iterative inference, while the LS method adopted in the perfect backhaul case can not utilize the prior information of symbol constellation. By contrast, in the high SNR region, the BER performance of the proposed scheme is restricted by the quantized resolution, thus resulting in a worse BER than perfect backhaul.  
	
Fig.~\ref{Fig13} investigates the influence of receiver antenna size and channel coding on the DD performance, given $G = 136$, $K_a = 15$, and $P = 3$. Alg. 4 is employed to perform DD in the MSCBP scenario, the quantization resolution is $\mathcal{B} = 2$\,bits and a low-density LDPC channel coding with $1/2$ code-rate is applied. As expected, channel coding can significantly improve the DD performance. For example, with the array size of $N_x = N_y = 10$ and at $\text{SNR}=20$\,dB, the BER of the uncoded case is about $2\times 10^{-4}$ while the BER of the coded case is around $4 \times 10^{-5}$. Obviously, increasing the array size can also dramatically improve the BER performance, as a larger receiver antenna array can provide more observations for performance enhancement. In the coded situation, for example, an $N_x = N_y  = 14$ array can harvest an SNR gain of nearly 8\,dB at the BER of $10^{-4}$, compared with an $N_x = N_y = 10$ array.
	
\section{Conclusions}\label{S6}

The random access problem in mMIMO-based satellite communication systems has been investigated in this paper. 	Considering the vast geographic distribution region of target UTs, a TSP-MCS has been designed with tolerance to imperfect synchronization. Specifically, we have constructed a multi-satellite system where a TS is utilized to perform JADCE at edge satellite nodes. Exploiting the sparse feature of TSLs and sporadic transmission of UTs' traffic, our system acquires an initial estimate through the OAMP-MMV algorithm, after which a 2D-ESPRIT method is applied for enhanced parameterized CE. Moreover, we have proposed a majority voting method to enhance AD performance by aggregating information from multiple edge nodes. To deal with the high spatial correlation among UTs' channels, a multi-satellite cooperative linear DD algorithm and a multi-satellite cooperative Bayes dequantization DD algorithm have been proposed for perfect backhaul and quantized backhaul, respectively. Simulation results have verified the effectiveness of our proposed scheme in terms of CE, AD, and DD for quasi-synchronous random access satellite systems. One deficiency of the proposed scheme is that it is limited to the case where the total number of potential UTs is fixed, and it can not adapt to the variation of newly enrolled/exited UTs. One possible solution is to investigate unsourced multiple access methods in LEO satellite constellations. Extensions to the hybrid mMIMO receiver and the designs of receiving beams at satellites for improved performance are also interesting directions.

\begin{appendices}
\section{Auxiliary matrices for 2D-ESPRIT algorithm}\label{apd1}
Define the exchange matrix $\bm{\Pi}_N\in \mathbb{R}^{N\times N}$ and the selection matrix $\bm{J}_{2}\in \mathbb{R}^{\left(N-1\right)\times N}$ as 
\begin{equation}
	\bm{\Pi}_{N} = 
	\left[
	\begin{matrix}
		&  &   & 1 \\
		&  & \begin{turn}{90} $\ddots$ \end{turn} &  \\
		& 1  &   &   \\
		1 &   &   &  
	\end{matrix}
	\right], 
	\bm{J}_{2} = 
	\left[
	\begin{matrix}
		1 & 0 & \dots & 0 & 0 \\
		0 & 1 & \dots & 0 & 0 \\
		\vdots & \vdots & \ddots & \vdots & \vdots \\
		0 & 0 & \dots & 1 & 0 
	\end{matrix}
	\right]. 
\end{equation}
For even-dimensional and odd-dimensional matrices, $\bm{Q}_{2K}$ and $\bm{Q}_{2K+1}$ are respectively defined as
\begin{subequations}
	\begin{align}
		\bm{Q}_{2K} &= 
		\frac{1}{\sqrt{2}}
		\left[
		\begin{matrix}
			\bm{I}_{k} & j \bm{I}_{k}  \\
			\bm{\Pi}_{k} & -j \bm{\Pi}_{k}  
		\end{matrix}
		\right],
		\\
		\bm{Q}_{2K+1} &= 
		\frac{1}{\sqrt{2}}
		\left[
		\begin{matrix}
			\bm{I}_{k} & \bm{0} & j \bm{I}_{k}  \\
			\bm{0}^{T} & \sqrt{2} & \bm{0}^{T}  \\
			\bm{\Pi}_{k} & \bm{0} & -j \bm{\Pi}_{k}  
		\end{matrix}
		\right]. 
	\end{align}
\end{subequations}
Other auxiliary matrices are defined as follows.
\begin{subequations}
	\begin{align}
		\bm{K}_{\mu_{1}} &= 
		\bm{I}_{M_{y}^{\mathrm{sub}}} \otimes \mathcal{R}e\left\{\bm{Q}_{M_x^{\mathrm{sub}}-1}^{H}\bm{J}_2\bm{Q}_{M_x^{\mathrm{sub}}}\right\}, 
		\\
		\bm{K}_{\mu_{2}} &= 
		\bm{I}_{M_{y}^{\mathrm{sub}}} \otimes \mathcal{I}m\left\{\bm{Q}_{M_x^{\mathrm{sub}}-1}^{H}\bm{J}_2\bm{Q}_{M_x^{\mathrm{sub}}}\right\}, 
		\\
		\bm{K}_{\nu_{1}} &= 
		\mathcal{R}e\left\{\bm{Q}_{M_y^{\mathrm{sub}}-1}^{H}\bm{J}_2\bm{Q}_{M_y^{\mathrm{sub}}}\right\} \otimes \bm{I}_{M_{x}^{\mathrm{sub}}}, 
		\\
		\bm{K}_{\nu_{2}} &= 
		\mathcal{I}m\left\{\bm{Q}_{M_y^{\mathrm{sub}}-1}^{H}\bm{J}_2\bm{Q}_{M_y^{\mathrm{sub}}}\right\} \otimes \bm{I}_{M_{x}^{\mathrm{sub}}}. 
	\end{align}
\end{subequations}
\end{appendices}


\begin{thebibliography}{10}
\providecommand{\url}[1]{#1}
\csname url@samestyle\endcsname
\providecommand{\newblock}{\relax}
\providecommand{\bibinfo}[2]{#2}
\providecommand{\BIBentrySTDinterwordspacing}{\spaceskip=0pt\relax}
\providecommand{\BIBentryALTinterwordstretchfactor}{4}
\providecommand{\BIBentryALTinterwordspacing}{\spaceskip=\fontdimen2\font plus
\BIBentryALTinterwordstretchfactor\fontdimen3\font minus
  \fontdimen4\font\relax}
\providecommand{\BIBforeignlanguage}[2]{{%
\expandafter\ifx\csname l@#1\endcsname\relax
\typeout{** WARNING: IEEEtran.bst: No hyphenation pattern has been}%
\typeout{** loaded for the language `#1'. Using the pattern for}%
\typeout{** the default language instead.}%
\else
\language=\csname l@#1\endcsname
\fi
#2}}
\providecommand{\BIBdecl}{\relax}
\BIBdecl

\bibitem{SAGIN_survey} 
J.~Liu, \emph{et~al.}, ``Space-air-ground integrated network: A survey,'' \emph{IEEE Commun. Surveys Tuts.}, vol.~20, no.~4, pp.~2714--2741, 4th Quart. 2018.

\bibitem{ShanzhiChen} 
S.~Chen, S.~Sun, and S.~Kang, ``System integration of terrestrial mobile communication and satellite communication -- The trends, challenges and key technologies in B5G and 6G,'' \emph{China Commun.}, vol.~17, no.~12, pp.~156--171, Dec. 2020.

\bibitem{NTN} 
M.~Hosseinian, \emph{et~al.}, ``Review of 5G NTN standards development and technical challenges for satellite integration with the 5G network,'' \emph{IEEE Aerosp. Electron. Syst. Mag.}, vol.~36, no.~8, pp.~22--31, Aug. 2021.

\bibitem{3gpp.36.300} 
3GPP, ``Evolved Universal Terrestrial Radio Access (E-UTRA) and Evolved Universal Terrestrial Radio Access Network (E-UTRAN); Overall description,'' 3rd Generation Partnership Project (3GPP), Technical Specification (TS) 36.300, 01 2022, version 16.7.0.

\bibitem{LSC_Mag} 
S.~Liu, \emph{et~al.}, ``LEO satellite constellations for 5G and beyond: How will they reshape vertical domains?'' \emph{IEEE Commun. Mag.}, vol.~59, no.~7, pp.~30--36, Jul. 2021.

\bibitem{nb-iot} 
Y.-P.~E.~Wang, \emph{et~al.}, ``A primer on 3GPP narrowband Internet of Things,'' \emph{IEEE Commun. Mag.}, vol.~55, no.~3, pp.~117--123, Mar. 2017.

\bibitem{IoT_app1} 
F.~Guo, \emph{et~al.}, ``Enabling massive IoT toward 6G: A comprehensive survey,'' \emph{IEEE Internet Things J.}, vol.~8, no.~15, pp.~11891--11915, Aug. 2021.

\bibitem{LEO_IoT_Access} 
Z.~Qu, \emph{et~al.}, ``LEO satellite constellation for Internet of Things,'' \emph{IEEE Access}, vol.~5, pp.~18391--18401, Aug. 2017.

\bibitem{LL_TSP} 
L.~Liu and W.~Yu, ``Massive connectivity with massive MIMO -- Part~I: Device activity detection and channel estimation,'' \emph{IEEE Trans. Signal Process.}, vol.~66, no.~11, pp.~2933--2946, Jun. 2018.

\bibitem{Ke_JSAC} 
M.~Ke, \emph{et~al.}, ``Massive access in cell-free massive MIMO-based Internet of Things: Cloud computing and edge computing paradigms,'' \emph{IEEE J. Sel. Areas Commun.}, vol.~39, no.~3, pp.~756--772, Mar. 2021.

\bibitem{OAMP_MMV} 
Y.~Cheng, L.~Liu, and L.~Ping, ``Orthogonal AMP for massive access in channels with spatial and temporal correlations,'' \emph{IEEE J. Sel. Areas Commun.}, vol.~39, no.~3, pp.~726--740, Mar. 2021.

\bibitem{YK_Mei} 
Y.~Mei, \emph{et~al.}, ``Compressive sensing-based joint activity and data detection for grant-free massive IoT access,'' \emph{IEEE Trans. Wireless Commun.}, vol.~21, no.~3, pp.~1851--1869, Mar. 2022.

\bibitem{Zhang_IOTJ} 
Z.~Zhang, \emph{et~al.}, ``User activity detection and channel estimation for grant-free random access in LEO satellite-enabled Internet of Things,'' \emph{IEEE Internet Things J.}, vol.~7, no.~9, pp.~8811--8825, Sep. 2020.

\bibitem{YouLi_JSAC} 
L.~You, \emph{et~al.}, ``Massive MIMO transmission for LEO satellite communications,'' \emph{IEEE J. Sel. Areas Commun.}, vol.~38, no.~8, pp.~1851--1865, Aug. 2020.

\bibitem{KXLi_Tcomm} 
K.-X.~Li, \emph{et~al.}, ``Downlink transmit design for massive MIMO LEO satellite communications,'' \emph{IEEE Trans. Commun.}, vol.~70, no.~2, pp.~1014--1028, Feb. 2022.

\bibitem{XJ_Ding} 
X. Ding, \emph{et~al.}, ``Spectrum reconstruction via deep convolutional neural networks for satellite communication systems,'' \emph{IEEE Trans. Wireless Commun},  vol. 70, no. 9, pp. 5989-6001, Sept. 2022. 

\bibitem{JH_Zhao} 
J.~Zhao, \emph{et~al.}, ``Computation offloading and resource allocation for cloud assisted mobile edge computing in vehicular networks,'' \emph{IEEE Trans. Veh. Technol.}, vol.~68, no.~8, pp.~7944--7956, Aug. 2019.

\bibitem{offload} 
Q.~Tang, \emph{et~al.}, ``Computation offloading in LEO satellite networks with hybrid cloud and edge computing,'' \emph{IEEE Internet Things J.}, vol.~8, no.~11, pp.~9164--9176, Jun. 2021.


\bibitem{MSC_RA} 
B.~Zhao, G.~Ren, and H.~Zhang, ``Multisatellite cooperative random access scheme in low earth orbit satellite networks,'' \emph{IEEE Syst. J.}, vol.~13, no.~3, pp.~2617--2628, Sept. 2019.

\bibitem{DM_MIMO} 
M.~Y.~Abdelsadek, G.~K.~Kurt, and H.~Yanikomeroglu, ``Distributed massive MIMO for LEO Satellite networks,'' \emph{IEEE Open J. Commun. Soc.}, vol.~3, pp.~2162--2177, Nov. 2022.

\bibitem{XY_Zhou} 
X.~Zhou, \emph{et~al.}, ``Active terminal identification, channel estimation, and signal detection for grant-free NOMA-OTFS in LEO satellite Internet-of-Things,'' \emph{IEEE Trans. Wireless Commun},  \textit{early access}, Oct. 2022.
\bibitem{UserGrouping} 
O.~Kodheli, \emph{et~al.}, ``An uplink UE group-based scheduling technique for 5G mMTC systems over LEO satellite,'' \emph{IEEE Access}, vol.~7, pp.~67413--67427, May 2019.
\bibitem{Zhen_TDS} 
Z.~Gao, \emph{et~al.}, ``Priori-information aided iterative hard threshold: A low-complexity high-accuracy compressive sensing based channel estimation for TDS-OFDM,'' \emph{IEEE Trans. Wireless Commun.}, vol.~14, no.~1, pp.~242--251, Jan. 2015.

\bibitem{Zhen_BC} 
Z.~Gao, C. Zhang, and Z. Wang, ``Robust preamble design for synchronization, signaling transmission, and channel estimation,'' \emph{IEEE Trans. Broadcast.}, vol.~61, no.~1, pp.~94--104, Mar. 2015.

\bibitem{2DESPRIT} 
M.~D.~Zoltowski, M.~Haardt, and C.~P.~Mathews, ``Closed-form 2-D angle estimation with rectangular arrays in element space or beamspace via unitary ESPRIT,'' \emph{IEEE Trans. Signal Process.}, vol.~44, no.~2, pp.~316--328, Feb. 1996.


\bibitem{Starlink} 
S.~J.~Mazlouman and K.~Schulze, ``Distributed phase shifter array system and method,'' U.S. Patent 0\,241\,122, Aug. 23, 2018.

\bibitem{3gpp.38.811} 
3GPP, ``Technical Specification Group Radio Access Network; Study on New Radio (NR) to support non-terrestrial networks,'' 3rd Generation Partnership Project (3GPP), Technical Report (TR) 38.811, 09 2020, version 15.4.0.

\bibitem{Doppler_Comp1} 
A.~Guidotti, \emph{et~al.}, ``Satellite-enabled LTE systems in LEO constellations,'' in \emph{Proc. ICC  Workshops 2017} (Paris, France), May~21-25, 2017, pp.~876--881.

\bibitem{Doppler_Comp2} 
O.~Kodheli, A.~Guidotti, and A.~Vanelli-Coralli, ``Integration of satellites in 5G through LEO constellations,'' in \emph{Proc. GLOBECOM 2017} (Singapore), Dec.~4-8, 2017, pp.~1--6.

\bibitem{OAMP} 
J.~Ma and L.~Ping, ``Orthogonal AMP,'' \emph{IEEE Access}, vol.~5, pp.~2020--2033, Jan. 2017.

\bibitem{Meng_AMP_init} 
S.~Wu, \emph{et~al.}, ``Block expectation propagation for downlink channel estimation in massive MIMO systems,'' \emph{IEEE Commun. Lett.}, vol.~20, no.~11, pp.~2225--2228, Nov.. 2016.

\bibitem{Ke_TSP} 
M.~Ke, \emph{et~al.}, ``Compressive sensing-based adaptive active user detection and channel estimation: Massive access meets massive MIMO,'' \emph{IEEE Trans. Signal Process.}, vol.~68, pp.~764--779, Jan. 2020.

\bibitem{Liao_Tcom} 
A.~Liao, \emph{et~al.}, ``Closed-loop sparse channel estimation for wideband millimeter-wave full-dimensional MIMO systems,'' \emph{IEEE Trans. Commun.}, vol.~67, no.~12, pp.~8329--8345, Dec. 2019.

\bibitem{dft_s_ofdm} 
A.~Sahin, \emph{et~al.}, ``Flexible DFT-S-OFDM: Solutions and challenges,'' \emph{IEEE Commun. Mag.}, vol.~54, no.~11, pp.~106--112, Nov. 2016.

\bibitem{CSI_feedback1} 
X.~Rao and V.~K.~N.~Lau, ``Distributed compressive CSIT estimation and feedback for FDD multi-user massive MIMO systems,'' \emph{IEEE Trans. Signal Process.}, vol.~62, no.~12, pp.~3261--3271, Jun. 2014.

\bibitem{CSI_feedback2} 
C.-K.~Wen, W.-T.~Shih, and~S.~Jin, ``Deep learning for massive MIMO CSI feedback,'' \emph{IEEE Wireless Commun. Lett.}, vol.~7, no.~5, pp.~748--751, Oct. 2018.

\bibitem{VI} 
S.~S.~Thoota and C.~R.~Murthy, ``Variational Bayes' joint channel estimation and soft symbol decoding for uplink massive MIMO systems with low resolution ADCs,'' \emph{IEEE Trans. Commun.}, vol.~69, no.~5, pp.~3467--3481, May 2021.

\bibitem{Bayes} 
C.-K.~Wen, \emph{et~al.}, ``Bayes-optimal joint channel-and-data estimation for massive MIMO with low-precision ADCs,'' \emph{IEEE Trans. Signal Process.}, vol.~64, no.~10, pp.~2541--2556, May 2016.

\bibitem{SpaceX} 
{Federal Communications Commission report}, (2016). ``Space{X} non-geostationary satellite system,''
  https://fcc.report/IBFS/SAT-LOA-20161115-00118/1158350.pdf.

\bibitem{RKinf} 
Y. Huang, \emph{et~al.},``Dynamic resource configuration for low-power IoT networks: A multi-objective reinforcement learning method,'' \emph{IEEE Commun. Lett.},  vol. 25, no. 7, pp. 2285-2289, Jul. 2021.

\bibitem{SOMP} 
J.-F.~Determe, \emph{et~al.}, ``On the noise robustness of simultaneous orthogonal matching pursuit,'' \emph{IEEE Trans. Signal Process.}, vol.~65, no.~4, pp.~864--875, Feb. 2017.

\bibitem{Oracle_LS} 
Z.~Gao, \emph{et~al.}, ``Spatially common sparsity based adaptive channel estimation and feedback for FDD massive MIMO,'' \emph{IEEE Trans. Signal Process.}, vol.~63, no.~23, pp.~6169--6183, Dec. 2015.

\end{thebibliography}
\end{document}